Increased resistance to photooxidation in Dion-Jacobson lead halide perovskites – implication for perovskite device stability

Zhilin Ren,[1#] Juraj Ovčar,[2#] Tik Lun Leung,[1,3,4#] Yanling He,[1] Yin Li,[1] Dongyang Li,[5] Xinshun Qin,[1] Hongbo Mo,[1] Zhengtian Yuan,[1] Jueming Bing,[3,4] Martin P. Bucknall,[6] Luca Grisanti,[2] Muhammad Umair Ali,[1] Peng Bai,[5] Tao Zhu,[5] Ali Ashger Syed,[1] Jingyang Lin,[1,7] Jingbo Wang, Abdul-Khaleed,[1] Wenting Sun,[1] Gangyue Li,[7] Gang Li,[5] Alan Man Ching Ng,[7] Anita W. Y. Ho-Baillie,[3,4] Ivor Lončarić,[2*] Jasminka Popović,[2*] Aleksandra B. Djurišić[1*]

[1]Department of Physics, The University of Hong Kong, Pokfulam, Hong Kong SAR

[2]Ruđer Bošković Institute, Bijenička 54, 10000 Zagreb, Croatia

[3]School of Physics, The University of Sydney, Sydney, NSW 2006, Australia

[4]Sydney Nano, The University of Sydney, Sydney, NSW 2006, Australia

[5] Department of Electrical and Electronic Engineering, Research Institute for Smart Energy (RISE), The Hong Kong Polytechnic University, 11 Yuk Choi Rd, Hung Hom, Hong Kong SAR

[6] Bioanalytical Mass Spectrometry Facility, Mark Wainwright Analytical Centre, UNSW Sydney, NSW 2052, Australia

[7]Department of Physics and Core Research Facilities, Southern University of Science and Technology, No. 1088, Xueyuan Rd., Shenzhen, 518055, Guangdong, P.R. China

[#]These authors contributed to this work equally.

**Abstract:** 2D metal halide perovskites have enabled significant stability improvements in perovskite devices, particularly in resistance to moisture. However, some 2D perovskites are even more susceptible to photooxidation compared to 3D perovskites. This is particularly true for more commonly investigated Ruddlesden-Popper (RP) perovskites that exhibit increased susceptibility to photoinduced degradation compared to Dion-Jacobson (DJ) perovskites. Comparisons between different RP and DJ perovskites reveal that this phenomenon cannot be explained by commonly proposed differences in superoxide ion generation, interlayer distance and lattice structural rigidity differences. Instead, the resistance to photooxidation of DJ perovskites can be attributed to decreased likelihood of double deprotonation events (compared to single deprotonation events in RP perovskites) required for the loss of organic cations and the perovskite decomposition. Consequently, DJ perovskites are less susceptible to oxidative degradation (both photo- and electrochemically induced), which leads to improved operational stability of solar cells based on these materials.

## Context and Scale

While photo/electrochemical stability of 3D metal halide perovskites has been extensively studied, the stability of 2D perovskites is not as well understood despite their common use to improve operational stability of perovskite devices. In this work, we investigate photooxidation of different 2D Ruddlesden-Popper (RP) and Dion-Jacobson (DJ) perovskites to elucidate mechanisms of their degradation under illumination. We demonstrate that improved stability of DJ compared to RP perovskites can be attributed to reduced loss of spacer cations, rather than differences in lattice spacing, structural rigidity, superoxide ion generation, or charge localization/oxidation of iodide. Increased stability of 2D DJ films results in increased stability of 3D/2D films and solar cells for three different 3D perovskite compositions, demonstrating generality of the approach and providing a strategy for stability improvement, namely the use of divalent organic cations and minimizing hole accumulation.

## INTRODUCTION

2D metal halide perovskites, in the form of 2D/3D or quasi-2D active layers, have been proposed to replace more commonly used 3D metal halide perovskites in photovoltaic and (opto)electronic devices with improved stability.[1-3] 3D perovskites with the formula $ABX_3$, where A is $Cs^+$ or a small organic cation (methylammonium (MA) or formamidinium (FA)), B is divalent metal (most commonly $Pb^{2+}$), and X is halide anion, are known to be unstable upon exposure to ambient (oxygen, moisture), heat, and illumination.[2] While 2D perovskites, containing bulky spacer cations separating layers of corner-sharing $[BX_6]^{4-}$ octahedra, typically exhibit improved stability compared to 3D perovskites, further improvements in stability and understanding of their degradation mechanisms are still required.[1,2]

The stability improvements in solar cells (SCs) and light-emitting diodes (LEDs) with 2D perovskites are commonly attributed to increased hydrophobicity and reduced ion migration.[1-3] While stability of 3D perovskites, including photooxidation, has been extensively studied,[4-12] investigations of the stability of 2D perovskites,[13-21] particularly under illumination and/or presence of oxygen,[13-15,20] are scarce. In 3D perovskites, it has been established that perovskite degradation involves hole trapping or accumulation.[22-25] Excess holes oxidize iodide, which starts the chain of electrochemical reactions (iodide oxidation, organic cation deprotonation) which ultimately lead to perovskite degradation (**Supplementary Note 1**). Hole accumulation is also responsible for accelerated performance degradation under open-circuit conditions[4] (**Supplementary Note 2**), as well as halide segregation in mixed perovskite materials,[22-28] driven by photo- and/or electrochemically-induced holes.[22,23] The electrochemical redox reactions are exacerbated by exposure to oxygen, since oxygen acts as an electron scavenger which causes the excess of photogenerated holes in the perovskite, which are the primary source of degradation.[22] The degradation is further accelerated by exposure to both oxygen and moisture.[29-32]

While 2D materials offer improved resistance to moisture, they can still be sensitive to photooxidation and/or degrade under illumination.[13,15,16,20] In some cases, 2D materials, such as 2D $PEA_2PbI_4$ (PEA denotes phenethyl ammonium), 2D $BA_2PbI_4$ (BA denotes butyl ammonium) and quasi-2D $BA_2MA_{n-1}Pb_nI_{3n+1}$ ($n$=2,3), are less stable compared to the corresponding 3D perovskite[4,16,20] and the degradation can be observed for oxygen levels as low as 100 ppm.[20] Due to the observed instability of 2D materials under illumination, their suitability for stability enhancements has been questioned recently.[24] It is therefore important to understand photo- and electrochemical behavior of these materials, as they remain commonly proposed strategies to mitigate ion migration and reduce perovskite degradation.[1-3,22]

## RESULTS AND DISCUSSION

### *Photochemical stability of 2D perovskites*

There are two distinct classes of 2D/quasi-2D perovskite materials: Ruddlesden-Popper (RP) with formula $C_2A_{n-1}B_nX_{3n+1}$ and Dion-Jacobson (DJ) with formula $CA_{n-1}B_nX_{3n+1}$, where C is the bulky spacer cation, and $n$ is the number of octahedral layers. As the offset between adjacent octahedral layers in RP and DJ halide perovskites can deviate from an ideal one observed in analogous 2D oxide perovskites, we classify different 2D perovskite materials as RP perovskites when $[BX_6]^{4-}$ octahedra are separated by a bilayer of monovalent organic cations bound with weak van der Waals forces, and as DJ perovskites when $[BX_6]^{4-}$ octahedra are separated by a single layer of divalent organic cations which form hydrogen bonds with adjacent octahedral layers.[19,21] While it was claimed that quasi-2D RP and DJ perovskites have comparable stability under illumination,[19] DJ perovskites consistently result in more stable SCs, LEDs, and photodetectors (**Supplementary Note 3**). Thus, we investigated the susceptibility of these materials to photooxidation to elucidate mechanisms of degradation and reasons for stability differences.

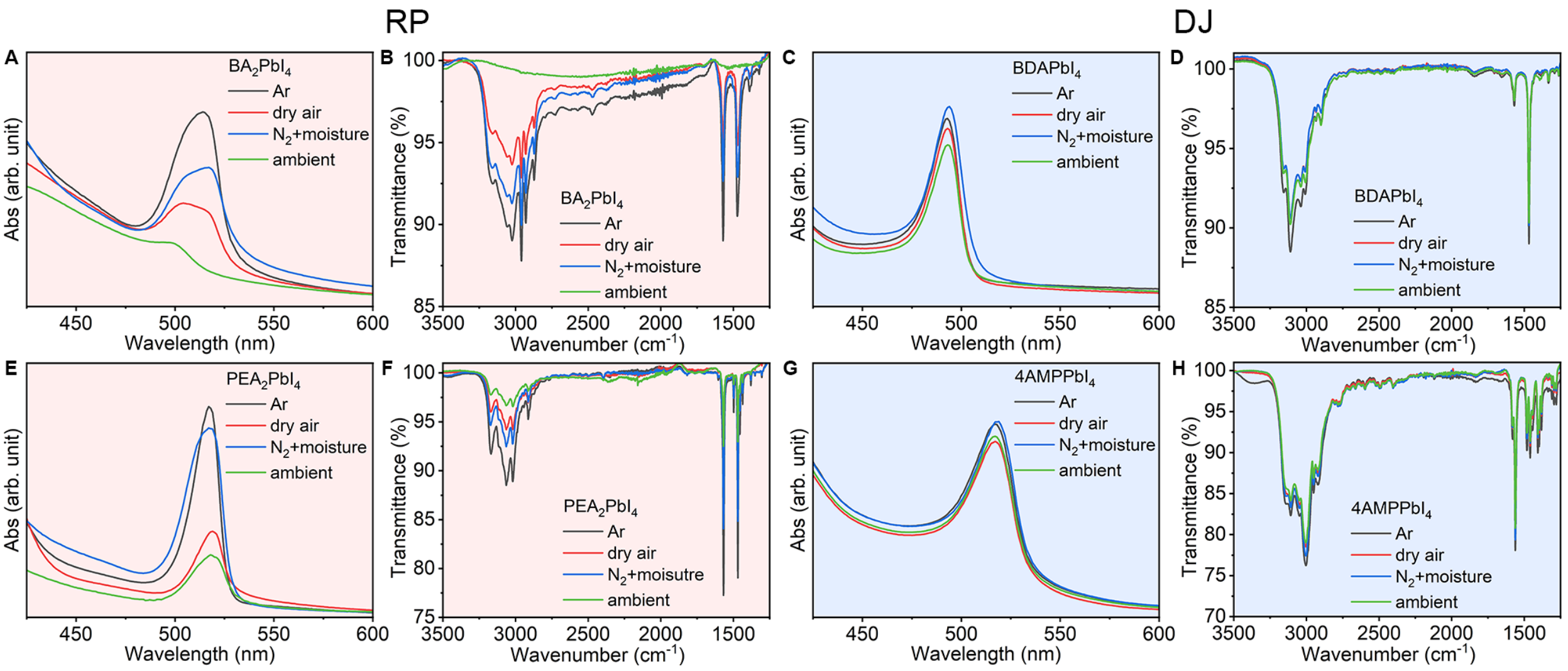

**Figure 1.** Absorption and FTIR spectra, respectively, of (A), (B) $BA_2PbI_4$ (C), (D) BDAPbI$_4$ (E), (F) $PEA_2PbI_4$ and (G), (H) 4AMPPbI$_4$ exposed to simulated solar illumination (100 mW/cm$^2$) in different atmospheres. For ambient atmosphere, RH was 60%, for $N_2$+moisture RH was 40%.

We observe dramatic differences in the photostability of six DJ vs. six RP 2D perovskites illuminated for 3 h (**Figures S1-S4**). DJ perovskites exhibit good photostability for both bromides and iodides. All RP iodide perovskites exhibit significant degradation, while RP bromide perovskites exhibit different trends depending on the spacer cation. In those RP bromide perovskites which degrade, the loss of spacer cation is observed (**Figure S5**). While all alkyl chain RP spacers exhibited poor stability, the presence of an aromatic ring does not guarantee stability (**Figure S6**). As iodide perovskites are more relevant for SCs and exhibit clear difference between RP (poor stability) and DJ (good stability) materials, we will focus on iodide perovskites.

There are two possible uses of RP or DJ perovskites in solar cells, as 3D/2D (3D perovskite capped with a 2D film) or quasi-2D (commonly $n$=3-5) absorber films. Here we will focus on stability of 2D films for applications in devices with 3D/2D absorber layers, since quasi-2D films typically contain multiple $n$ phases and exhibit complex behavior under illumination (**Figure S7 & S8**),[18,33-35] as both degradation and phase transformation/disproportionation can occur since both small cations and bulky spacers are mobile in RP perovskites.[34,35] The process is further complicated by moisture exposure which facilitates phase transformations.[36] The presence of moisture can further complicate the stability comparisons between RP and DJ perovskites, since quasi-2D DJ perovskites have lower moisture resistance[37] and can exhibit formation of new phases in ambient air (**Figure S9**). Thus, to isolate the factors responsible for stability differences between DJ and RP perovskites, we focus on 2D materials ($n$=1) since they cannot exhibit disproportionation, and perform experiments in different atmospheres to distinguish effect of oxygen and moisture, as shown in **Figure 1**. Quasi-2D perovskites exhibit similar trends on atmosphere composition (**Figure S10**), and thus degradation mechanisms elucidated for 2D materials should be generally applicable.

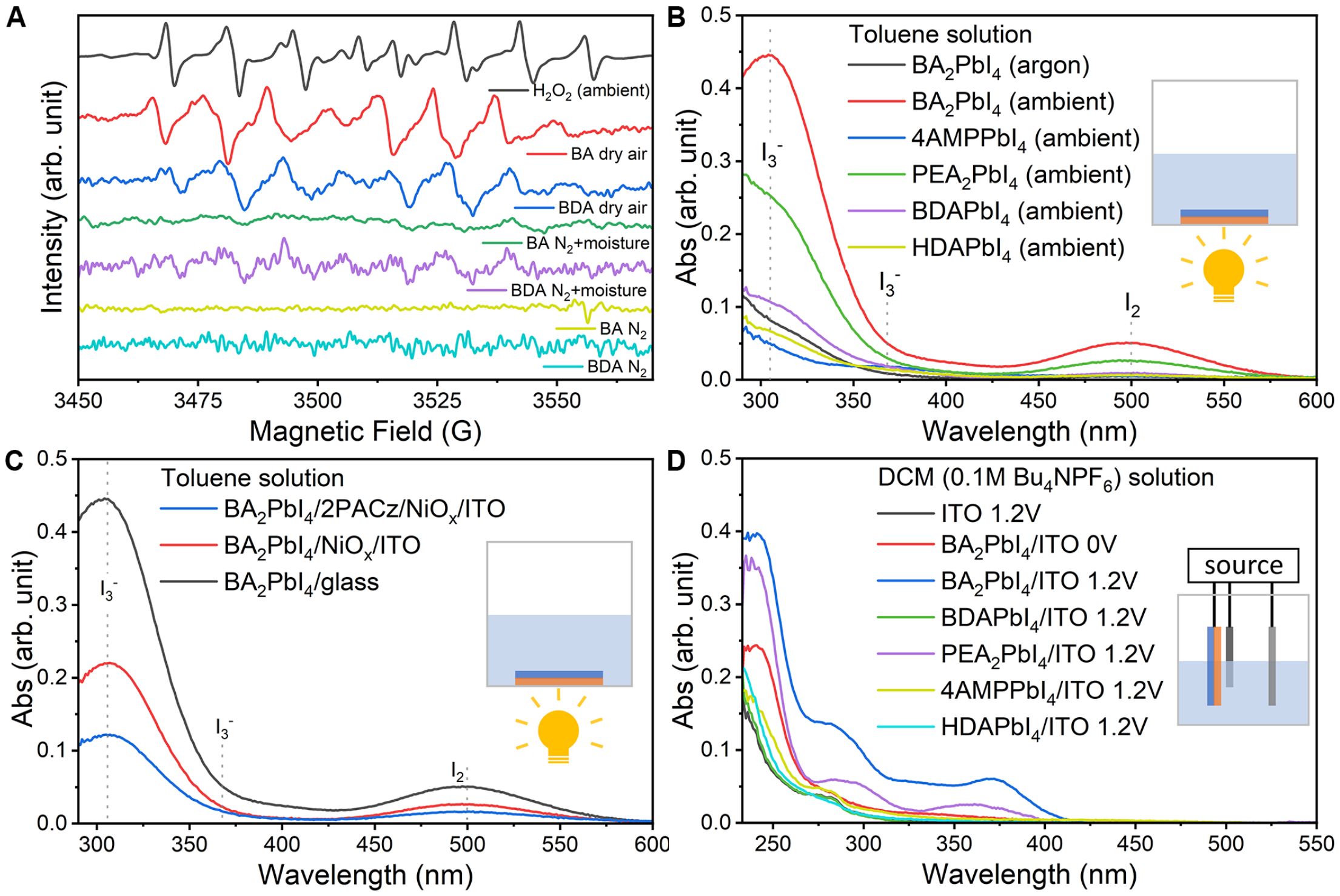

**Figure 2**. (A) EPR spectra of $BA_2PbI_4$ and $BDAPbI_4$ perovskite samples with 5-(Diethoxyphosphoryl)-5-methyl-1-pyrroline-N-oxide (DEPMPO) under illumination in dry air, $N_2$ and $N_2$+$H_2O$ are also shown. EPR spectrum of DEPMPO with 5% w/w of $H_2O_2$ illuminated in ambient (multiplied by 0.02 for scale) is also shown for reference of DEPMPO probe peaks. (B) absorption spectra of toluene solutions after 1 h exposure of immersed perovskite films on glass substrates $BA_2PbI_4$, $4AMPPbI_4$, $PEA_2PbI_4$, $BDAPbI_4$ and $HDAPbI_4$ to simulated solar illumination (1 Sun) in ambient (HDA denotes hexane-1,6-diammonium). The absorption spectrum of toluene solution of perovskite film $BA_2PbI_4$ exposed to illumination in Ar atmosphere is also shown, and the inset shows schematic diagram of the measurement. (C) Absorption spectra of toluene solutions of immersed $BA_2PbI_4$ films on different substrates (ITO, $NiO_x$/ITO, 2PACz/$NiO_x$/ITO, where ITO denotes indium tin oxide and 2PACz denotes [2-(9H-carbazol-9-yl)ethyl]phosphonic acid)) after exposure to simulated solar illumination (1 Sun); peaks corresponding to iodine $I_2$ and triiodide $I_3^-$ absorption are labelled. (D) Absorption spectra of DCM solution with 0.1M $Bu_4NPF_6$ after 20 s immersion of the 2D perovskite samples with bias of 1.2 V. The absorption spectrum after 20 s immersion of $BA_2PbI_4$ without bias is also shown, consistent with previously established window of stability (<30 min)[24] in absence of bias or illumination. The insets in panels b), c) and d) show schematics diagrams of the measurements.

For detailed investigations of stability differences, we selected four commonly studied 2D perovskite materials,[1] including two RP (BA-based and PEA-based) and two DJ (BDA-based and 4AMP-based) perovskites, where BDA denotes butane-1,4-diammonium and 4AMP denotes 4-(amino methyl) piperidinium). BA and BDA were chosen due to their similar structure, while PEA and 4AMP were chosen as commonly used spacers with more complex structure to examine generality of observed differences. Unlike DJ perovskites, both RP perovskites exhibit prominent decrease in the absorption in ambient and dry air, with the worst stability observed in ambient air due to accelerated degradation in the presence of both oxygen and water.[29-32] The degradation of RP perovskites clearly involves the loss of organic cations, as evidenced by the reduction of FTIR bands related to N-H stretching vibrations (region ~3100-3400 $cm^{-1}$),[5] $CH_3$ vibrations (~2800-3000 $cm^{-1}$),[36] N-H bending vibrations (~1650 $cm^{-1}$),[8] and C-H scissoring (~1470 $cm^{-1}$),[8] similar to photooxidative degradation of $MAPbI_3$ due to deprotonation of MA.[8] Degradation under illumination is also observed in a RP perovskite single crystal (**Figure S11**), starting from the edges consistent with previous reports.[13,33]

Since superoxide ion ($O_2^-$) was proposed to participate in photooxidation reaction of 3D perovskites[4-11,29] and quasi-2D perovskites,[15,29] we performed electron paramagnetic resonance (EPR) spectroscopy measurements (**Figure 2A** and **Figure S12**) for both types of perovskites illuminated under different environments. Results show that there is no clear relationship between $O_2^-$ generation and perovskite degradation as significant $O_2^-$ is still generated in DJ perovskites ($4AMPPbI_4$ and $BDAPbI_4$) with negligible degradation, in agreement with $O_2^-$ production in $MAPbBr_3$ which is not susceptible to photooxidation.[10,11]

Since $O_2^-$ production simply indicates transfer of photogenerated electrons to oxygen molecules, which causes excess holes in the perovskite,[22] we hypothesize that DJ perovskites are more tolerant than RP perovskites to hole accumulation and investigate their electrochemical stability. Since iodide oxidation by excess holes under illumination leads to expulsion of iodide into the solution in 3D perovskites (**Supplementary Note 1**), UV-VIS absorptions of RP and DJ perovskites in toluene solution were measured (**Figure 2B**). $I_2$ (at ~520 nm[20]) and $I_3^-$ species (at ~290 nm and ~360 nm[22,28]) were observed in RP perovskites in the presence of oxygen signifying iodide expulsion from RP perovskites,[22] in agreement with film photostability trends and similar to previous report for $BA_2MA_{n-1}Pb_nI_{3n+1}$ ($n$ = 1, 2, 3).[20] Negligible expulsion was observed in RP perovskites in argon or DJ perovskites in any environment. The photooxidative release of halides occurs due to the weakening of the Pb-X bond upon optical excitation.[27] Resulting iodine oxidation products include HI, $I^*$ radicals, $I_2$, triiodide ($I_3^-$), and interstitial iodine defects ($I_i^n$, where n=-1, 0, +1).[27] The degradation process likely involves deprotonation of organic cation, since significant suppression of iodine expulsion is observed in films with high Cs content.[28] Similar to 3D organic precursor MAI,[8] both RP and DJ precursors (BAI, PEAI, BDAI, and 4AMPI) release iodide species into the solution (**Figure S13**) indicating that stability difference is related to the incorporation of spacer cation in the perovskite structure.

***Electrochemical stability of 2D perovskites***

The involvement of excess holes is investigated on $BA_2PbI_4$ films (as they are susceptible to photooxidation) and confirmed by decreased release of iodide species for films on hole transport layer (HTL) (**Figure 2C**), due to efficient hole extraction from perovskite (evidenced by reduced PL emission, **Figure S14 & S15**). In contrast, films on electron transport layer (ETL) degrade faster (**Figure S14**). Illumination of $BA_2PbI_4$ films on different substrates also results in differences in outgassing products (**Figure S16)**, namely butylamine, $NH_3$, iodobutane, and $CH_3I$ gases. HI could not be observed due to the reaction between HI and column stationary phase.[38] The lowest outgassing was observed for films on indium tin oxide (ITO), which can extract both electrons and holes. For the samples on ETL (excess holes), we observe more outgassing of oxidation product $CH_3I$, while on self-assembled monolayer (excess electrons) we observe more outgassing of reduction product butylamine. These results strongly suggest that the degradation process is photoelectrochemical, similar to $MAPbI_3$ (**Supplementary Note 1**). The electrochemical nature of the process is also confirmed by iodide expulsion into solution under bias[22-24] (**Figure 2D**), where significant expulsion occurs for RP but not for DJ perovskites. Consistent with this, devices with 2D RP perovskites after bias exhibit obvious corrosion of the electrode (absent for DJ perovskites, **Figure S17**) and lower I/Pb ratio in the perovskite film after bias (**Table S2**). Cyclic voltammetry (CV) measurements were used to investigate the deprotonation reaction in 2D perovskites with different spacers by biasing the perovskite with a reduction potential in the electrochemical cell (**Figure S18**), and the onset of rise in cathodic current for monoammonium spacers (BA and PEA) is found to be at -0.7 V vs Pt which is less negative than the onset for diammonium spacers (BDA and 4AMP). More importantly, RP perovskites degraded completely after scan, different from DJ perovskites (**Figure S19**). These results indicate the possibility that a difference in the reduction reaction (deprotonation of organic cation), rather than oxidation reaction (oxidation of iodide and formation of interstitial iodine defects, $I_2$, and/or $I_3^-$) could suppress the degradation pathway in DJ perovskites.

***Mechanism of degradation under illumination and/or bias***

Possible mechanisms for observed increased resistance of DJ perovskites to photooxidation are listed in **Figure 3A.** Extensive calculations were performed to examine different possibilities, as described **Supplementary Notes 4-6**, and depicted in **Figures S20-S27.** We found that increased photostability of DJ perovskites cannot be correlated to their structural rigidity (**Figure S21, Supplementary Note 5**). In addition, hole-doping does not result in significant changes in radial distribution functions (**Figures S22 & S23**) or phonon mode softening (**Figure S24 & S25**) in both RP and DJ perovskites. Furthermore, the increased photostability of DJ perovskites cannot be generally correlated to shorter interlayer distance,[37] since $DDDAPbI_4$ with interlayer spacing of 16.02 Å is more stable than $BA_2PbI_4$ with interlayer spacing of 13.75 Å (**Figure 4**). RP and DJ perovskites also exhibit similar tendencies to localize excess holes on iodine (**Figures S26 &S27**, **Supplementary Note 6**), as expected since valence band is mostly made of iodine 5p orbitals.

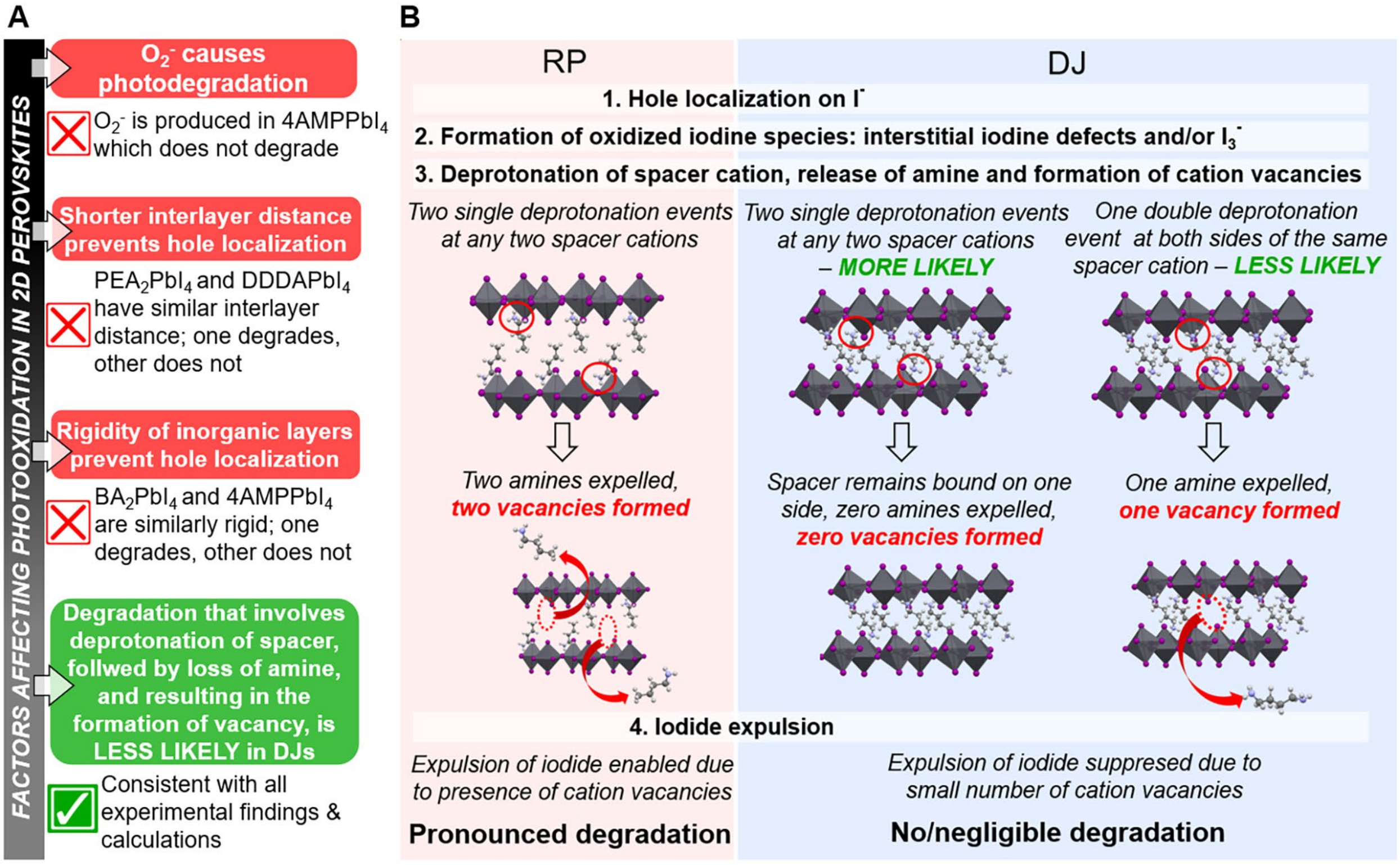

**Figure 3**. (A) Schematic illustration of the possible mechanisms contributing to increased photostability of DJ perovskites (B) Schematic illustration of the proposed process highlighting differences in the formation of cation vacancies in DJ and RP perovskites. In RP perovskites, organic amine is readily lost after deprotonation of the ammonium cation leading to the formation of cation vacancies. This results in the enhancement of ion migration, facilitating degradation of the perovskite layer. On the other hand, in the case of DJ perovskites a higher density of excess holes would be needed for simultaneous deprotonation of the cation on both sides to create cation vacancy, which makes this double deprotonation event less likely. As the electrochemical reactions are reversible in the absence of loss of volatile reaction products, singly deprotonated cations which remain bound can revert to protonated state. Consequently, no vacancies are formed thus the release of halides is suppressed, preventing the degradation of DJ materials.

We then examined the energetics of organic spacer vacancy formation in neutral and positively charged systems (see **Supplementary Note 4** for details). Since forming a charged spacer vacancy would be extremely energetically unfavourable, we investigate the existence of a singly (RP) or doubly (DJ) deprotonated state to neutralize the spacer (**Fig. S27-29, Tables S3&S4)**. By performing geometry relaxations starting from various geometries including a deprotonated spacer (**Tables S2 & S3, Figures S28 &S29** and Supplementary Data for computational methods), we find that introducing either a hole or an interstitial halide defect into a neutral pristine system modifies the potential energy surface to create a locally bound interhalide-proton protonated state, where the alkylammonium group is deprotonated **(Fig. S27)**. The interstitial iodine defects are found to lower the barrier for deprotonation and spacer vacancy formation. In addition, deprotonated state is unstable for $I_i^-$ defects for DJ perovskite, while in RP perovskites these defects result in negative spacer vacancy formation energy (**Figure S29**), indicating that $I_i^-$ defects can easily destabilize RP perovskite. Thus, we can conclude that differences in the likelihood of spacer vacancy formation, which requires single deprotonation on any RP spacer cation and double deprotonation on both sides of the same DJ spacer cation (less likely than two single deprotonation events at one side of any two different cations), are responsible for photostability differences between DJ and RP perovskites, as illustrated in **Figure 3b**.

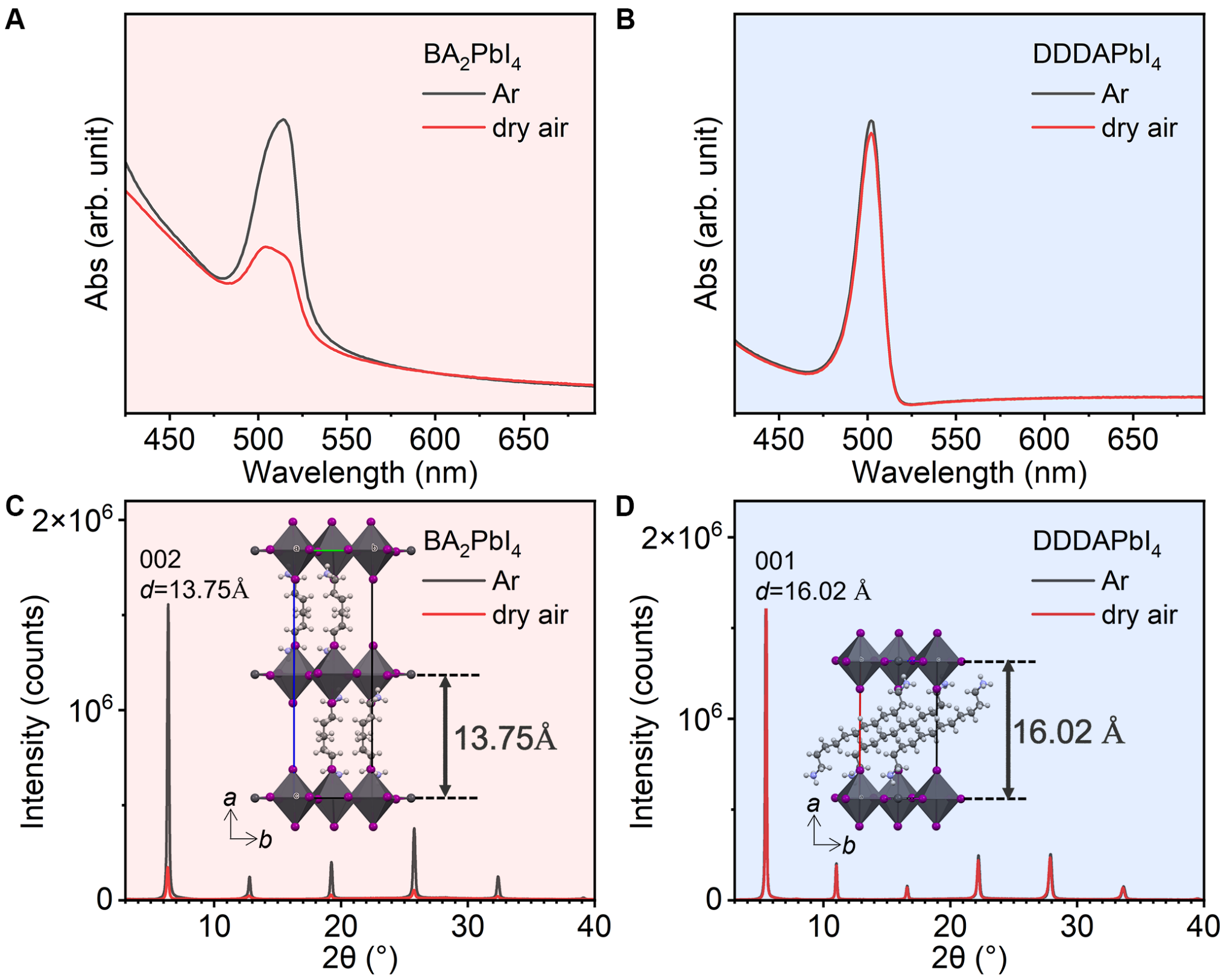

**Figure 4**. Absorption spectra of (A) $BA_2PbI_4$ and (B) $DDDAPbI_4$ and XRD patterns of (C) $BA_2PbI_4$ and (D) $DDDAPbI_4$ after simulated solar illumination (100 mW/cm$^2$) in argon and dry air. Insets show the interlayer spacings.

Thus, from the obtained experimental data and theoretical calculations, as well as the literature reports on photo/electrochemical stability of 3D perovskites (**Supplementary Notes 1, 7 & 8**), we propose that photooxidation of 2D perovskites involves the oxidation of iodide which results in the generation of mobile iodine species, and the reduction/deprotonation of organic cation, resulting in organic spacer vacancy formation and release of volatile degradation products. The first step in photooxidation process involves the localization of hole on $I^-$, weakening of Pb-I bond and the oxidation of $I^-$ by the hole into oxidized iodine species (interstitial iodine defect, $I_2$, and/or $I_3^-$).[39] Next, the deprotonation of spacer cation follows which becomes favorable in the presence of excess holes or interstitial iodide defects (**Supplementary Note 4**), in particularly $I_i^-$, consistent with the report that $I_3^-$ causes deprotonation of organic cation by forming strong hydrogen bonds with organic ammonium cations.[39] The significant role of iodide is also confirmed by rapid degradation of $BA_2PbI_4$ under illumination in the presence of excess iodine (**Figure S30**), similar to 3D perovskite degradation in presence of iodine.[6] As iodide oxidation occurs in all perovskites, to explain the differences in the degradation of RP and DJ perovskites, we need to consider the likelihood of the formation of cation vacancies after deprotonation, as illustrated in **Figure 3B.** Reduced cation vacancy formation in DJ perovskites results in suppressed ion diffusion since the ion diffusion in 2D perovskites proceeds layer-by-layer and the organic spacer cations in 2D perovskites serve ion-blocking function.[40] In agreement with this expectation, significantly lower activation energy for halide migration was reported for BA-based perovskite compared to BDA-based perovskite.[41]

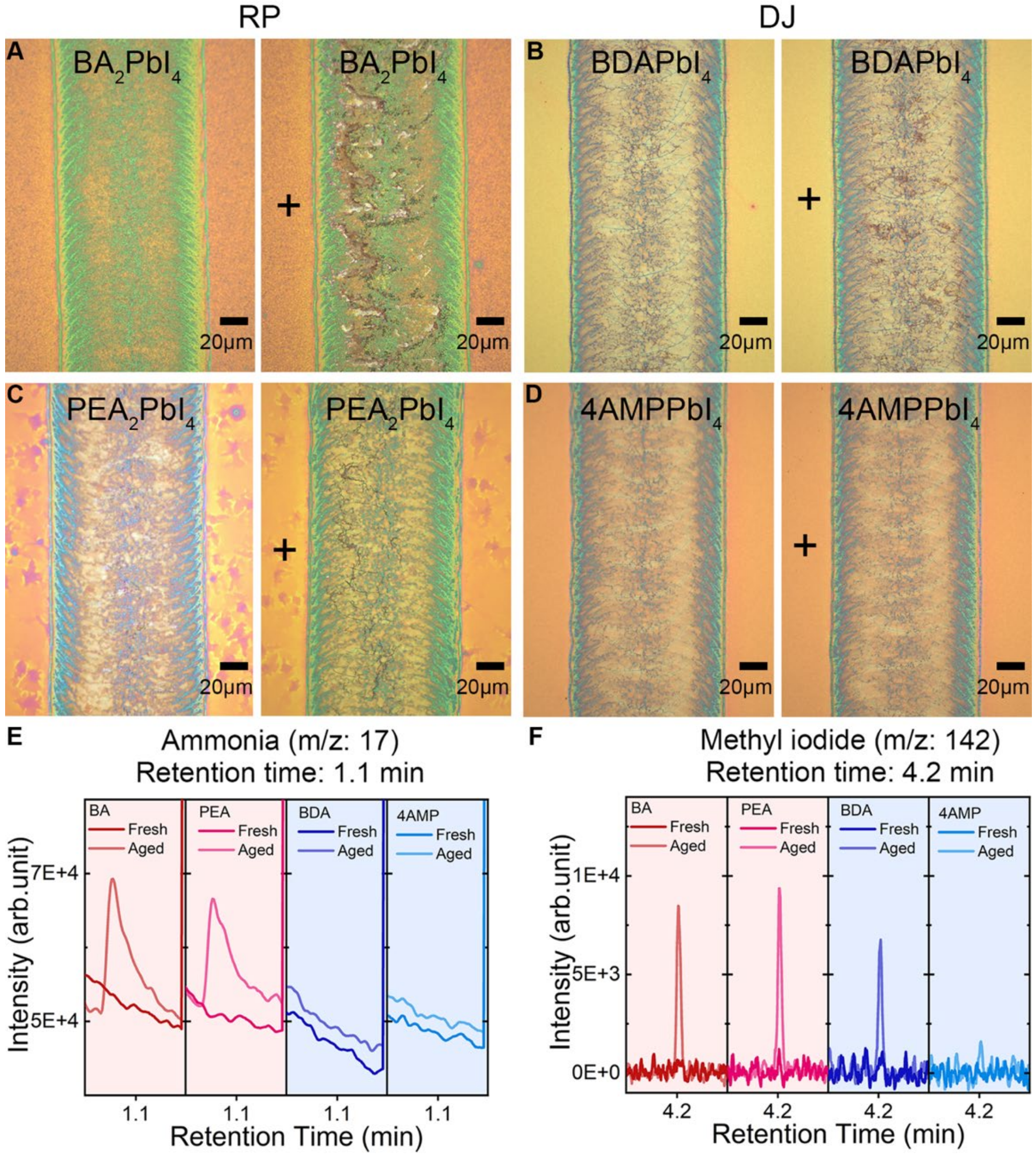

**Figure 5**. Microscope images of (A) $BA_2PbI_4$, (B) $BDAPbI_4$, (C) $PEA_2PbI_4$, and (D) $4AMPPbI_4$ lateral geometry devices with ITO electrodes separated by 110 μm gap before (left) and after (right) bias of 10 μA current for 20 min without illumination in ambient (~60% RH). Scale bar is 20 μm. GC-MS signal corresponding to (E) ammonia and (F) $CH_3I$ for fresh and aged samples of different 2D perovskites.

Consistent with proposed mechanism, we observe significant degradation after bias in ambient near positive electrode in lateral geometry devices for RP perovskites (**Figure 5A-D**). Encapsulated $BA_2PbI_4$ exhibits the same degradation pattern (**Figure S31**), but it degrades slower than the sample without encapsulation. RP perovskites also exhibit more significant outgassing under illumination **(Figure 5E&F, Supplementary Note 9**). From the GC-MS measurements **(Figure 5E&F, Supplementary Note 9**), we can observe that iodide oxidation (resulting in I-containing degradation products such as $CH_3I$) is more prominent in RP perovskites, but it occurs both in RP and DJ perovskites in agreement with theoretical calculations. On the other hand, we observe $NH_3$ outgassing (which is a decomposition product originating from spacer cations) only in RP perovskites consistent with reduced likelihood of spacer vacancy formation in DJ perovskites. The proposed degradation mechanism is also consistent with the observed changes in *in situ* Raman measurements (**Figures S32-S34**), which show distinctly different patterns for RP and DJ perovskites. For both perovskites, no significant changes are observed in low wavenumber region corresponding to the inorganic part of perovskite lattice,[42-44] while for RP perovskites we observe more significant changes in the regions corresponding to different vibrations of organic cation.[45-49] In the case of $BA_2PbI_4$, which has the worst stability, we also observe that new peaks (at ~402 $cm^{-1}$, 518 $cm^{-1}$ and 643 $cm^{-1}$), attributed to deformations of butylamine[45-47] and $HI_2^-$,[48] disappear after the second scan of the same area (**Figure S34**), consistent with the presence of loosely bound surface species.

Therefore, we can conclude that the degradation under illumination in 2D perovskites, similar to their 3D counterparts, is mainly driven by electrochemical redox reactions. Excess photogenerated holes, which result either from imbalances in charge extraction or from the presence of oxygen which serves as electron scavenger, result in oxidation of iodide and generation of oxidized iodine species (interstitial iodine defect, $I_2$, and/or $I_3^-$).[39] These oxidized iodine species participate in further redox reactions, including deprotonation of organic cation, triggering the chain reaction of degradation.[39] These reactions become irreversible due to loss of volatile reaction products, such as desorption of organic amine after organic ammonium cation deprotonation. As this occurs more readily in RP perovskites (where single deprotonation event can result in spacer cation vacancy) than in DJ perovskites (where two deprotonation events on the same spacer cation are needed to produce neutral amine which can then desorb), DJ perovskites exhibit improved photo/electrochemical stability, as evidenced by their improved stability under illumination (**Figure 1, Figure 4, Figures S1-S10**), their lack of significant expulsion of iodide in solution under illumination and bias (**Figure 2B** & **2D**), the lack of visible changes in the electrodes after bias (**Figure S17**), and the lack of degradation under C-V measurement (**Figure S19**) and bias in lateral devices (**Figure 5C** & **5D**), as summarized in (**Supplementary Note 11**). Critical role of holes in the degradation process due to electrochemical redox reactions in RP perovskites is demonstrated by differences in perovskite degradation and outgassing of reaction products for perovskite deposited on HTL (excess electrons) and ETL (excess holes) (**Figure 2C, Figures S14-16**) and the degradation near positive electrode in lateral bias devices (**Figure 5A** & **B)**, while the iodine-induced degradation (**Figure S30**) is consistent with the hypothesized role of oxidized iodine species in the degradation process. The loss of organic cations as a key step towards perovskite degradation is evident from FTIR measurements (**Figure 1B & F**, **Figure S5A, Figure S10B**), GC-MS measurements (**Figure S16A & B**, **Figure 5E**), and changes in Pb:I ratio after bias or illumination stability test (**Table S2**, **Table S8**).

***Device Stability***

Three different 3D perovskite compositions, CsFAMA triple cation perovskite, low Br perovskite, and MA-free perovskite, are considered to investigate generality of the approach. Overall photostability of different 3D/2D film combinations (**Figures S35-S38**) depends on 3D perovskite used, and for each 3D perovskite stability trends (DJ>RP) are consistent with those of 2D films. The performance of corresponding SC devices is summarized in **Figure 6**, **Figure S39-S41** and **Table S5**. The stability tests were performed without encapsulation in dry air to accelerate degradation due to photo/electrochemical redox reactions (under this condition, there will be higher hole accumulation since $O_2$ acts as electron scavenger,[22] and there will be increased loss of volatile decomposition products compared to devices with blanket encapsulation[38]). Slower degradation can be clearly observed for devices with 2D DJ capping layer for all 3D perovskite compositions, although the stability trends between spacers in the same family (BA vs. PEA, BDA vs. 4AMP) vary for different 3D perovskite compositions and can differ from trends observed in individual films. This can be attributed to possible differences in interfacial defects and charge accumulation in devices, which would be different for different 3D perovskites.

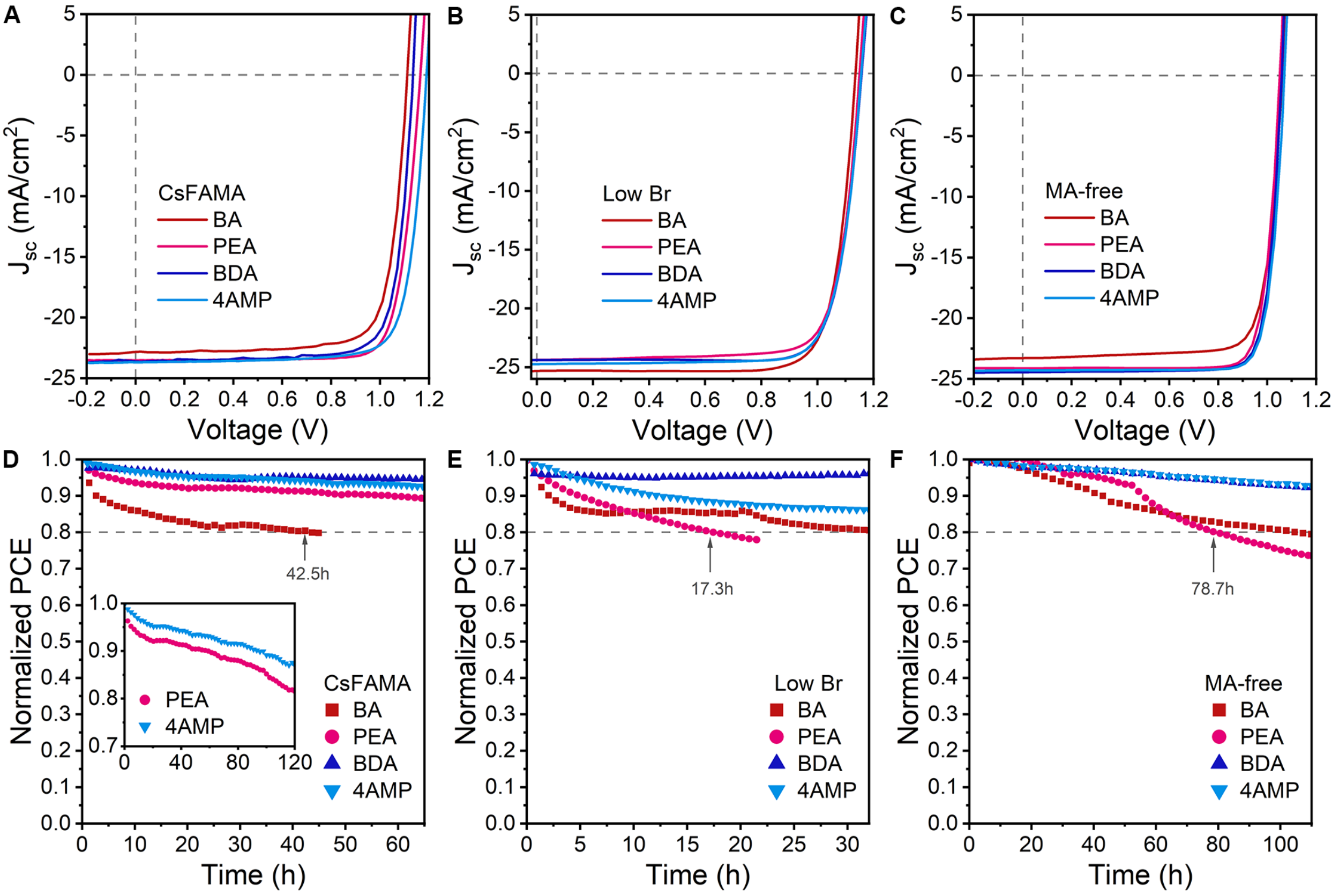

**Figure 6**. I-V curves (reverse scan, the best performance devices listed in **Table S5**) of solar cells with 3D/2D (BA, PEA, BDA, 4AMP) perovskite layer for different 3D perovskite compositions: (A) triple cation CsFAMA (B) Low Br and (C) MA-free; (D), (E), (F) Corresponding normalized PCE as a function of time of illumination (1 Sun, 100 mW/cm$^2$) in dry air without encapsulation. Temperature during stability test was 30ºC.

To further investigate this, additional characterizations were performed for different 2D capping layers for commonly used CsFAMA triple cation perovskite, which exhibits stable performance under illumination when encapsulated.[50] The obtained atomic force microscopy (AFM), Kelvin probe force microscopy (KPFM), and TRPL results, as well as composition investigations after tests are shown in **Figures S42-S44** and **Table S6-S8**, and summarized in **Supplementary Note 10**. We can observe that no consistent pattern between DJ and RP perovskites in charge extraction from TRPL decay times for 3D/2D perovskite between HTL and ETL (corresponding to full device structure without the electrode, **Figure S44B** and **Table S6**) or from the change in surface potential with illumination **Figure S43**. In the case of 4AMP, we observe positive change of surface potential and conventional hysteresis, while for BA, PEA, and BDA negative change of surface potential and reverse hysteresis are observed (hysteresis in all the devices is small, **Figure S39G**). For BA, we observe the lowest $J_{sc}$ values and the longest TRPL decay times, consistent with less efficient charge extraction, and these devices also exhibit the shortest $T_{80}$ lifetime under illumination (**Figure 7D**). For low Br 3D perovskite, we observe higher $J_{sc}$ values for BA-based devices than for PEA, and longer $T_{80}$ for BA than for PEA. The obtained results are consistent with dominant contribution of photo/electrochemical redox reactions to degradation under illumination, with 3D/2D interface in part affecting the degradation rate due to differences in interfacial defects and/or charge accumulation. Devices with RP and DJ capping layers show similar stability when stored in the dark in inert environment (**Figure S45**), also consistent with degradation under illumination dominated by photo/electrochemical redox reactions. This is further evidenced by the degradation patterns of encapsulated devices under forward bias with $J_{sc}$, under reverse bias and illumination (**Figure S46**) and OC stability testing (**Figure S47**), as well as observable electrode damage under OC testing **(Figure S47)**, where clear metal diffusion to the opposite electrode is observed, consistent with redox-reactions driven degradation leading to metal deposition on the opposite electrode due to migration of metal-iodide complexes followed by metal reduction.[51]

**Conclusions**

In this work, we demonstrate that DJ perovskites have increased resistance to photooxidation due to reduced formation of organic cation vacancies under illumination and/or bias, resulting in improved stability due to reversibility of photo/electrochemical reactions in the absence of loss of volatile reaction products, which makes them more suitable for applications in high stability devices. While the intrinsic stability difference related to reduced likelihood of formation of cation vacancies remains valid for all DJ materials, to maximize this advantage it is necessary to optimize device architecture to minimize detrimental hole accumulation during operation, as well as minimize the defects in the perovskite films.

**Data availability**. The data that support the plots within this paper and other findings of this study are available from the corresponding authors upon reasonable request.

## Acknowledgements

This work was supported by the Seed Funding for Basic Research, the Seed Funding for Strategic Interdisciplinary Research Scheme, and Research Output Prize of the University of Hong Kong, RGC CRF project 7018-20G and NSFC project 6207032617. The authors thank the Materials Characterization and Preparation Center (MCPC) for some characterizations in this work. J. B. acknowledges the support by the Australian Renewable Energy Agency (ARENA) via project 2020 RND003. A H.-B. is supported by the Australian Research Council (ARC) Future Fellowship FT210100210.

## Author contributions

A. B. D., J. P., and T. L. L. conceived the project, and A. B. D., J. P., A. M. C. N., A. H. B. and I. L. supervised the project. J.O. performed all computations with help and guidance from L.G. and I.L., Z. L. R. performed preparation and characterization of different perovskite thin films, with some of the samples prepared and/or characterized by T. L. L. and Z. Y. In addition, Z. L. R. and A. A. S. performed degradation experiments in solution under illumination and/or bias, Z. L. R. performed lateral bias experiments, T. L. L. performed cyclic voltammetry and KPFM

measurements, J. B. performed GC-MS measurements, M. P. B. contributed to the GC-MS methodology and assisted with the measurements. and M. U. A. performed TRPL measurements. Y. L. H. performed EDX, EPR, *in situ* absorption and *in situ* XRD measurements, as well as Raman measurements together with G.Y. L. X. S. Q. prepared and characterized charge injection devices and samples for EDX. W. T. S. prepared $BA_2PbI_4$ single crystal, while J. Y. L., A.-K., and J. W. prepared $NiO_x$ nanoparticles for solar cell fabrication. Y. L., Z. T. Y., D. Y. L., H. M., P. B. and T. Z. prepared and characterized solar cells under the guidance of G. L. and A. B. D.

## Competing interests

The authors declare no competing interests.

## Additional information

**Supplementary information** is available for this paper at https://xxxx.

**Reprints and permissions information** is available at www.nature.com/reprints.

**Correspondence and requests for materials** should be addressed to A.B.D., J. P. or I. L.

**Publisher's note**: Springer Nature remains neutral with regard to jurisdictional claims in published maps and institutional affiliations.

# Supplementary information

**Increased resistance to photooxidation in Dion-Jacobson lead halide perovskites – implication for perovskite device stability**

Zhilin Ren,[1#] Juraj Ovčar,[2#] Tik Lun Leung,[1,3,4#] Yanling He,[1] Yin Li,[1] Dongyang Li,[5] Xinshun Qin,[1] Hongbo Mo,[1] Zhengtian Yuan,[1] Jueming Bing,[3,4] Martin P. Bucknall,[6] Luca Grisanti,[2] Muhammad Umair Ali,[1] Peng Bai,[5] Tao Zhu,[5] Ali Ashger Syed,[1] Jingyang Lin,[1,7] Jingbo Wang,[1] Abdul-Khaleed,[1] Wenting Sun,[1] Gangyue Li,[7] Gang Li,[5] Alan Man Ching Ng,[7] Anita W. Y. Ho-Baillie,[3,4] Ivor Lončarić,[2*] Jasminka Popović,[2*] Aleksandra B. Djurišić[1*]

[1]Department of Physics, The University of Hong Kong, Pokfulam, Hong Kong SAR

[2]Ruđer Bošković Institute, Bijenička 54, 10000 Zagreb, Croatia

[3]School of Physics, The University of Sydney, Sydney, NSW 2006, Australia

[4]Sydney Nano, The University of Sydney, Sydney, NSW 2006, Australia

[5]Department of Electrical and Electronic Engineering, Research Institute for Smart Energy (RISE), The Hong Kong Polytechnic University, 11 Yuk Choi Rd, Hung Hom, Hong Kong SAR

[6]Bioanalytical Mass Spectrometry Facility, Mark Wainwright Analytical Centre, UNSW Sydney, NSW 2052, Australia

[7]Department of Physics and Core Research Facilities, Southern University of Science and Technology, No. 1088, Xueyuan Rd., Shenzhen, 518055, Guangdong, P.R. China

[#]These authors contributed to this work equally.

**Table of Contents**

# METHODS

## Materials

N,N-dimethylformamide (DMF, anhydrous, 99.9%) and dimethyl sulfoxide (DMSO, anhydrous, 99.9%) were purchased from Alfa Aesar. Toluene (anhydrous, 99.8%) %), iodine (anhydrous, beads, 99.999%), $H_2O_2$ (30% (w/w) in $H_2O$) and nickel(II) nitrate hexahydrate (99.999%) were purchased from Sigma-Aldrich. Chlorobenzene (anhydrous, 99.5%) was purchased from Aladdin. Lead bromide ($PbBr_2$, ≥98%), lead iodide ($PbI_2$, ≥98%), tetrabutylammonium hexafluorophosphate ($Bu_4NPF_6$), (2-(9H-carbazol-9-yl)ethyl)phosphonic acid (2PACz, >98.0%), hydrobromic acid (HBr, 47%), [2-(3,6-Dimethoxy-9H-carbazol-9-yl)ethyl]phosphonic Acid (MeO-2PACz), dichloromethane (DCM, anhydrous, >99%) and ethanolamine (2-Aminoethanol) were purchased from Tokyo Chemical Industry Co., Ltd. (TCI). Methylammonium bromide/iodide (MABr/I), n-butylammonium bromide/iodide (n-BABr/I), n-pentylammonium bromide/iodide (PentyABr/I), benzylammonium bromide/iodide (BZABr/I), phenethylammonium bromide/iodide (PEABr/I), 4-fluoro-phenethylammonium bromide/iodide (F-PEABr/I), 2-thiopheneethylammonium bromide/iodide (TEABr/I), 2-thiophenemethylammonium iodide (TMAI), butane-1,4-diammonium dibromide/diiodide ($BDABr_2/I_2$) were purchased from Greatcell Solar. 4-(aminomethyl)piperidinium dibromode/diiodide ($4AMPI_2$), 1,4-Phenyldimethylammonium dibromide/diiodide ($PDMABr_2/I_2$), 2-thiophenemethyl ammonium bromide (TMABr), hexane-1,6-diammonium dibromide/ diiodide ($HDABr_2/I_2$), 1,10-decanediammonium dibromide/diiodide ($DDABr_2/I_2$), and 1,12-dodecanediammonium dibromide/diiodide ($DDDABr_2/I_2$) were purchased from Xi'an Yuri Solar Co., Ltd. 135-Tri[(3-pyridyl)-phen-3-yl]benzene (TmPyPB, >99.5%) and 8-Hydroxyquinolinolato-lithium (Liq, >99.9%) were purchased from Luminescence Technology Corp. Aluminum pellets (Al, 99.999%) were purchased from Kurt J. Lesker. Zinc magnesium oxide (ZnMgO) nanoparticles were purchased from Guangdong Poly OptoElectronics Co., Ltd. 5-(Diethoxyphosphoryl)-5-methyl-1-pyrroline-N-oxide (DEPMPO, 99%) was purchased from Abcam. Isopropyl alcohol (IPA, 99.5%), ethanol (99.9%) and acetone (99.5%) were purchased from Anaqua. ITO substrates with 100~200nm $MgF_2$ antireflection coating were purchased from Liaoning YouXuan Technology Co., Ltd. All materials above were used as received.

## Preparation of perovskite thin films

Precursor solutions of 2D/quasi-2D RP perovskites were prepared by dissolving a monovalent spacer salt, methylammonium salt and lead halide with molar ratios of 2:0:1 for n=1, 2:1:2 for n=2 and 2:2:3 for n=3 in DMF/DMSO solvent (4:1 in volume, unless specified otherwise in figure caption, Figures S1-S4). All precursor solutions have the $Pb^{2+}$ concentration of 0.2 M for 2D (unless specified otherwise in Figure caption) and 0.5 M for quasi-2D perovskites. Precursor solutions of DJ perovskites were prepared in a similar way except the molarity of divalent spacer salt was halved. Substrates were cleaned sequentially with detergent, water, and IPA for 10 mins each in a sonication bath. $O_2$ plasma treatment (1 minute) was employed to improve wetting. Perovskite thin films were deposited by spin-coating the precursor solution on a substrate at 4000 rpm for 30 s in an argon filled glovebox. The film was then annealed at 80 °C for 10 mins. 2D perovskite film samples for Raman characterization were prepared by drop-casting. 60 µL of 0.5M precursor solution was dropped onto 20 × 20 $mm^2$ quartz substrates followed by annealing at 80ºC for 15 mins. 3D perovskites were prepared in the same method as described in *Device fabrication*. Perovskite thin films were also deposited on different charge transporting layers.

## Preparation of charge transporting layers

The $NiO_x$ thin film[1,2] was deposited on ITO following the method in our previous work.[1] The ethanolamine solution (1:50 v: v in ethanol) was spin-coated onto as prepared $NiO_x$/ITO substrate at 4000 rpm for 30 s, followed by annealing at 100 °C for 5 mins. 2PACz dissolved in IPA with a concentration of 0.5 mg/ml was spin-coated on $NiO_x$ layer at 4000 rpm for 30s inside glovebox, followed by annealing at 100 ºC for 10 mins. ZnMgO nanoparticles (25 mg/ml in ethanol) were spin-coated on cleaned ITO/glass substrates at 4000 rpm for 30 s, followed by annealing at 110 ºC in air for 15 mins.

### Electrochemical cell

The cell design was based on a previously reported method.[3] Perovskite film deposited on an ITO substrate was immersed into $Bu_4NPF_6$ solution (0.1 M in DCM) in a three-necks electrochemical cell. Platinum foil and AgCl/Ag metal plate were used as the counter electrode and the reference electrode, respectively. Positive bias was applied on the perovskite electrode using Keithley 2400 source meter.

### Device fabrication

There are three solar cell architectures fabricated in this work with three different 3D perovskite compositions, so called CsFAMA perovskite ($Cs_{0.03}(FA_{0.83}MA_{0.17})_{0.97}Pb(I_{0.83}Br_{0.17})_3$),[1,2,4] low Br perovskite ($Cs_{0.05}(FA_{0.98}$ $MA_{0.02})_{0.95}Pb(I_{0.98}Br_{0.02})_3$)[5] and MA-free perovskite $Cs_{0.1}FA_{0.9}PbI_{2.9}Br_{0.1}$.[6] For CsFAMA solar cell device, 25 × 25 $mm^2$ ITO/$NiO_x$/ethanolamine /2PACz were prepared as described above. Deposition of 3D CsFAMA perovskite films followed the method used in our previous work.[2] Then the perovskite films were immediately annealed at 110 ºC for 40 mins. After cooling to room temperature (RT), 2D perovskite precursor (PEAI and BAI 1 mg/mL, $BDAI_2$ and $4AMPI_2$ 0.75 mg/mL in IPA) solution was then dynamically spin-coated on the top of 3D perovskite at 5000 rpm for 30s and annealed at 100 ºC for 5 mins. After that, a PCBM layer was deposited on top of 2D perovskite by spin-coating PCBM in CB (20 mg/ml) solution with 1200 rpm for 30 s. After annealing at 100 ºC for 10 mins, BCP in IPA (0.5 mg/ml) was spin-coated on the top with 4000 rpm for 30s. Samples were then transferred into a thermal evaporator for depositing 80 nm Ag. Device active area was 15 $mm^2$ and the mask-limited device area was 8 $mm^2$. The dimensions of both device areas and mask areas were determined by an optical microscope.

For low Br solar cell device, $NiO_x$ (20 mg/ml in DI water) was spin-coated on top of 15 × 15 $mm^2$ clean ITO at 5000 rpm for 30 s and annealed at 150 ºC for 10 mins. Afterwards, self-assembled monolayer (Me-4PACz, 0.3 mg/ml in EtOH) was spin-coated to the substrates at 5000 rpm for 30 s, followed by annealing at 100 ºC for 5 min. After cooling down to RT, perovskite precursor solution with a molar concentration of 1.73 M was prepared elsewhere with a stoichiometric composition of $Cs_{0.05}(FA_{0.98}MA_{0.02})_{0.95}$ $Pb(I_{0.98}Br_{0.02})_3$ by adding corresponding CsI, MACl, FAI, $PbI_2$ and $MAPbBr_3$ into mixed DMF/DMSO solution (8:1). After stirring for 2 h at RT, 45 µL precursor was dripped onto substrates and spin-coating at 1000 rpm 10 s and 5000 rpm for 30 s, while antisolvent (CB) was dripped after 17 s before the end of the process. The 2D perovskite fabrication method is the same as the CsFAMA perovskite. PCBM (20 mg/ml in CB) and BCP (1 mg/ml in EtOH) are further coated on ITO/ $NiO_x$/ Me-4PACz /3D/2D at 1000 rpm and 4000 rpm and annealed at 65 ºC for 5 mins. The devices were transferred to thermal evaporator for the deposition of 100 nm silver to finish the whole device. Device active area was 10 $mm^2$ and the mask-limited device area was 7.36 $mm^2$.

For MA-free solar cell device, $NiO_x$ (20 mg/ml in DI water) was spin-coated on top of 25 × 25 $mm^2$ clean ITO at 4000 rpm for 30 s and annealed at 150 ºC for 10 mins. Afterwards, self-assembled monolayer (Me-4PACz, 0.5 mg/ml in EtOH) was spin-coated onto the substrates at 5000 rpm for 30 s, followed by annealing at 100 ºC for 10 min. After cooling down to RT perovskite precursor solution with a molar concentration of 1.4 M was prepared with a stoichiometric composition of $Cs_{0.1}FA_{0.9}PbI_{2.9}Br_{0.1}$ by adding corresponding CsI, FAI, $PbI_2$ and $PbBr_2$ into mixed DMF/DMSO solution (4:1). After stirring overnight at RT, 70 µL precursor was dripped onto substrates and spin-coated at 2000 rpm 10 s and 4000 rpm for 30 s, while antisolvent (CB) was dripped after 10 s before the end of the process. The 2D perovskite fabrication method is the same as the CsFAMA perovskite. PCBM (20 mg/ml in CB) and BCP (0.5 mg/ml in IPA) are further coated on ITO/ $NiO_x$/ Me-4PACz /3D/2D at 1000 rpm and 4000 rpm and annealed at 100 ºC for 10 mins. The devices were transferred to thermal evaporator for the deposition of 100 nm silver to finish the whole device. Device active area was 15 $mm^2$ and the mask-limited device area was 4 mm$^2$. For charge injecting device, $25 \times 25$ mm$^2$ ITO/$NiO_x$/2PACz/2D perovskite substrate were fabricated as described in "Preparation of perovskite thin films" and "Preparation of charge transporting layers". Afterwards, TmPyPB (40 nm), Liq (2 nm) and Al (100 nm) were deposited on top of the as prepared substrate by thermal evaporation. For encapsulated devices, encapsulant (PIB tape/cover glass) was pressed on top of solar cell (backside) and annealed at 80 ºC for 10 min, after which encapsulant was tightly fixed with device substrate. For lateral devices, laser-etching was used to remove the ITO and create 110 μm gaps between ITO contacts. The 2D perovskite films were spin-coated on the substrates in the same way as described in Preparation of perovskite thin films. Gaps between ITO electrical contacts were fully covered by perovskites.

## Material/thin film characterizations

XRD patterns of thin film were measured by Rigaku MiniFlex 600-C X-ray Diffractometer with Co (**Figure S4**, panels HDA and 4AMP) and Cu (all remaining XRD patterns) radiation source. Absorption spectra were measured with Agilent Cary 60 UV-Vis spectrometer. The *in-situ* XRD patterns were measured by Rigaku SmartLab 9KW equipped with a quartz window. Cu Kɑ was used as the X-ray target. The spectrum was recorded in the dark first and then illumination from the solar simulator with Xenon lamp (Zolix Instruments Co., Ltd) was introduced through the quartz window during the repeated measurement with a total scan time of 4 mins. FTIR spectra were obtained from Spectrum Two ATR-FTIR spectrometer (PerkinElmer). PL emission of perovskite was recorded by a PDA-512 USB fiberoptic spectrometer (Control Development Inc.) with a He-Cd laser (325 nm) as the excitation source. To investigate photooxidation, perovskite thin films were exposed to standard AM 1.5G illumination (ABET Sun 2000 solar simulator with Xe lamp source). For controlled environments, films were placed in a vessel with a continuous flow of specified gases. Inert, water-only (RH~40%) and oxygen-only (RH<5%) environments were generated by circulating argon, nitrogen thought water and dry air, respectively. After the light exposure, changes of perovskite thin films were characterized by measuring their absorption spectra, FTIR spectra and XRD patterns. Iodine expulsion experiment was carried out by ageing the perovskite in toluene solvent under illumination from a solar simulator. Photooxidation can then be monitored by measuring the content of triiodide in toluene. EPR (Electron Paramagnetic Resonance) equipped with X-band microwave (EMXPlus-10/12, Bruker) was used to detect the oxygen radicals with the following parameters: frequency 9.847 GHz, time constant 1.28 ms and microwave power 6.325 mW. Thin film samples packed in the vacuum bags were placed in a glovebox. Before turning on the light, the glovebox was purged with $N_2$ gas for 3 times and then the chamber was sealed and pressurized with $N_2$ gas. The 20 μL of the liquid spin trap DEPMPO was drop-casted on the thin film samples which were spin-coated on quartz. The simulated solar illumination (Zolix Instruments CO., LTD) was turned on and the sample was illuminated for 15 mins. Then, the reacted liquid DEPMPO was extracted by a capillary tube. The solution in the capillary tube was measured at RT. For experiments in dry air environment, $N_2$ gas was changed to compressed air. The EPR reference spectrum was measured with 5% (w/w) $H_2O_2$ aqueous solution with 1 second illumination time. The XPS measurement were carried out on a PHI 5000 Versaprobe III form ULVAC-PHI with Al Kɑ anode (1.4866 V) and the $C_{1s}$ peak at 284.6 eV was used as the energy reference. The Raman spectroscopy was performed using a LabRAM HR Evolution, Horiba Ltd. ranging from 50 to 200 cm$^{-1}$ and 300 to 1600 cm$^{-1}$. The sample was excited with a 785 nm laser. The spectrum was acquired under dark with signal accumulation time of 100 s and scan times of 2. Then the samples were illuminated under simulated solar spectrum for 10 mins, 20 mins, 30 mins, 40 mins and 50 mins separately and the new spectra were recorded using the same parameters as dark measurement after every illumination. Gas Chromatography-Mass Spectrometry (GC-MS) measurements were conducted using Thermo Trace DSQ GC-MS instrument. Each stressed quartz vial (Yuanfeng, China) was incubated at 95 °C for 5 minutes and gaseous products were then sampled with a heated (150 °C) 2.5 mL glass headspace autosampler syringe. A volume of Helium gas (0.4 ml), equal to the amount to be withdrawn, was pre-filled into the vial to balance the pressure inside. Sampled gas was then injected into the programmable temperature vaporizer (PTV) inlet (200 °C). PTV was equipped with a Merlin Microseal mechanical septum (Supelco, Bellefonte, PA) and a 1 mm internal diameter glass inlet liner, operated at an elevated temperature in constant temperature split mode at a certain split ratio. Restek Rtx-Volatile Amine column (30 m x 0.32 mm) was connected to the PTV inlet. Peaks were identified by comparing them to reference spectra from NIST 2011 / Wiley 9 Combined Mass Spectral Library using the NIST Mass Spectral Search program (version 2.0g) with a detector gain of $3 \times 10^5$. Gas chromatograms were smoothed with 5-point adjacent-averaging by the OriginPro 2022 software. Atomic force microscopy (AFM)/ Kelvin probe force microscopy (KPFM) images were obtained by a Neaspec s-SNOM system. The surface potential signal was extracted by scanning the perovskite/HTL/ITO sample with a PtIr5 coated AFM tip (Arror EFM, Nanoworld) in a frequency modulation mode. The lighting LED (10 mW/$cm^2$) in the sample compartment was turned on to study the impact of illumination on the surface potential. Cyclic voltammetry (C-V) measurement was done in ambient by a Biologic VSP potentiostat with a 2-electrode configuration, which has a perovskite/ITO sample (1 $cm^2$) as an active electrode and a Pt foil (2 $cm^2$) as a counter electrode. Electrolyte used was 0.1 M $Bu_4NPF_6$ in DCM. Electrochemical cell was shielded securely with a metal foil to avoid any photocurrent. Time resolved photoluminescence (TRPL) measurements were carried out on a FLS1000 Photoluminescence Spectrometer (Edinburgh Instruments) with 375 nm laser diode as excitation source operating at 1 MHz. For bias degradation test in lateral devices, 10 mA current was applied using Keithley 2400 sourcemeter. The width and length of the gap between ITO contacts were 110 µm and 10 mm, respectively.

## Device characterization

For CsFAMA and MA-free solar cell devices, J-V curves were measured with a programmable Keithley 2400 source measure unit. The scan range was from -0.2 V to 1.2 V for both forward scan (from -0.2 to 1.2 V) and reverse scan (1.2 V to -0.2 V) with 0.01 V step size and 10 ms delay time.  Solar cell performance was evaluated by measuring J-V curve under standard AM 1.5G illumination (ABET Sun 2000) for encapsulated devices in ambient condition at RT (~ 25ºC), and the illumination intensity is calibrated by a certified Enli PVM silicon standard reference cell. For low Br solar cell devices, both forward (from -0.02 to 1.4 V) and reverse (from 1.4 to -0.02 V) scan were performed with 0.02 V step size and 1 ms delay time. The J-V curves were obtained using Keithley 2400 source measure unit under standard AM 1.5 G illumination (Enli Technology Co. Ltd., Taiwan) inside glovebox at RT (~ 25ºC). The intensity was calibrated with a certified standard KG-5 Si diode before measurement.  For all solar cell device architectures, EQE spectra were measured with a QE-R 3011 EQE system (Enli Technology Co. Ltd., Taiwan) using 210 Hz chopped monochromatic light ranging from 300 to 850 nm. Stability measurements were performed on devices without encapsulation at MPP in a flowing dry air environment (RH<5%) under 100 mW $cm^{-2}$ simulated solar illumination (Sunbrick™ Solar Simulator, G2V). Temperature during stability measurements was 30ºC.

## Computational methods

All density functional theory (DFT) calculations were performed using the CP2K software package,[7] using GTH-PBE pseudopotentials,[8-10] PBE+D3[11,12] exchange-correlation functional and triple-zeta valence gaussian basis sets with two sets of polarization functions.[13] A 4-level "Quickstep" multi-grid[7] was used for real-space integration with a planewave cutoff of 360 Ry at the finest level of the multi-grid and a planewave cutoff of 40 Ry of the reference grid covered by a Gaussian with unit standard deviation. The Brillouin zone was sampled at the Γ-point.  We constructed $2 \times 2 \times 1$ ($BA_2PbI_4$, $PEA_2PbI_4$) or $2 \times 2 \times 2$ ($4AMPPbI_4$, $HDAPbI_4$ and $BDAPbI_4$) supercells of experimentally determined crystal structures.[14-18] The geometries of all structures were relaxed until the respective force on each atom was less than $4.5 \times 10^{-4}$ hartree/bohr. For $BA_2PbI_4$ and $4AMPPbI_4$, molecular dynamics simulations in the NVT ensemble[19] were run at T=300 K for at least 4.5 ps, following 2 ps of equilibration and using 1 fs time step. The phonon density of states was calculated as the Fourier transform of the velocity autocorrelation function.

# PHOTO/ELECTROCHEMICAL STABILITY OF 3D PEROVSKITES

## SUPPLEMENTARY NOTE 1 – The Critical Role of Holes

Excess charge carriers were proposed to be the cause of degradation in 3D halide perovskite materials and devices.[19] Destabilization of the perovskite by trapped charges is further exacerbated by the presence of water (which can deprotonate organic cations) and oxygen.[20] More specifically, hole accumulation resulting in iodide oxidation and causing either photoinduced segregation or photooxidative degradation has been proposed by multiple previous works.[21-33]

### Why are holes contributing to the degradation more significantly than electrons?

It was proposed that the degradation of perovskite involves simultaneous anodic and cathodic reactions, namely oxidation of halide anion and deprotonation of organic cation.[25] However, experimental data indicate that excess holes play a more significant role than electrons in photooxidation process. The reasons for more significant role of holes are as follows:

a) **Excess electrons are not necessary for deprotonation of the organic cation**, due to the formation of $I_3^-$, which has been shown to readily deprotonate methylammonium and formamidinium cations, with the process regenerating $I_2$ which can then react with $I^-$ to form $I_3^-$ and continue the cycle.[34] It has been proposed that reactions leading to iodine expulsion during illumination include: $I^- + h^+ \rightarrow I_i^0$ , $I_i^0 + h^+ \rightarrow I_i^+$ , $I_i^+ + I_i^+ \leftrightarrow I_2$, and $I_2 + I^- \leftrightarrow I_3^-$ , and $I_2$ and $I_3^-$ have been detected in the solution absorption spectra.[31] Thus, holes are sufficient to cause cascading degradation reaction causing both the loss of iodine and the loss of organic cations. The ability of holes to trigger self-catalyzing degradation pathways is in agreement with the observation that the degradation extent under open-circuit condition was dependent on the accumulated illumination dose (product of illumination intensity and time), while under short circuit condition it was dependent on illumination intensity in agreement with simple photogeneration of defects which facilitate ion migration.[35]

b) **Iodide participates in both anodic and cathodic reactions**, which reduces the impact of excess electrons on organic cation deprotonation. In other words, some of the excess electrons are consumed in the reduction of different iodide species. Several iodine species can coexist in the perovskite,[21,29,32] which can react with both electrons and holes, as well as iodide vacancies and lattice iodide, to form different reaction products,[21,29] such as neutral, positively and negatively charged interstitial iodine, molecular iodine $I_2$, and triodide $I_3^-$.[29] The proposed anodic and cathodic reactions involved in the generation of differently charged interstitial iodine are as follows:[25]

   **Anodic reactions**
   Lattice $I^- + h^+ \rightarrow I_i^0 + V_I^+$
   Lattice $I^- + 2h^+ \rightarrow I_i^+ + V_I^+$
   **Cathodic reactions:**
   $I_i^+ + 2e^- \rightarrow I_i^-$
   $I_i^0 + e^- \rightarrow I_i^-$
   $I_i^- + V_I^+ \rightarrow$ Lattice $I^-$

c) **Organic cation vacancies accelerate halide migration but do not contribute to the creation of new mobile halides.** While organic cation vacancies facilitate halide migration (different mechanisms proposed, such as the dependence of activation barrier for halide migration on hydrogen bonding[36] and reduced steric hindrance and formation of new halide vacancies due to the formation of antisite defects, i.e. halides occupying organic cation vacancy site[37]), they will not facilitate the oxidation of iodide as Pb-I bond is not affected. In addition, the perovskite lattice is tolerant to the existence of organic cation vacancies, since the *in situ* electron microscopy study demonstrated the formation of relatively stable intermediate MA-deficient phase $MA_{0.5}PbI_3$ during the decomposition of $MAPbI_3$ to $PbI_2$.[38]

d) **Halide vacancies could facilitate formation of cation vacancies.** The oxidation of iodide, resulting in the formation of interstitial neutral iodine and iodide vacancy, is expected to weaken the bonding between organic cation and perovskite octahedra (since the perovskite is held together by ionic interactions between organic and inorganic portion and hydrogen bonding between $NH_3^+$ and iodide.[39] The weakened hydrogen bonding would then facilitate deprotonation of the organic cation.

## Evidence for the role of excess holes in the degradation of the 3D perovskite

a) **Studies on bias thresholds for electrochemical reactions** (see **SUPPLEMENTARY NOTE 2** – Stability under different bias conditions). The experimental observations of the fastest degradation under open circuit operating conditions, with open circuit voltage typically higher than the threshold for electrochemical degradation, are consistent with the hypothesis that electrochemical reactions play a significant role in ion migration. In addition, significant degradation in the dark was observed under bias conditions resulting in significant hole injection.[32,40] Holes have also been implicated in reverse bias degradation,[26,41] where hole injection under reverse bias condition resulted in increased concentration of I-related defects, and degradation could be mitigated by insertion of hole blocking layer.[41]

b) **The effect of charge transport layers on the degradation of perovskite films and devices.** The choice of charge transport layer and its position (below or above the perovskite) affected perovskite film photostability, and device stability was affected by the architecture used.[42] Slower photoinduced segregation was observed for perovskite deposited on HTLs, which was attributed to lower hole accumulation in the perovskite,[43] or by coating the perovskite film with an HTL.[44] Photosegregation, attributed to hole accumulation in the perovskite, was observed on ETL but not on insulating substrate.[24] In addition, differences in the degradation of perovskite deposited on ETL and HTL in different environments were observed.[45] Finally, it has been shown that the perovskite degradation on different ETLs proceeds at different rates, and starts at different interfaces in the device.[46,47] Devices on $TiO_2$ degrade from ETL/perovskite interface, while devices with $C_{60}$ degrade from HTL/perovskite interface, which was attributed to differences in charge extraction and the presence of trapped charges.[46,47]

c) **Studies showing destabilizing effects of positive charge under various conditions.** For example, $MAPbI_3$ crystal was found more unstable with positive charge than with negative charge.[45] It was also shown that the perovskite degrades significantly faster when exposed to positive nitrogen ions compared to negative hydrogen ions.[46]

d) **Studies demonstrating iodide expulsion under illumination and/or bias.** Decrease in the perovskite film absorption and/or iodide expulsion into solution occurs under illumination[27,31,48] and under bias.[29,31,48,49] Expulsion occurs due to oxidation of iodide by the holes, and it is also accompanied by the loss of organic cation MA.[29] As $I_2$ is not stable in the lattice, it tends to migrate to the surface and then it can be expelled in solution.[29] The expulsion of iodide under illumination can also be mediated by applied bias (enhanced by positive bias, suppressed by negative bias),[23] which confirms electrochemical nature of the process. This phenomenon occurs in organic-inorganic 3D perovskites such as $MAPbI_{1.5}Br_{1.5}$,[27,29] as well as some RP materials (BA-based).[48] In $BA_2PbBr_2I_2$, expulsion of both iodine (fast) and bromine (slow) was observed,[48] while in 3D perovskites typically just the expulsion of iodine is observed, in agreement with the lack of stability of both $BA_2PbI_4$ and $BA_2PbBr_4$ under illumination.[48] In addition to iodide expulsion in solution, iodine escape from the perovskite samples under illumination could be detected by iodine presence on Si substrates placed 1 mm above samples in nitrogen environment.[50]

e) **Investigations of halide migration.** Acceleration of ion migration by hole injection was found in simulations.[51]

## SUPPLEMENTARY NOTE 2 – Stability under different bias conditions

PSC stability under open circuit (OC), short circuit (SC), and maximum power point (MPP) testing conditions has been extensively studied in the literature, with two or more conditions compared. In all reported cases, the fastest degradation under 1 Sun illumination is observed under OC condition, which was significantly faster compared to SC and/or MPP condition.[19,35,52-58] In studies comparing all three conditions, commonly OC>SC>MPP is observed,[19,35,52,53] although in one case OC>MPP>SC trend was reported.[54] The OC>MPP degradation trend is sufficiently general that it was observed over 160 variants of the devices (device stack design, treatments or additives in inverted devices).[58] The trend OC>SC>MPP degradation trend is expected if photoelectrochemical reactions are responsible for the degradation, as MPP condition is expected to correspond to the lowest charge accumulation in the devices. Possible reasons for variations in trends between MPP and SC could occur due to differences in defect concentrations and/or charge accumulations in devices with different perovskite compositions and/or device architectures. **However, significant acceleration of the degradation under OC condition is indisputable in all literature reports.**

Other important differences between OC and SC conditions were also reported. The two conditions exhibited significant differences in the recovery in the dark, with slower recovery and/or more pronounced irreversible degradation observed under OC condition.[55,59] In addition, the two conditions exhibited different dependences on illumination dose (product of intensity and time) and illumination intensity.[35] Degradation under SC condition was found to be intensity-dependent, indicating that photogeneration of vacancies and other defects which facilitate ion migration was the dominant process.[35] In contrast, under OC condition, the degradation was dependent on accumulated illumination dose, indicating participation of additional processes, which could include defect formation by reaction with accumulated charges.[35]

Finally, OC condition also corresponds to the highest bias, which typically exceeds previously reported thresholds for $MAPbI_3$ degradation by electrochemical reactions. Two distinct bias thresholds have been reported for electrochemical reactions in $MAPbI_3$/$MAPbX_3$. The lower threshold, in the range ~0.8-1.1 V,[25,40,60,61] was attributed to the oxidation of iodide and iodide/triiodide/iodine electrochemistry.[25,61] Higher threshold voltage, in the range ~1.1-1.2 V,[25,40,60,61] corresponded to significant increase in integrated charge[60] and degradation acceleration,[40] and it was attributed to the coexistence of iodide oxidation (anodic process, occurring at the lower threshold voltage) and reduction/deprotonation of methylammonium (cathodic process, occurring at the higher threshold voltage).[25,40,61] These redox reactions trigger the degradation of the perovskite, as well as significant migration of the metal (Au) in the form of gold-iodide complexes towards the opposite electrode where gold is reduced to gold metal.[61]

# PHOTOCHEMICAL STABILITY OF 2D PEROVSKITES

## *SUPPLEMENTARY NOTE 3: Summary of device stability comparisons in literature*

**Table S1**. Stability of different devices (solar cells, LEDs, and photodetectors) with RP and DJ perovskite in different form (quasi-2D perovskites, as capping layers in 3D/2D structure, or cation used as additive to 3D perovskite) has been compared in literature.[62-73] In all reports, devices with DJ perovskite exhibited better stability compared to RP perovskite.

| Device type | RP | DJ | Stability comparison | Ref. |
|---|---|---|---|---|
| Solar cell, 3D+additive | PA | PDA | DJ>RP | 62 |
| Solar cell, quasi-2D | BA | PDA | DJ>RP | 63 |
| Solar cell, RP quasi-2D +2D (RP or DJ) | BA | BDA | DJ>RP | 64 |
| Solar cell, 3D/2D | PEA | ODA | DJ>RP | 65 |
| Solar cell, quasi-2D | BA | 3AMP | DJ>RP (film on FTO under illumination, no direct device performance comparison) | 66 |
| Solar cell, quasi-2D | PA | PDA | DJ>RP | 67[a] |
| Solar cell, 3D/2D | HA | HDA | DJ>RP (film under illumination, no direct device performance comparison) | 68 |
| Photodetector, 2D | iBA | DMPDA | DJ>RP | 69 |
| Photodetector, 3D/2D | EA, BA | EDA | DJ>RP | 70 |
| Photodetector, quasi-2D | iBA | EDA | DJ>RP | 71 |
| Blue LED, quasi-2D | AA | DMP | DJ>RP | 72 |
| Infrared LED, quasi-2D | BAB | PEA | DJ>RP | 73 |

EA denotes ethyl ammonium; EDA denotes ethylene diammonium; BA denotes butyl ammonium, iBA denotes iso-butylammonium; PDA denotes propane-1,3-diammonium; PA denotes n-propyl ammonium; BDA denotes 1,4-butanediammonium; DMPDA denotes N,N-dimethyl-1,3-propanediamonnium; DMP denotes 1,5-Diamino-2-Methylpentane; AA denotes pentabasic amylammonium; PEA denotes phenethyl ammonium; ODA denotes octyldiammonium; 3AMP denotes 3-(aminomethyl) piperidinium; BAB denotes 1,4-bis(aminomethyl)benzene, HDA denotes 1,6-hexanediammonium, HA denotes hexyl ammonium.

[a]*Comparison is particularly significant, as it involves stability comparisons of PA- vs. PDA-based devices under illumination for 3000 h in glove box (95% of PCE retained for DJ, vs. 60% for RP), damp heat for 168 h (95% of PCE retained for DJ, vs. 60% for RP), and ambient storage (40-70% RH) for 4000 h (95% of PCE retained for DJ, vs. 25% for RP).*

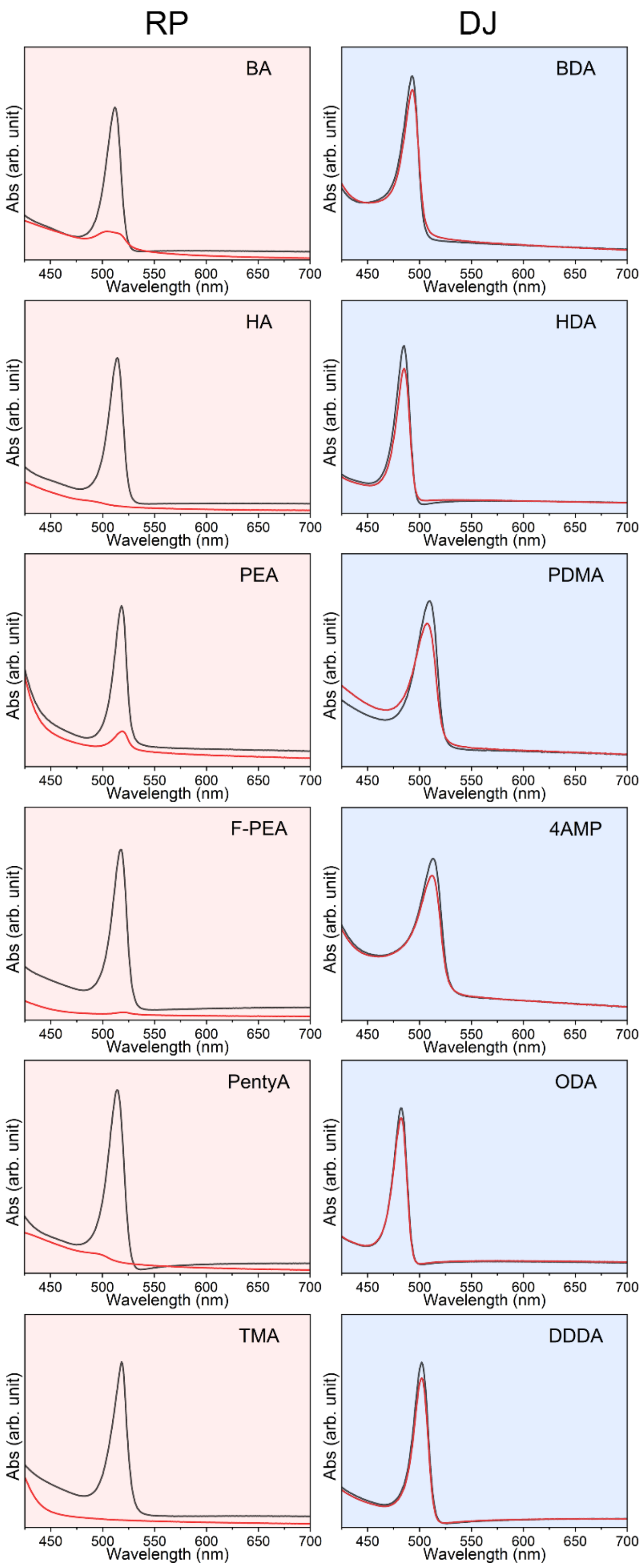

**Figure S1**. Absorption spectra of $n$ = 1 2D lead iodide perovskite samples before (black lines) and after (red lines) 3h simulated solar illumination (100 mW/cm$^2$) in dry air with different monoammonium (BA, HA, PEA, F-PEA, PentyA, and TMA) and diammonium (BDA, HDA, PDMA, 4AMP, ODA, DDDA) spacers. To obtain high quality films, ODAPb$I_4$ and DDDAPb$I_4$ were prepared from pure DMF solutions, while PDMAPb$I_4$ was prepared using DMF:DMSO=3:1.

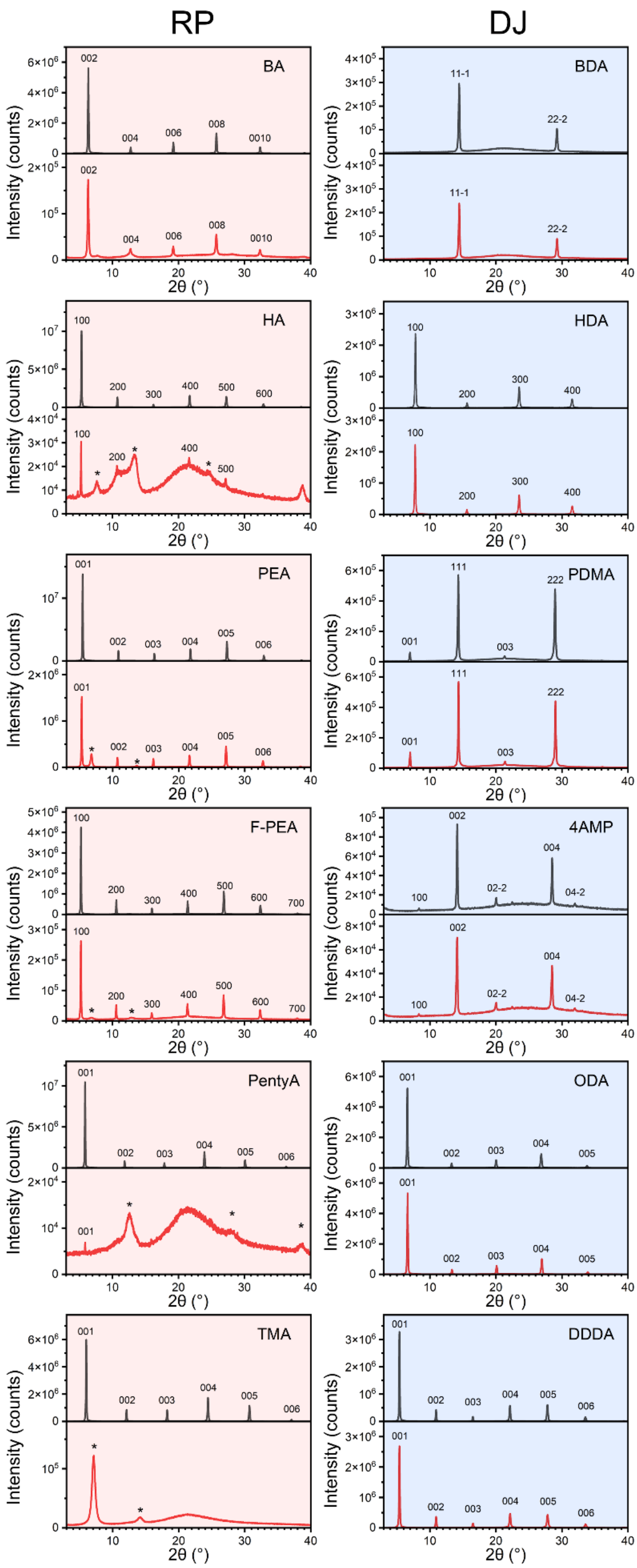

**Figure S2**. XRD patterns of $n$ = 1 2D lead iodide perovskite samples before (black lines) and after (red lines) 3h simulated solar illumination (100 mW/cm$^2$) in dry air with different monoammonium BA, HA, PEA, F-PEA, PentyA, and TMA) and diammonium (BDA, HDA, PDMA, 4AMP, ODA, DDDA) spacers. Asterisk indicates peaks corresponding to $PbI_2$. $ODAPbI_4$ and $DDDAPbI_4$ were prepared from pure DMF solutions, while $PDMAPbI_4$ was prepared using DMF:DMSO=3:1.

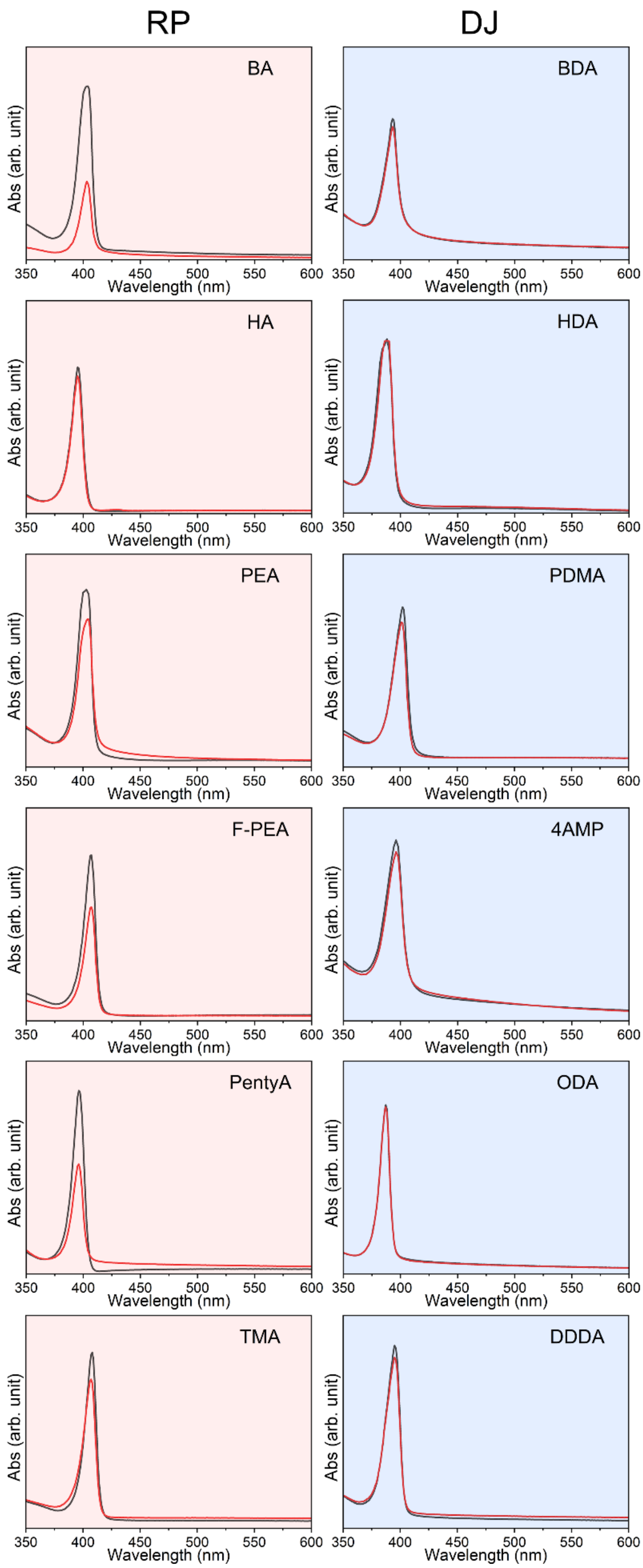

**Figure S3**. Absorption spectra of $n = 1$ 2D lead bromide perovskite samples before (black lines) and after (red lines) 3h simulated solar illumination (100 mW/cm$^2$) in dry air with different monoammonium (BA, HA, PEA, F-PEA, PentyA, and TMA) and diammonium (BDA, HDA, PDMA, 4AMP, ODA, DDDA) spacers. $DDDAPbBr_4$ and $PDMAPbBr_4$ were prepared using DMF:DMSO=1:1.

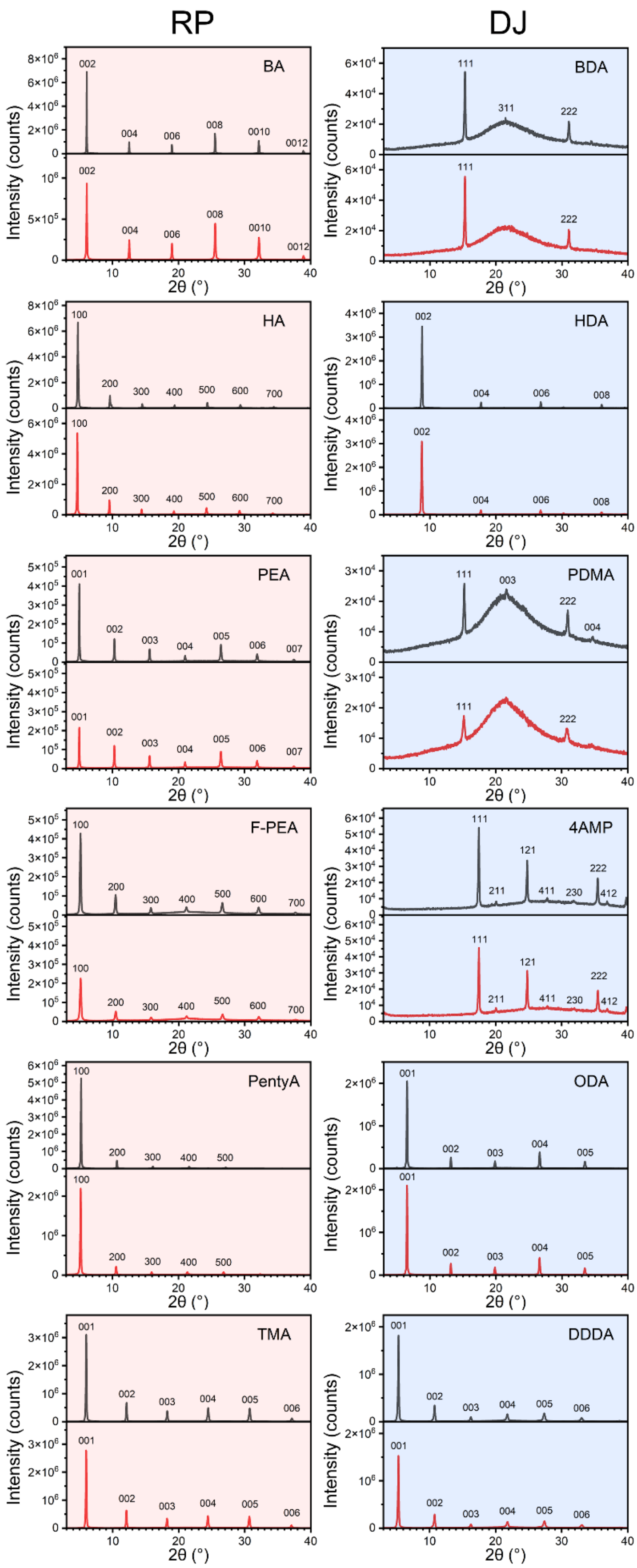

**Figure S4**. XRD patterns of *n* = 1 2D lead bromide perovskite samples before (black lines) and after (red lines) 3h simulated solar illumination (100 mW/cm$^2$) in dry air with different monoammonium (BA, HA, PEA, F-PEA, PentyA, and TMA) and diammonium (BDA, HDA, PDMA, 4AMP, ODA, DDDA) spacers. $DDDAPbBr_4$ and $PDMAPbBr_4$ were prepared using DMF:DMSO=1:1.

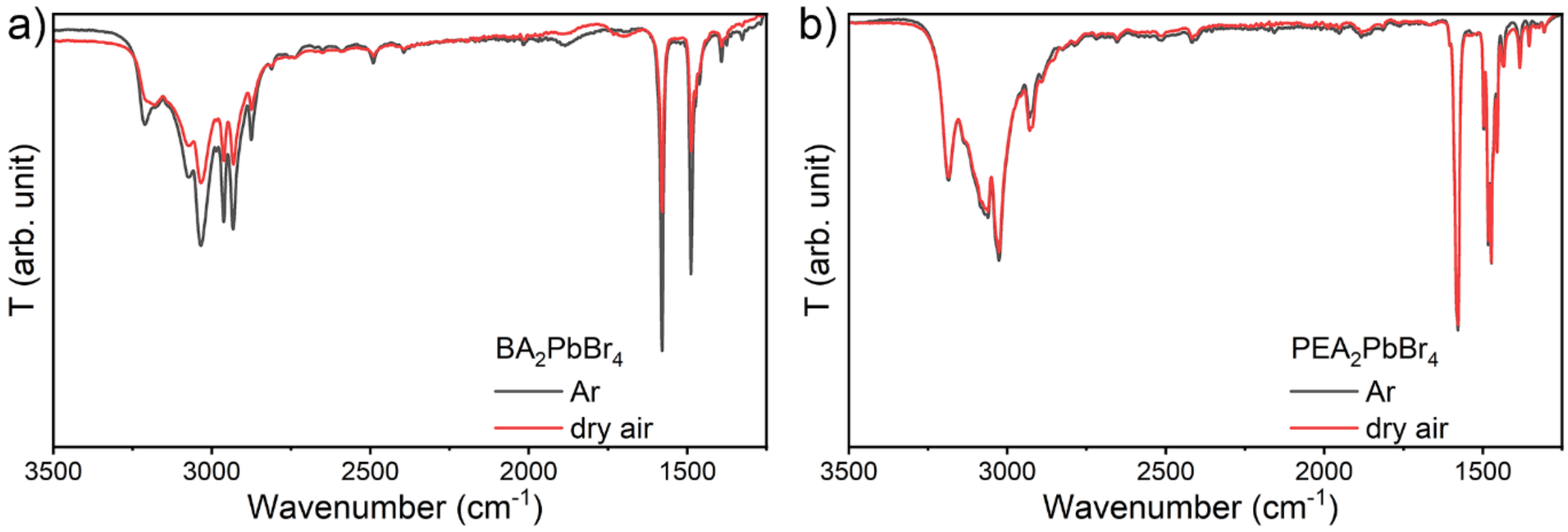

**Figure S5**. FTIR spectra of 2D perovskite samples after 3 h of simulated solar illumination in argon and oxygen/dry air. a) $BA_2PbBr_4$, b) $PEA_2PbBr_4$.

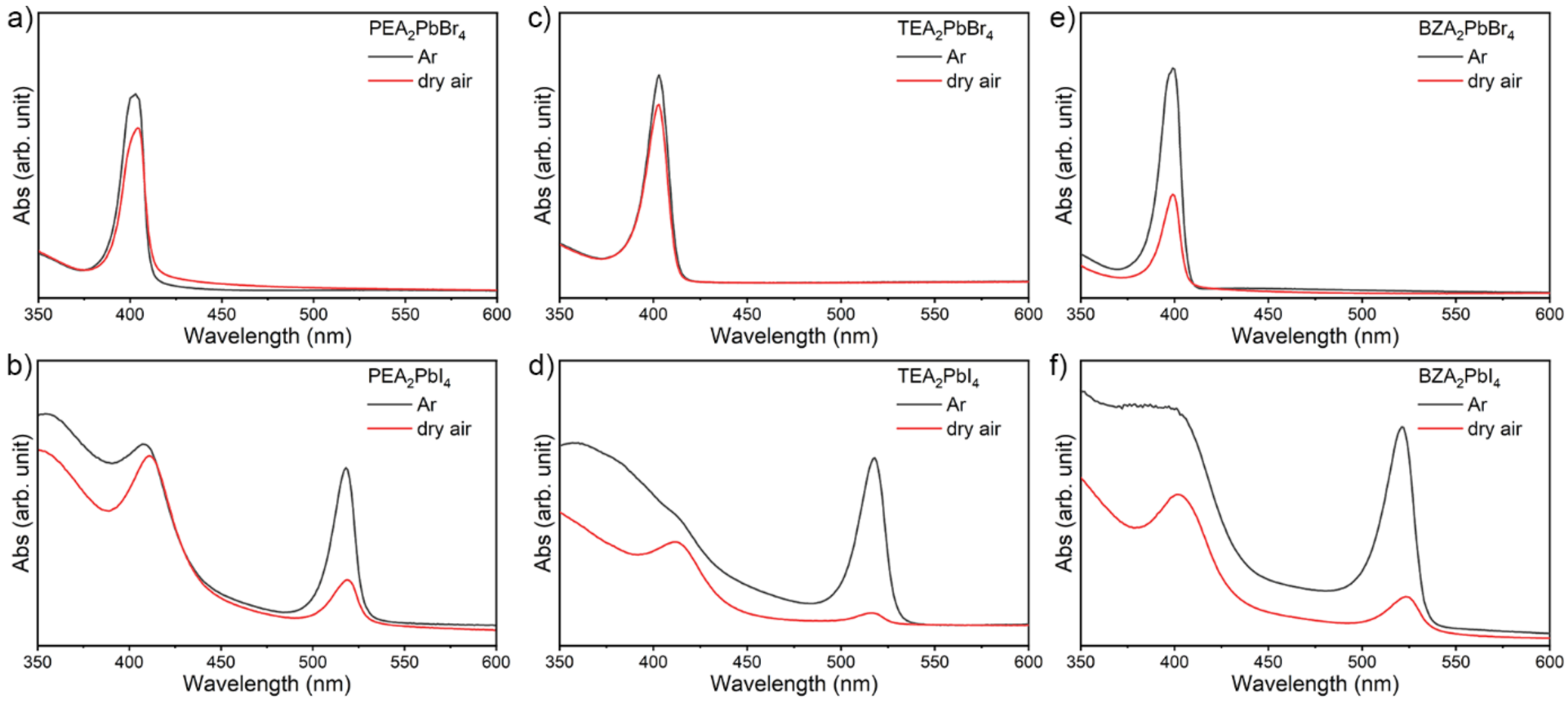

**Figure S6.** Absorption spectra of $n$ = 1 2D lead bromide and lead iodide perovskite samples with different monoammonium (PEA, TEA and BZA) spacers after 3 h of simulated solar illumination in argon and dry air.

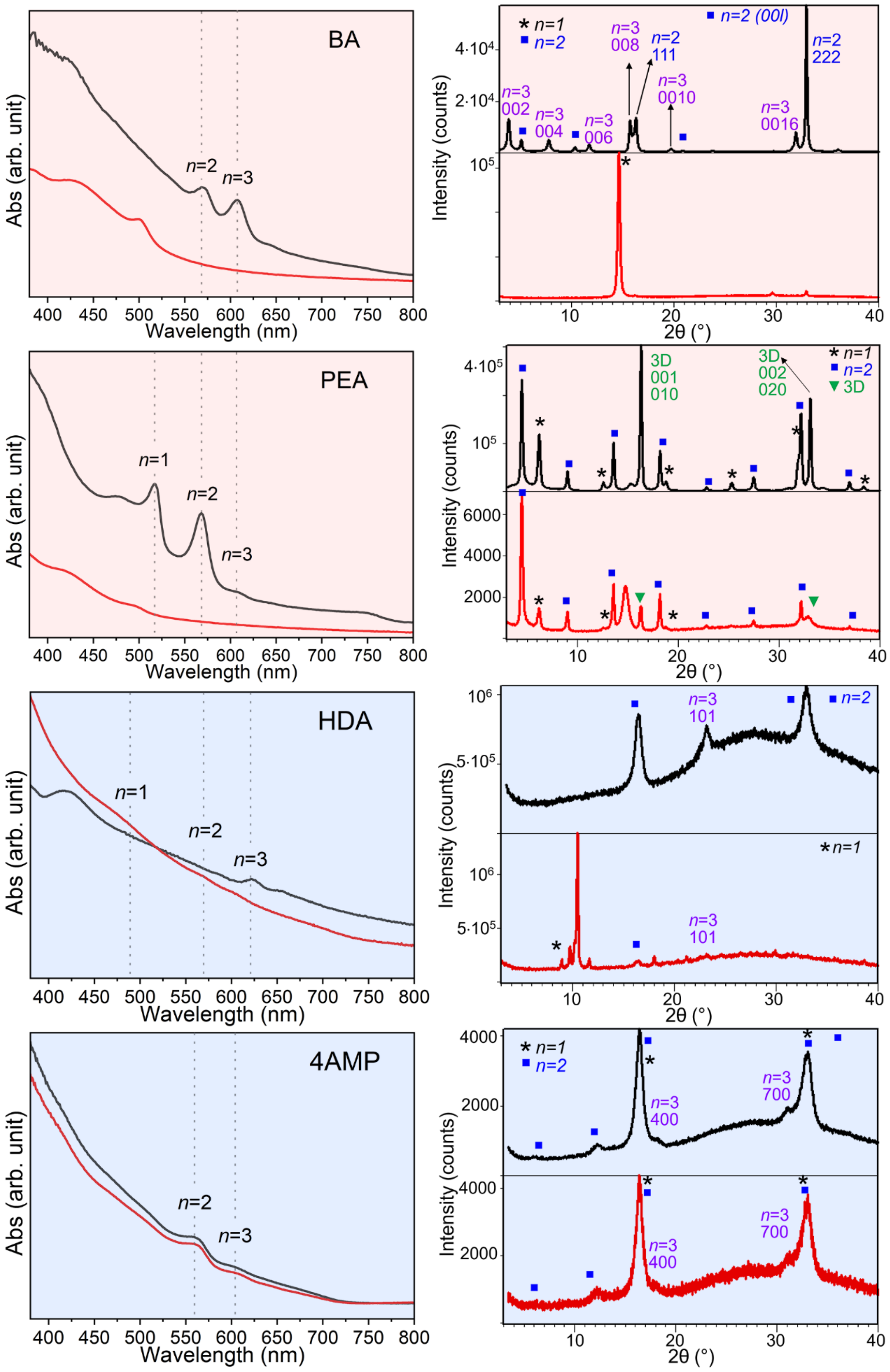

**Figure S7.** Absorption spectra (left) and XRD patterns (right)) of *n* = 3 quasi-2D MA-based lead iodide perovskite samples before (black lines) and after (red lines) 3h simulated solar illumination (100 mW/cm$^2$) in ambient air (RH ~50-55%) with different monoammonium (BA, PEA) and diammonium (4AMP, HDA) spacers.

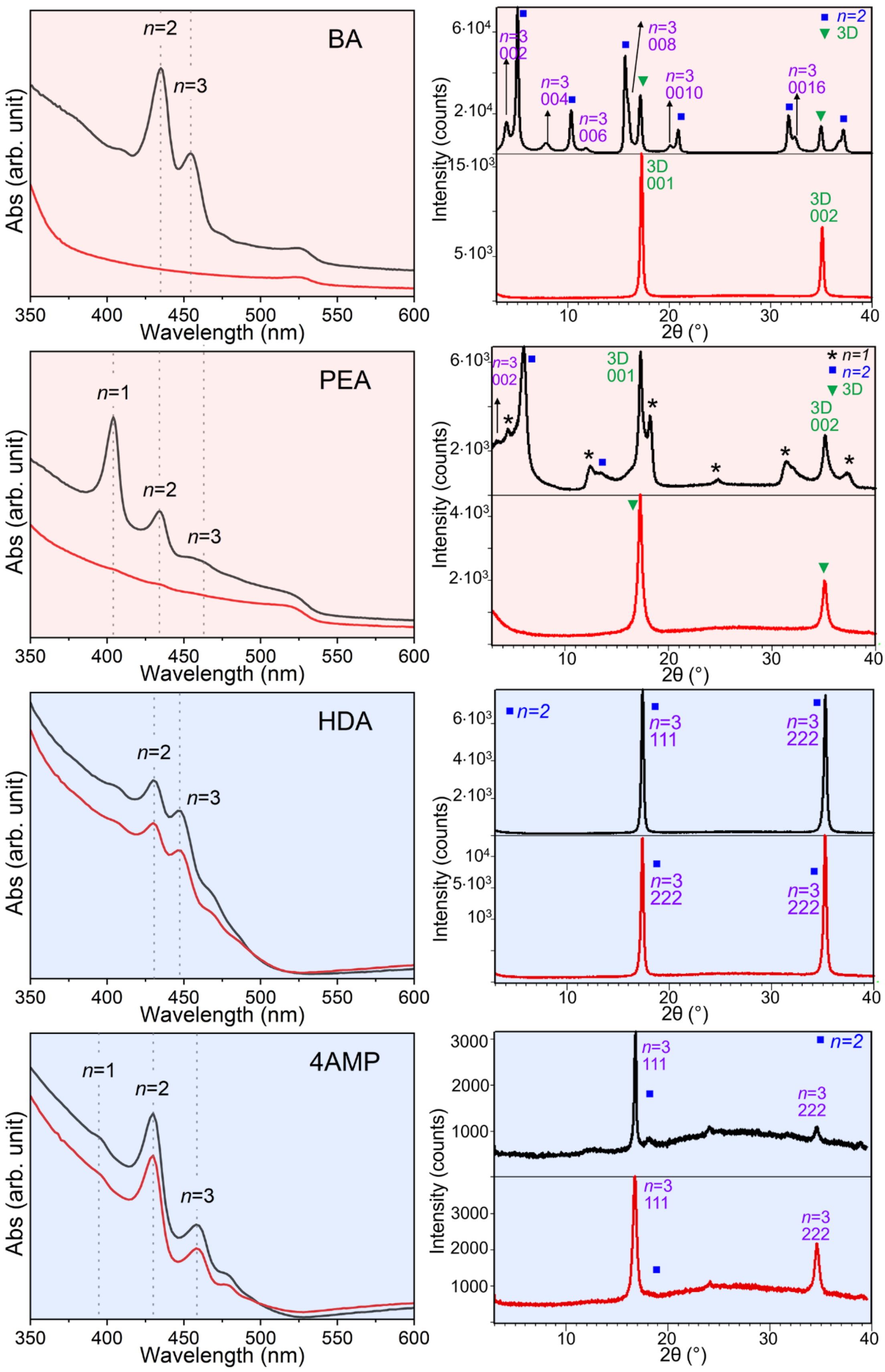

**Figure S8.** Absorption spectra (left) and XRD patterns (right)) of $n$ = 3 quasi-2D MA-based lead bromide perovskite samples before (black lines) and after (red lines) 3h simulated solar illumination (100 mW/cm$^2$) in ambient air (RH ~50-55%) with different monoammonium (BA, PEA) and diammonium (4AMP, HDA) spacers.

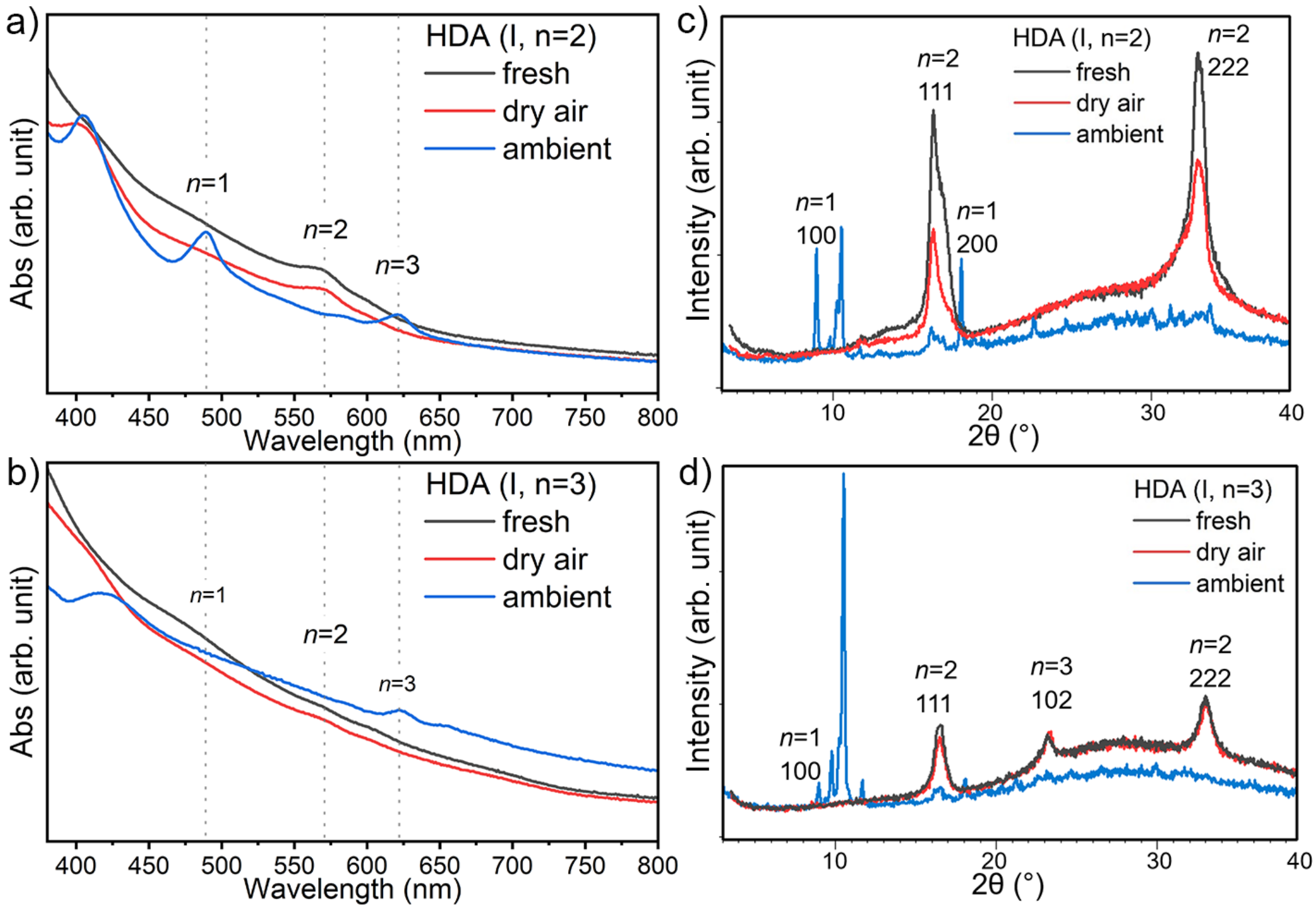

**Figure S9**. **a), b)** Absorption spectra and **c), d)** XRD patterns of quasi-2D $n = 2, 3$ HDA lead iodide perovskite samples before illumination and after 3 hours simulated solar illumination in ambient (RH 60%) and dry air.

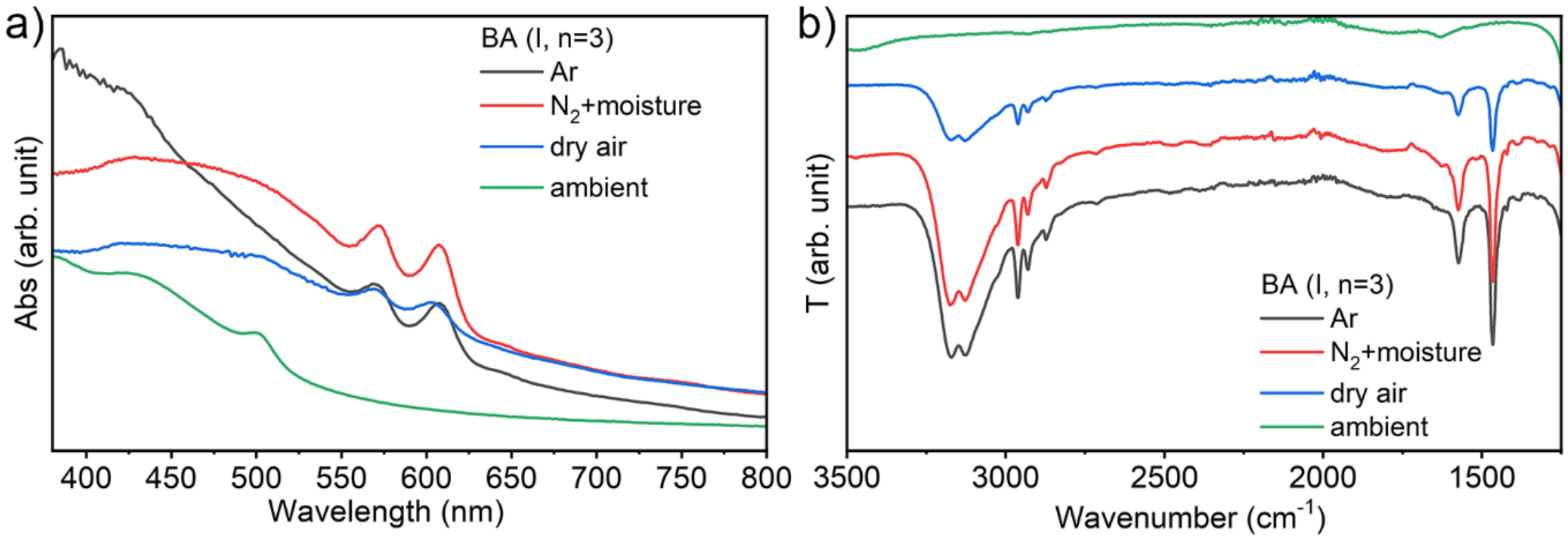

**Figure S10. a)** Absorption and **b)** FTIR spectra of $n = 3$ quasi-2D BA lead iodide perovskite samples after 3h simulated solar illumination (100 mW/cm$^2$) in different environments. For ambient, RH was 60%, for $N_2$+moisture RH was 40%.

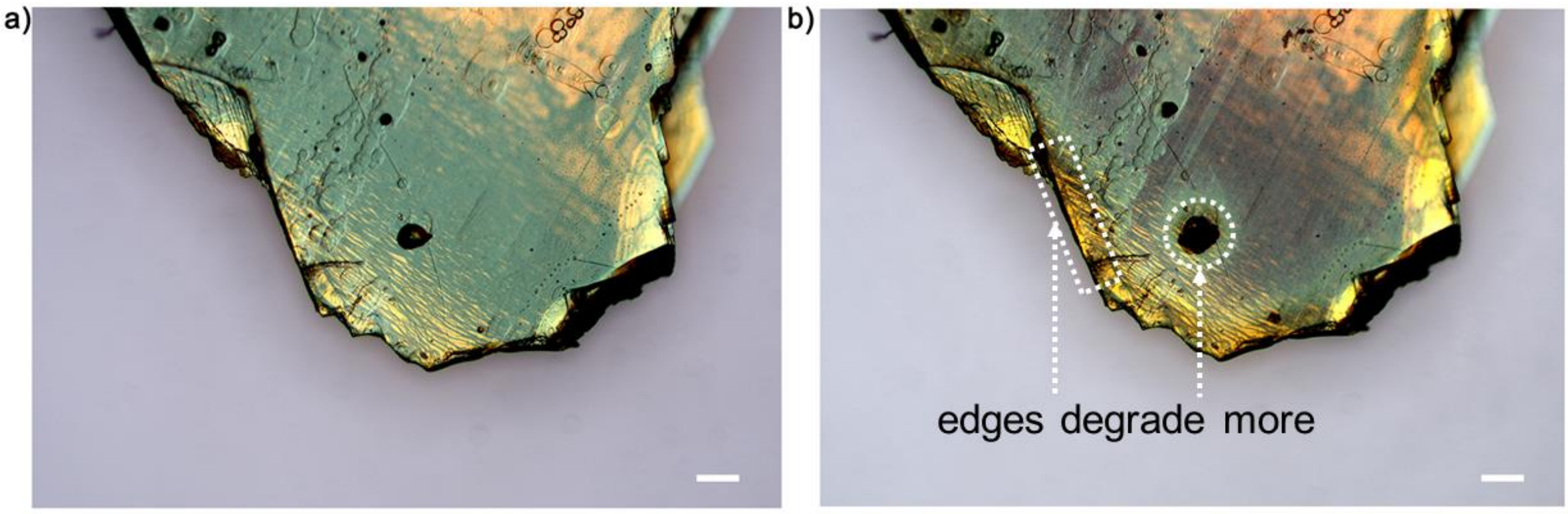

**Figure S11.** Photos of $BA_2PbI_4$ single crystal after 1 sun illumination in ambient **a)** 0 min **b)** 30 mins. Scale bar is 10 µm.

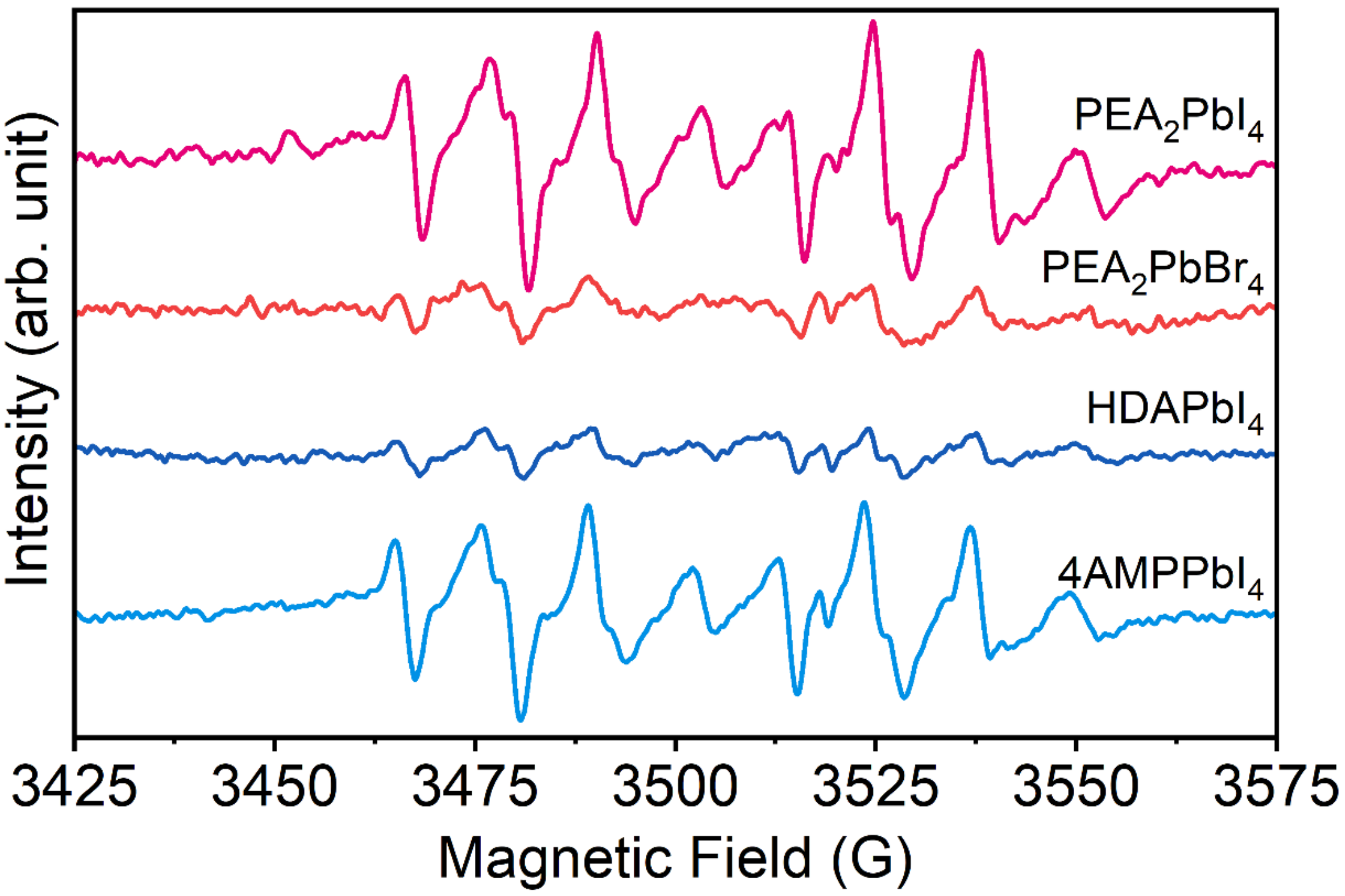

**Figure S12**. EPR spectra of different perovskite samples with DEPMPO under illumination in dry air ($PEA_2PbI_4$, $PEA_2PbBr_4$, $HDAPbI_4$, and $4AMPPbI_4$).

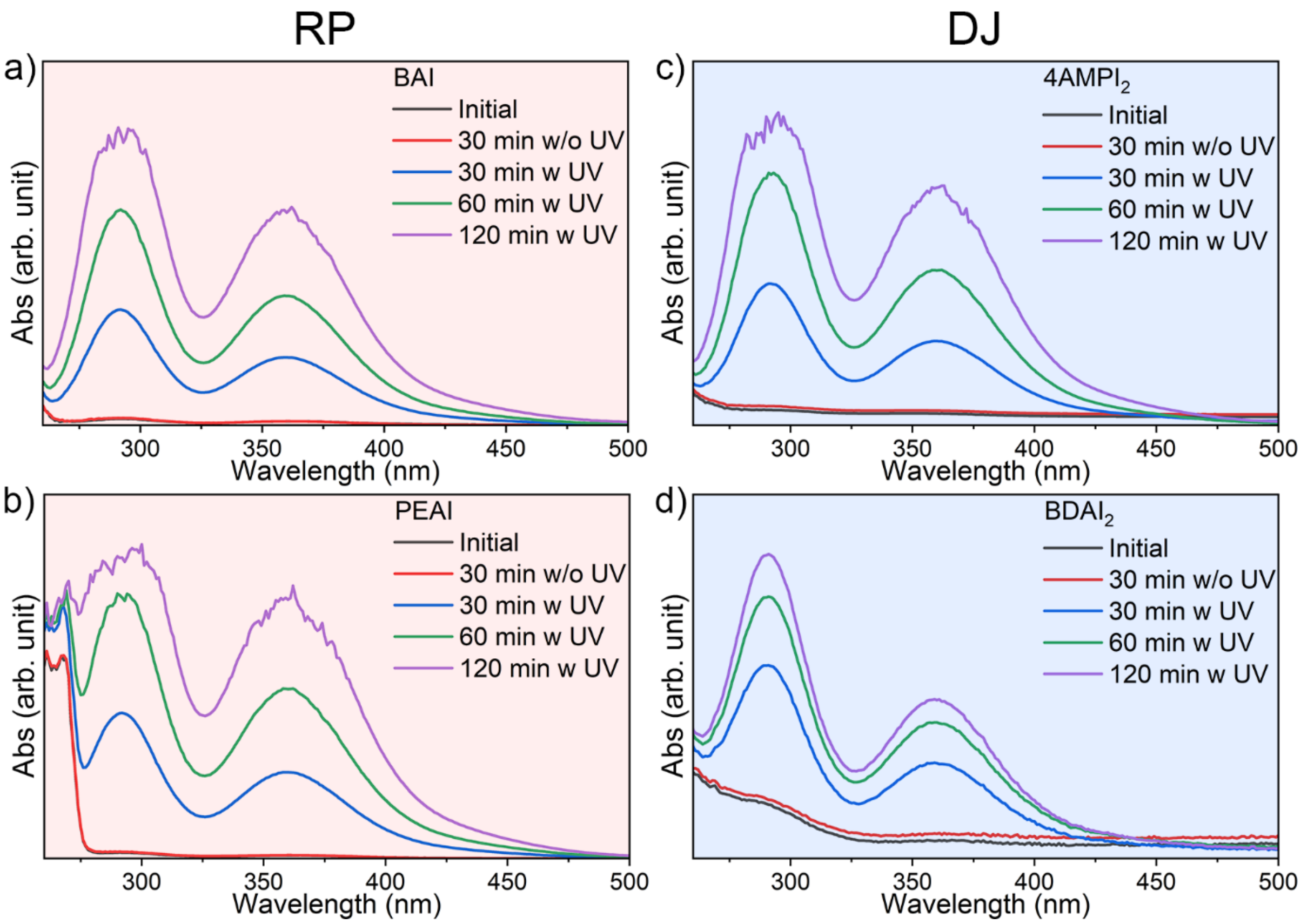

**Figure S13.** Absorption of 0.1M IPA solution of different organic ammonium iodides under UV illumination (365 nm, 5 mW). **a)** BAI **b)** PEAI **c)** $4AMPI_2$ **d)** $BDAI_2$.

# ELECTROCHEMICAL STABILITY OF 2D PEROVSKITES

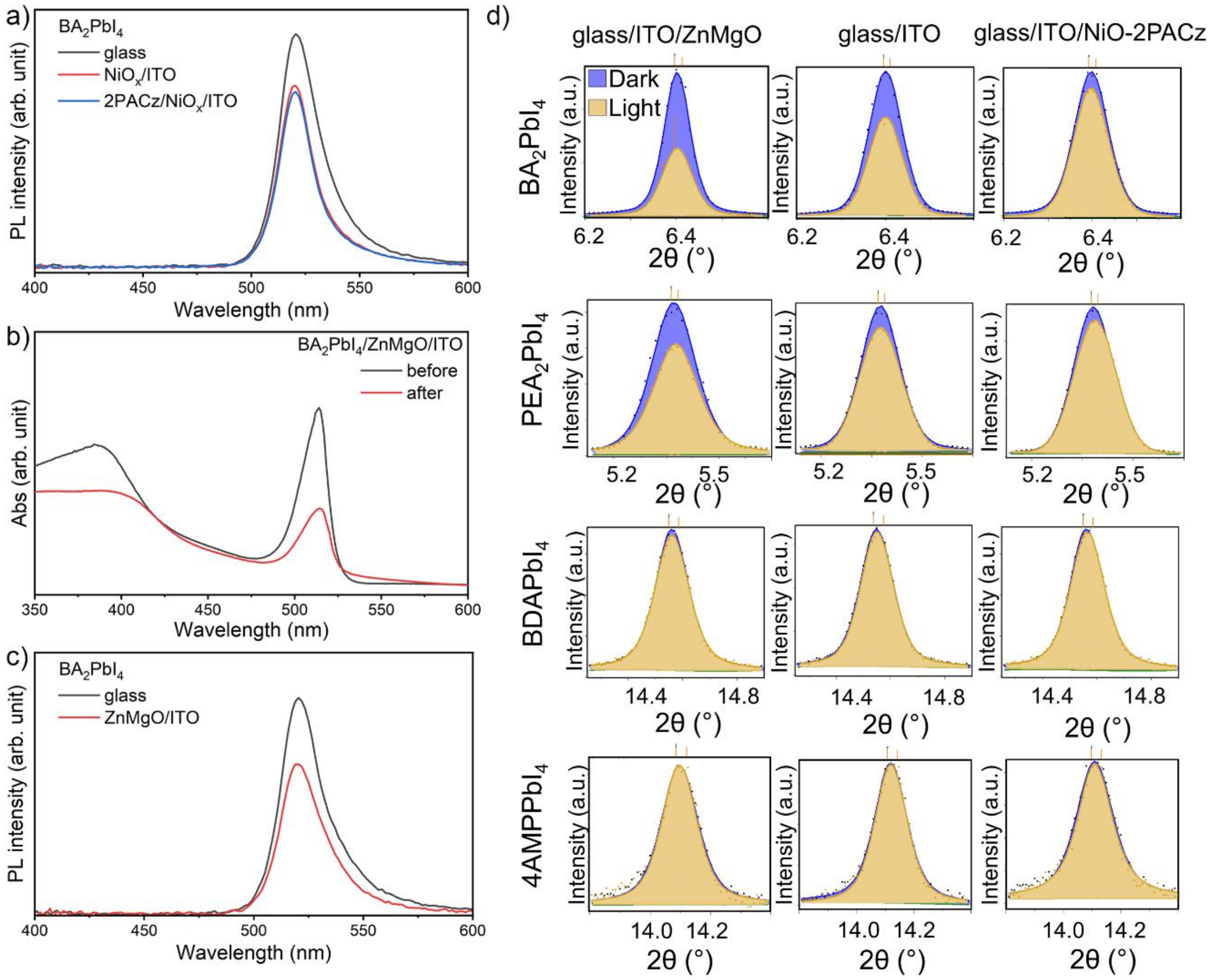

**Figure S14**. **a)** PL spectra of $BA_2PbI_4$ samples on different substrates before illumination **b)** absorption spectra of $BA_2PbI_4$ on ZnMgO/ITO before and after illumination in Ar; **c)** PL spectra of $BA_2PbI_4$ samples on ITO and ITO/ZnMgO. **d)** *In situ* XRD (dominant diffraction reflection) of $BA_2PbI_4$, $4AMPPbI_4$, $PEA_2PbI_4$, and $BDAPbI_4$ on ITO/ZnMgO, ITO, and ITO/$NiO_x$/2PACz.

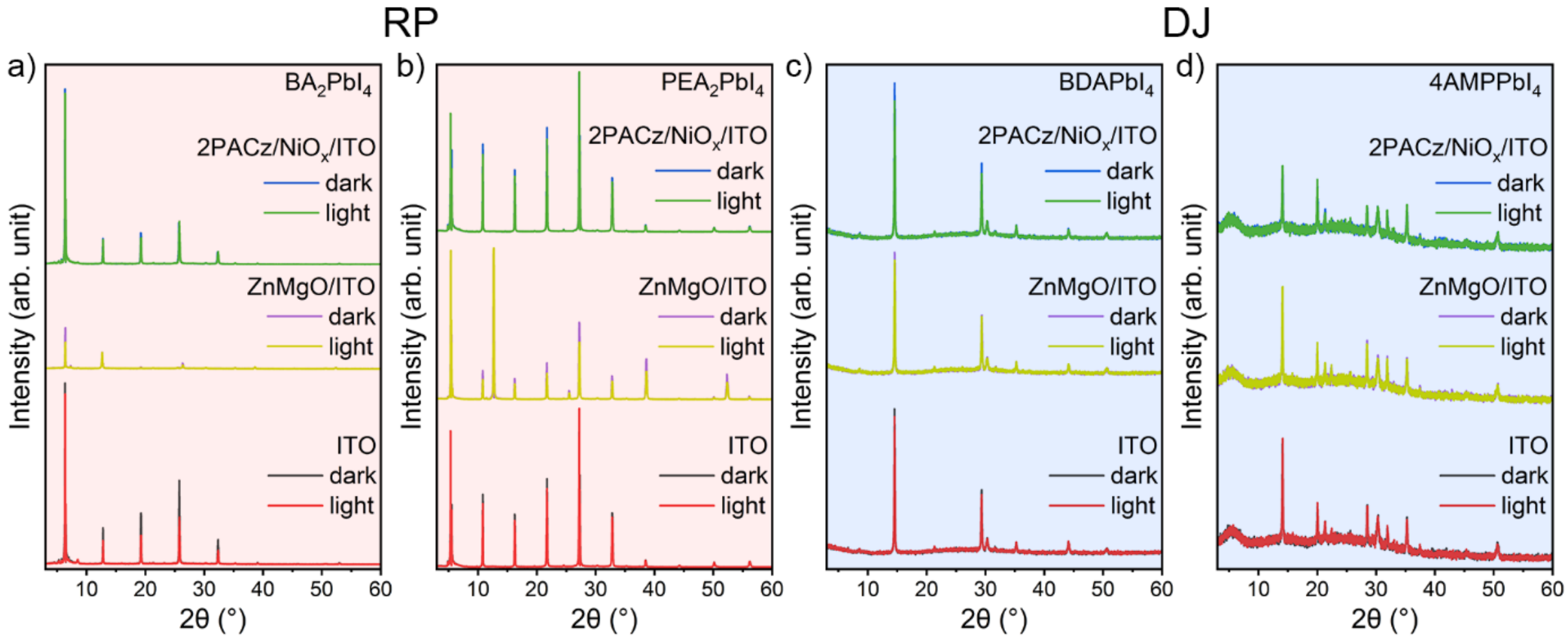

**Figure S15**. *In situ* XRD patterns of **a)** $BA_2PbI_4$ **b)** $PEA_2PbI_4$ **c)** $BDAPbI_4$ and **d)** $4AMPPbI_4$ on ITO, ITO/$NiO_x$/2PACz, and ITO/ZnMgO.

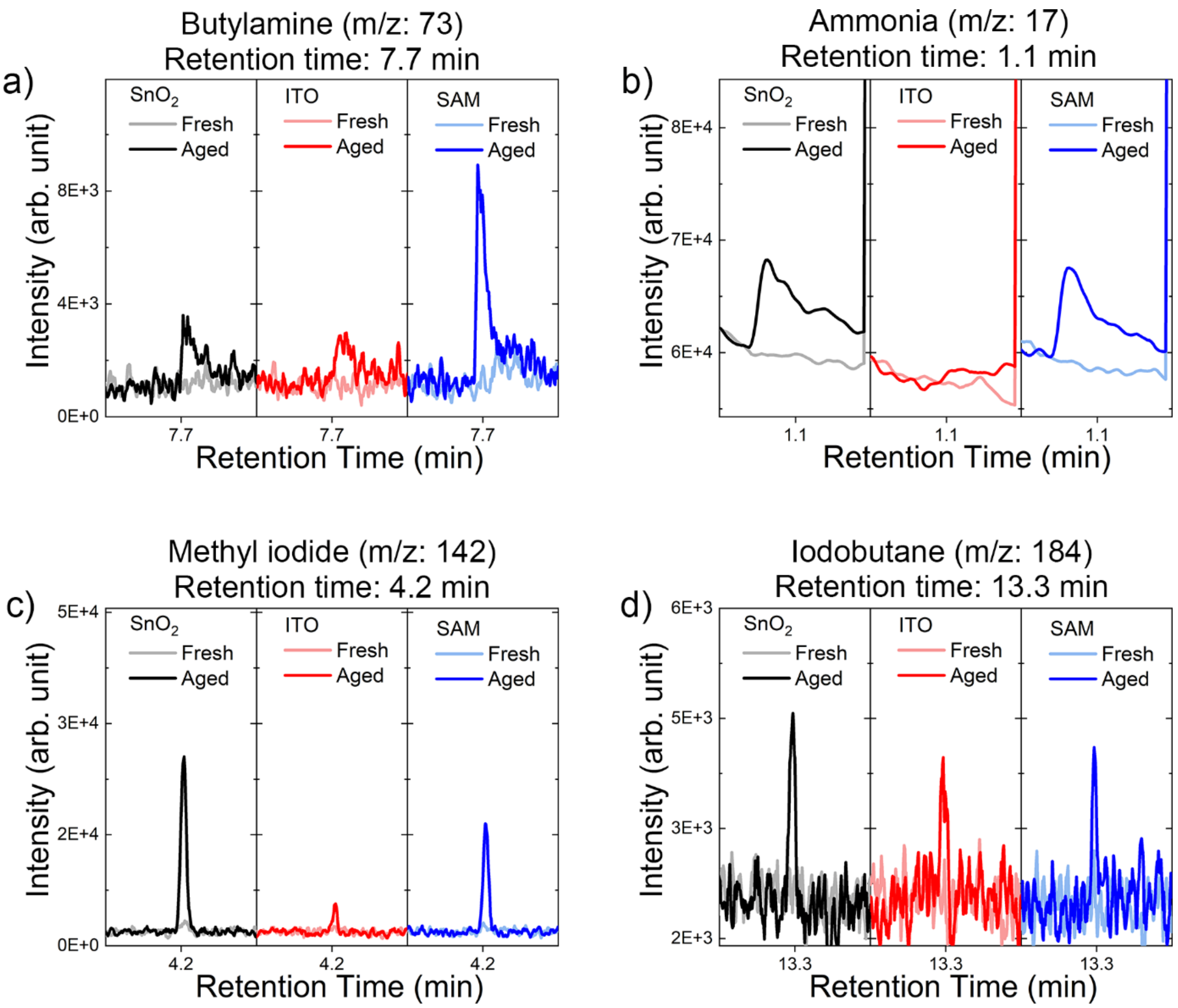

**Figure S16.** GC-MS signal traces for fresh and aged samples for $BA_2PbI_4$ on different substrates, where SAM denotes MeO-2PACz. Hydrogen iodide is not observed due to the reaction between HI and the column stationary phase.[74] While there are multiple possible degradation pathways (see **SUPPLEMENTARY NOTE 9**), amine/$NH_3$ molecules are a result of reduction reactions, while I-containing molecules are a result of oxidation reactions. We can observe a clear difference in trends on different substrates, with more significant outgassing of butylamine (reduction product) on hole transport layer (SAM) which would result in excess electrons in the perovskite, and more significant outgassing of methyl iodide (oxidation product) on electron transport layer ($SnO_2$) which would result in excess holes in the perovskite.

**Table S2**. Pb and I content in $BA_2PbI_4$ and $4AMPPbI_4$ before and after bias determined by EDX.

| Sample | Element | Atomic % | I/Pb |
|---|---|---|---|
| 4AMP fresh | Pb<br>I | 0.16<br>0.65 | 4.06 |
| 4AMP 8V | Pb<br>I | 0.13<br>0.50 | 3.85 |
| 4AMP 16V | Pb<br>I | 0.16<br>0.61 | 3.81 |
| BA fresh | Pb<br>I | 0.17<br>0.67 | 3.94 |
| BA 8V | Pb<br>I | 0.48<br>0.16 | 3.00 |

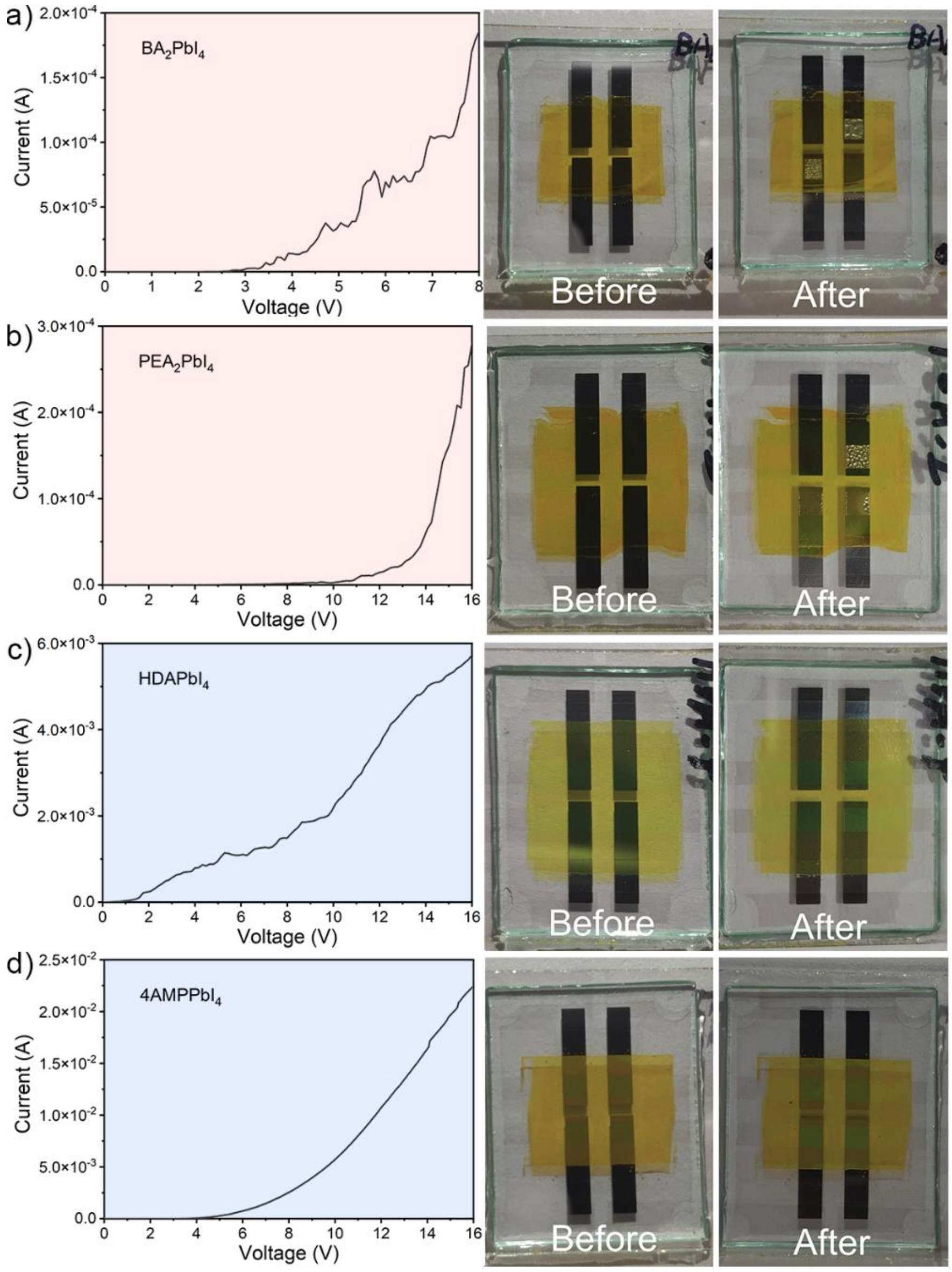

**Figure S17**. I-V curves (ITO positive bias) of ITO/$NiO_x$/ 2PACz /perovskite /TmPyPB /Liq/Al devices (left) and photos before and after bias (right) for **a)** $BA_2PbI_4$ **b)** $4AMPPbI_4$ **c)** $PEA_2PbI_4$ **d)** $HDAPbI_4$.

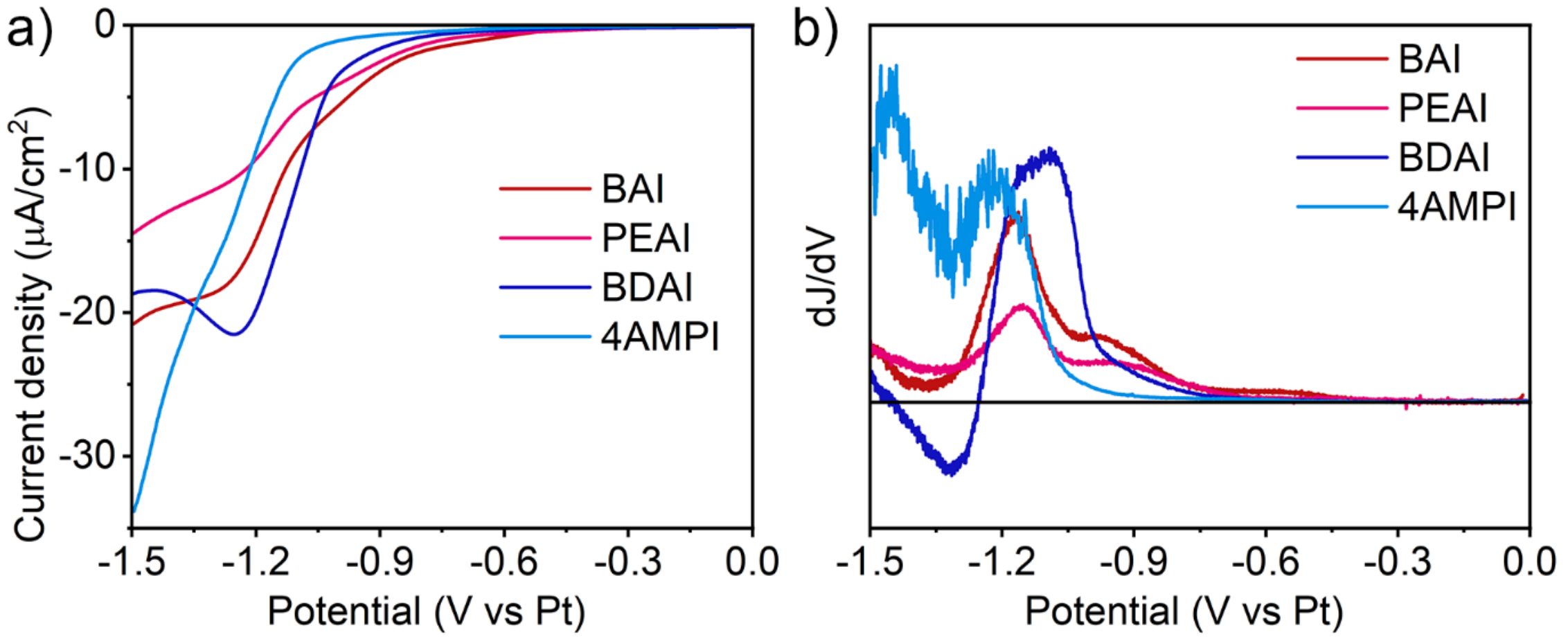

**Figure S18. a)** Cyclic voltammograms and **b)** corresponding derivatives of $n = 1$ 2D lead iodide perovskite samples with different monoammonium (BA, PEA) and diammonium (BDA, 4AMP) spacers. The samples were scanned from 0 to -1.5 V vs Pt with a scan rate of 5 mV/s in the electrochemical cell.

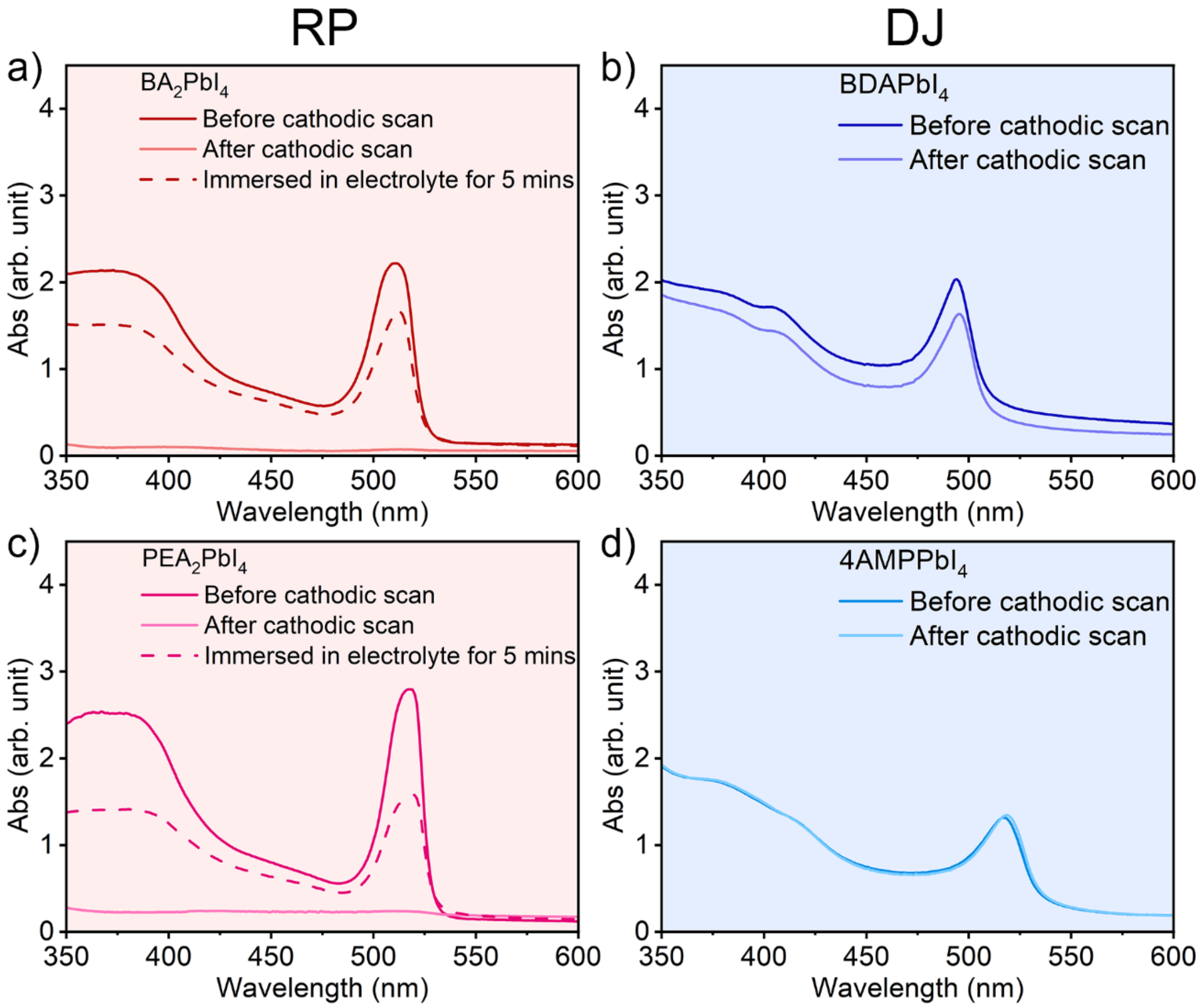

**Figure S19.** Absorption spectra of $n = 1$ 2D lead iodide perovskite samples before (black lines) and after (red lines) the cathodic scan in the electrochemical cell with different monoammonium (BA, PEA) and diammonium (BDA, 4AMP) spacers. RP films were prepared with precursor solution concentration of 0.5M to obtain thicker film, otherwise films degrade too fast to be measured. Since RP perovskites completely degraded, changes on immersion into electrolyte without bias for the equal duration as cyclic voltammetry measurement (5 min) were also tested (black dashed line). It can be observed that while there is reduction in the absorption without bias, bias is necessary to achieve complete degradation of the films.

# MECHANISM OF DEGRADATION UNDER ILLUMINATION AND/OR BIAS

## SUPPLEMENTARY NOTE 4 *Deprotonation and spacer vacancy formation energies*

We first investigated the energetics of organic spacer vacancy formation in neutral and hole-doped $BA_2PbI_4$, $PEA_2PbI_4$, $HDAPbI_4$ and $BDAPbI_4$. $HDAPbI_4$ and $BDAPbI_4$ were selected as representative for DJ perovskites as HDA and BDA are symmetric molecules with respect to the positions of its amine groups, thereby eliminating specifics which may arise by investigating deprotonation in e.g. $4AMPPbI_4$.

The spacer vacancy formation energy was calculated as:

$$E_f^{vac} = E_{vac} + E_{mol} - E_{dep},$$

where $E_{vac}$ is the energy of a system with a removed deprotonated organic spacer, $E_{dep}$ is the energy of a system containing a deprotonated organic spacer and $E_{mol}$ is the energy of the deprotonated spacer in vacuum. Furthermore, we have estimated the energy required to deprotonate an organic spacer:

$$E_f^{dep} = E_{dep} - E_{init},$$

where $E_{init}$ is the energy of the initial structure in the ground state.

**The calculation procedure is described below.**

1. Starting from the experimental structures, $2 \times 2 \times 1$ ($BA_2PbI_4$, $PEA_2PbI_4$) and $2 \times 2 \times 2$ ($HDAPbI_4$, $BDAPbI_4$) supercells were created. The total charge (0 or +1) was fixed and the structures were relaxed as described in Computational Methods.
2. An organic spacer was randomly selected to be deprotonated. We performed deprotonation of the chosen organic spacer by removing an H atom from the amine group and placing it to a nearby position in the inorganic frame between two iodide atoms (see **Figure S28a** for the tried positions in the inorganic frame**)**. The two iodide atoms were brought closer together to a distance of 4.9 A (axial deprotonation position) or 4.2 A (equatorial deprotonation position), which are values observed during *ab initio* MD. Note that, to be neutralized, the organic spacers in RP structures need to be deprotonated once, while the ones in DJ structures need to be deprotonated twice. The structure with a deprotonated spacer was relaxed, with the energy of the relaxed structure being labeled $E_{dep}$.
3. If the spacer remained deprotonated after relaxation (H atom remained bound to the inorganic frame), we removed the deprotonated spacer from the structure and performed another relaxation, with the corresponding energy of this relaxed structure being labeled $E_{vac}$.

All relaxed structures are given in the Supplementary Data in exyz format. The summary of the results is given in **Table S3** and plotted on **Figure S28.**

The main qualitative result is that, for all systems, we find a locally bound deprotonated state only upon doping the system with a hole, while for neutral systems the H atom returns to the organic spacer during the relaxation, i.e. the deprotonated state is unstable.

We find that the energy required to deprotonate a spacer once by moving an H atom to an axial position is similar for all systems ($E_f^{dep} \approx 2.0$ eV), except for $PEA_2PbI_4^+$, in which case the axial deprotonated state is found to be unstable. The formation energy of deprotonating to an equatorial position is also found to have a similar value across different systems ($E_f^{dep} \approx 2.45$ eV), being consistently higher than the energy required to deprotonate to an axial position. Deprotonating HDA and BDA on both sides costs approximately 2.1 times the energy required for a single deprotonation, indicating that deprotonation is a process dominated by the local environment.

Creating a neutral organic spacer vacancy in $BA_2PbI_4$ following deprotonation to the equatorial position is much more energetically favorable ($E_f^{vac}$ = 1.01 eV) compared to creating a vacancy following deprotonation to an axial position ($E_f^{vac}$ = 5.39 eV). As expected, the organic spacer vacancy formation energies following a single deprotonation for $HDAPbI_4$ and $BDAPbI_4$ are very large ($E_f^{vac}$ > 9.0 eV), as HDA and BDA remain positively charged and bound to the inorganic perovskite frame on one side. Deprotonating $HDAPbI_4$ and $BDAPbI_4$ twice leads to organic spacer formation energies comparable to the ones obtained after deprotonating $BA_2PbI_4$ and $PEA_2PbI_4$ to the equatorial position once, with $E_f^{vac}$ being in the range of 1.0-1.5 eV.

From an electrochemical point of view, oxidation of lattice iodide by an excess hole results in the formation of interstitial iodide defects, which can have different charge as a result of participating in electrochemcial reactions with photogenerated charge carriers. Therefore, we have investigated the energetics of organic spacer vacancy formation in systems containing an additional iodine atom, with the total charge of the system being fixed to $Q_{tot}$ = {-1, 0, +1}, modelling the presence of an $I_i^-$, $I_i^0$ or $I_i^+$ defect respectively. We have tried 3 different positions at which we placed the additional I atom (see **Figure S29a**). After relaxing the defect-containing system, we have calculated the organic spacer vacancy formation energy using a procedure analogous to the one described in steps 2. and 3. given above for the pristine systems. The summary of the results is given in **Table S4** and plotted on **Figure S29b**. Note that on **Figure S29b**, we show only the results for single deprotonation of RP perovskites and double deprotonation of DJ perovskites, as single deprotonation of DJ perovskites leads to large values of $E_f^{vac}$, as is the case for considerations of pristine systems described above.

For $Q_{tot}$ = -1, we find a bound deprotonated state only for the RP perovskites $BA_2PbI_4$ and $PEA_2PbI_4$, while for $HDAPbI_4$ and $BDAPbI_4$, the twice-deprotonated state is unstable. For $Q_{tot}$ = 0 and $Q_{tot}$ = +1, we find at least one stable deprotonated state for all systems. We find that the calculated $E_f^{dep}$ values are all lower than the corresponding $E_f^{dep}$ values calculated for the pristine systems, meaning that the presence of a defect generally lowers the barrier toward deprotonation.

Interestingly, we find that, while the deprotonated state is unstable for $Q_{tot}$ = -1 for the DJ perovskites, we find two cases for RP perovskites in which deprotonation with $Q_{tot}$ = -1 leads to negative $E_f^{vac}$. This can be interpreted as the $I_i^-$ defects easily destabilizing the RP perovskites, while the DJ systems remain stable, as no bound deprotonated state which would lead to organic spacer vacancy formation is found. For all systems, we find cases with $E_f^{vac}$ being lower than the values calculated for pristine systems, meaning that the presence of an interstitial iodine defect may lower the barrier toward organic spacer vacancy formation, as well as the barrier toward deprotonation.

**From all the calculations described above, we can draw the following conclusions:**

(1) To form an organic spacer vacancy, it is first necessary to neutralize the spacer by deprotonation, as forming a charged spacer vacancy is extremely energetically unfavorable. To neutralize them, spacers in RP perovskites need to be deprotonated once, while the ones in DJ perovskites need to deprotonated twice.

(2) States with deprotonated spacers are found to be stable (bound) in two considered cases:

   a. Doping pristine systems with holes;

   b. Introducing interstitial iodine defects into the pristine systems.

Therefore, doping systems with holes has a detrimental effect on the system stability due to stabilization of states with deprotonated spacers, allowing for organic spacer vacancy formation. The presence of interstitial iodine defects results not only in stabilization of states with deprotonated spacers, but also lower barriers for organic spacer vacancy formation.

(3) From the considerations above and no observed difference in the dynamical behavior of neutral and hole-doped systems (see **SUPPLEMENTARY NOTES 5&6**), we conclude that the experimentally observed photostability of DJ perovskites is primarily the result of the necessity of double deprotonation of a particular spacer to destabilize the material, which is a less likely event compared to a single deprotonation event necessary to neutralize organic spacers in RP perovskites.

## **SUPPLEMENTARY NOTE 5** *Considerations of the effects of structural rigidity and non-covalent interlayer halide interactions on charge localization*

We hypothesize that hole localization is a necessary step in 2D perovskite photodegradation. Charge localization in metal halide perovskites is stabilized by distortions in the inorganic sublattice.[75] Therefore, we investigated whether DJ perovskites are generally more structurally rigid compared to RP perovskites to ascertain whether the experimentally observed trend of DJ phostability could be attributed to structural robustness which prevents charge localization.

It is known that the lattice thermal fluctuations in 2D-perovskites are reduced in $PEA_2PbI_4$[76-78] and $3AMPPbI_4$[77] compared to $BA_2PbI_4$. This has been attributed to enhanced stiffness in the organic layer, which stems from the π-π interactions between the organic cations in the case of $PEA_2Pb2I_4$ and the complete elimination of the van der Waals gap in $3AMPPbI_4$.

Lead atom deviations from the lead-iodide planes have been used as a measure of rigidity of the inorganic sublattice by extracting the standard deviations of Gaussian functions fitted to histograms of out-of-layer coordinate components of Pb atoms during MD.[19] We obtain $\sigma_{BA_2PbI_4} \approx 0.17$Å and a comparable value of $\sigma_{4AMPPbI_4} \approx 0.21$Å for MD simulations of systems doped with a single hole. Another measure of structural rigidity are average atomic displacements shown in **Figure S21**. These results show that, although DJ perovskites are often significantly more rigid compared to RP perovskites, it is not the general case and therefore cannot fundamentally resolve the observed differences in photostability trends.

It is also known that non-covalent interlayer halogen interactions enhance carrier delocalization.[18] Since the interlayer distance in DJ perovskites is usually shorter compared to RP perovskites, one may assume that this is the fundamental mechanism of hole delocalization, preventing photodegradation. However, this cannot be the general case, since $HDAPbI_4$ is photostable and its shortest interlayer I ⋯ I distance is 6.0 Å, well outside their Van der Waals diameter (4.4 Å). There are also examples of stable DJ perovskites ($DDDAPbI_4$) with comparable or larger interlayer spacing to that of unstable RP perovskites ($PEA_2PbI_4$, $BA_2PbI_4$). DDDA exhibits interlayer spacing of 16.02 Å which is close to that of PEA (16.52 Å) and significantly larger than that of 4AMP (10.53 Å), HDA (11.85 Å), as well as BA (13.75 Å). Thus, there is no clear correlation between interlayer spacing and stability under illumination. We therefore conclude that a hypothetical resistance to charge localization in DJ perovskites cannot explain the general trend of their photostability compared to RP perovskites.

## **SUPPLEMENTARY NOTE 6** *Charge localization*

We have investigated charge localization during dynamics by calculating the Mulliken charges[79] on each atom at each step of MD. For each atom i we calculate the average Mulliken charge $\langle Q_0^i \rangle$ during neutral MD. We then define the time dependent "extra charge" on each atom during the charged (+1) MD as: $Q_{+1}^{i,extra}(t) = Q_{+1}^{i}(t) - Q_0^i$. Perfect localization of a hole a single atom at time t would correspond to $Q_{+1}^{i,extra}(t) = 1$. A histogram of the extra charge distribution on iodine atoms for $BA_2PbI_4$ and $4AMPPbI_4$ is shown in **Figure S26**. It can be seen that no perfect localization event is observed with this method in either system, with no significant difference between the systems observed.

We have also calculated the difference between the ground state charge density of charged and neutral systems at the same selected geometry, shown in **Figure S27.** The geometries were selected as points from the MD with extremal $Q_{+1}^{i,extra}(t)$ The figure shows that hole localization is possible both for RP and DJ perovskites.

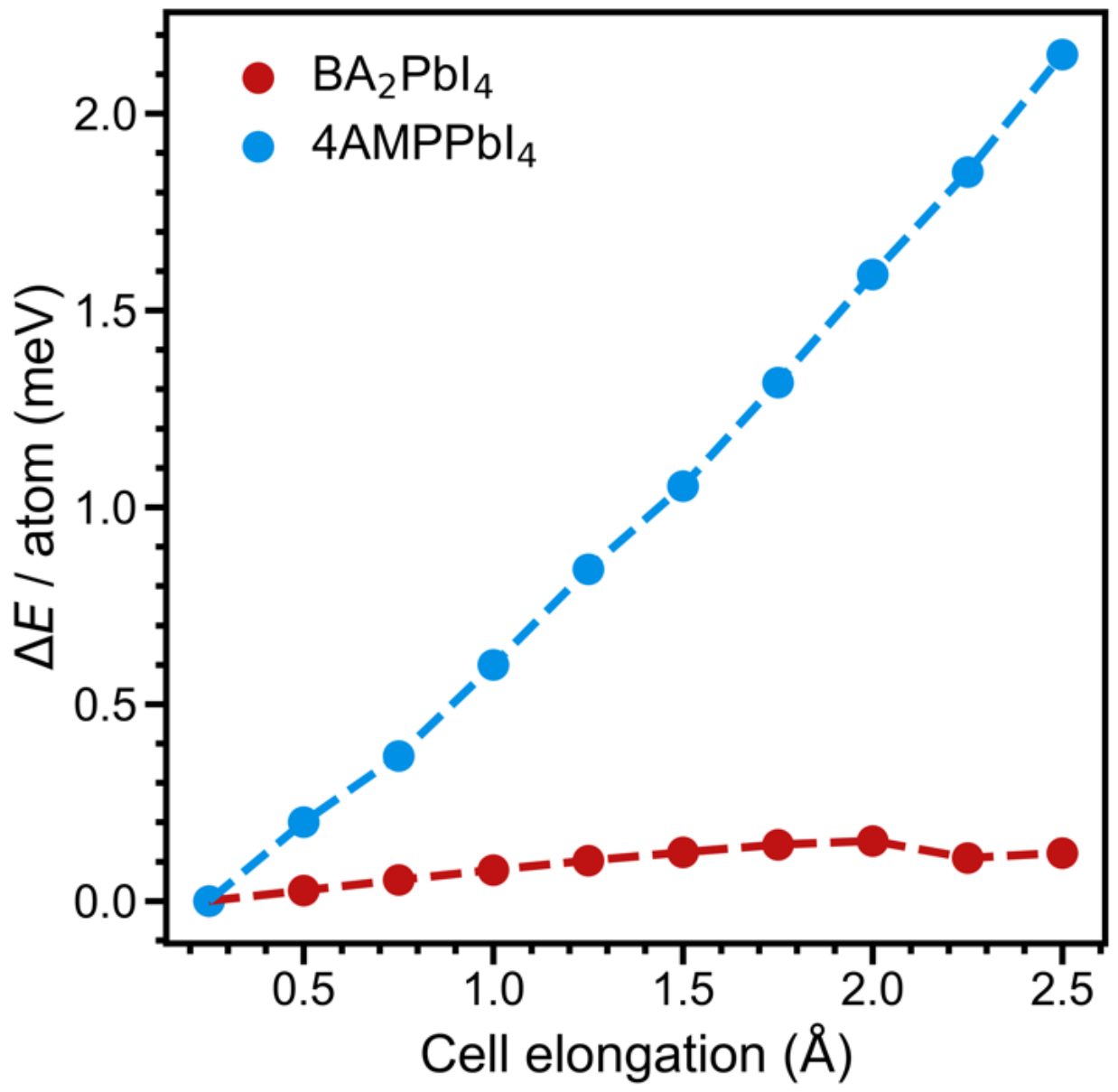

**Figure S20.** Delamination energy for $BA_2PbI_4$ and $4AMPPbI_4$. Starting from the experimental structures, we increased the length of the unit cell vector perpendicular to the inorganic layers in steps of 0.25 Å and relaxed the structure. Delamination energy shows an obvious difference between DJ and RP perovskites, as expected from the difference in the bonding for these two classes of materials. However, delamination (separation of whole layers) does not fit the pattern of degradation observed during illumination/bias: degradation from grain edges, changes in Pb:I ratios, and organic cation loss.

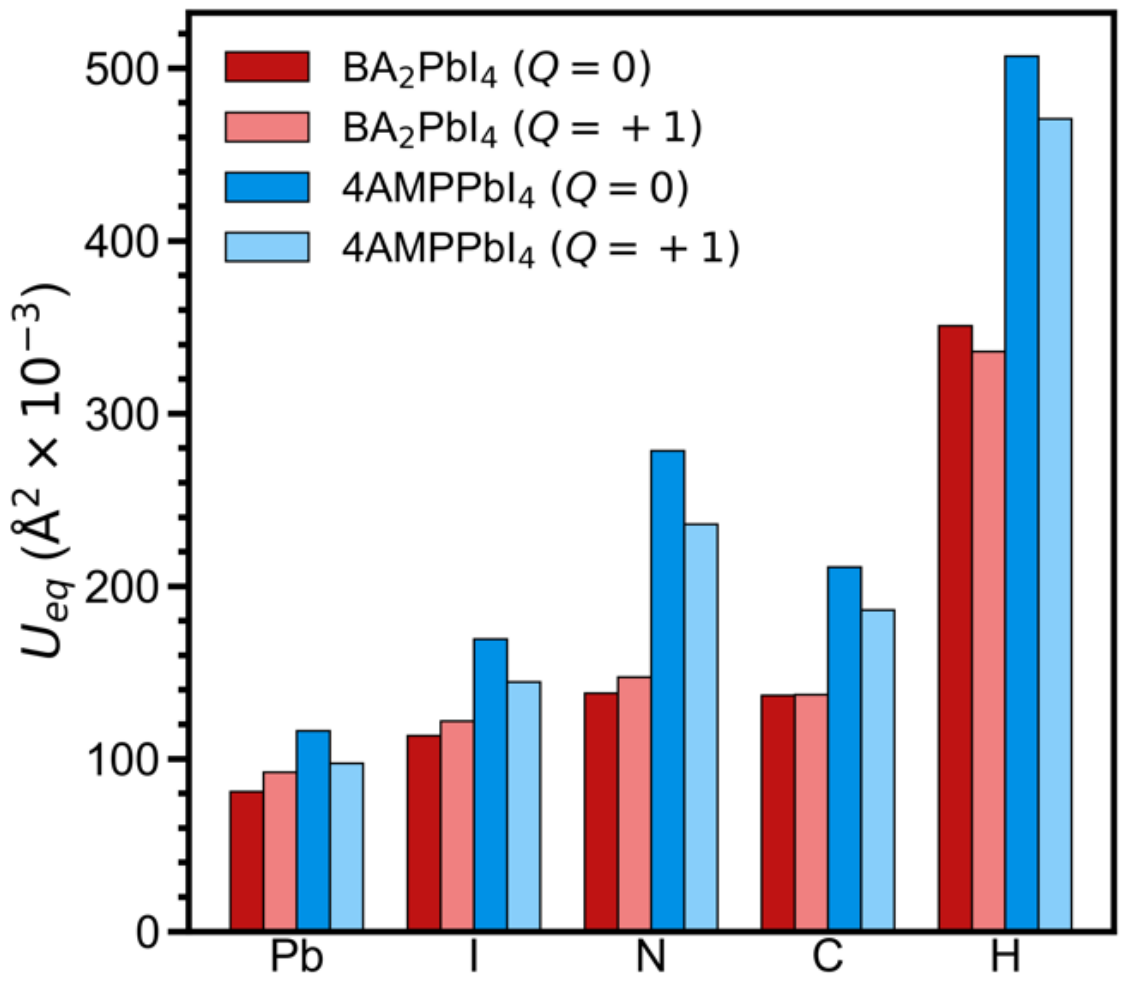

**Figure S21.** Average atomic displacements extracted from MD simulations of neutral ($Q = 0$) and doped ($Q = +1$) $BA_2PbI_4$ and $4AMPPbI_4$. It can be observed that the atomic displacements in $BA_2PbI_4$ are lower than those in $4AMPPbI_4$, indicating higher rigidity for less stable $BA_2PbI_4$. Doping the systems with a hole ($Q = +1$) slightly increases (decreases) the thermal fluctuations of Pb and I atoms in $BA_2PbI_4$ ($4AMPPbI_4$).

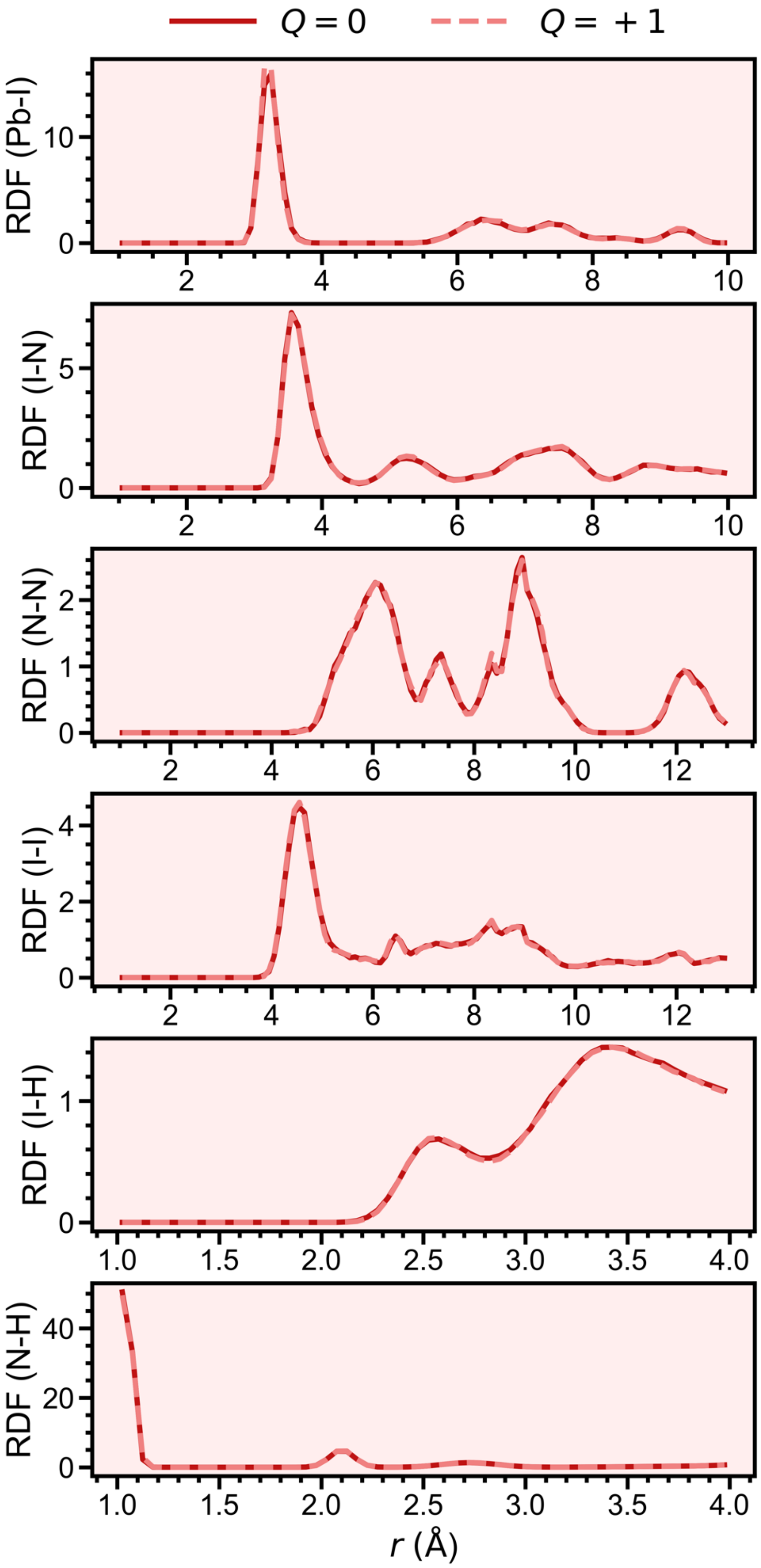

**Figure S22.** Radial distribution functions of different pairs of atomic species for $BA_2PbI_4$. From top to bottom, radial distribution functions for Pb-I, N-I, N-N, I-I, I-H, N-H are shown. Full (dashed) lines show radial distribution functions for a neutral (hole-doped) system.

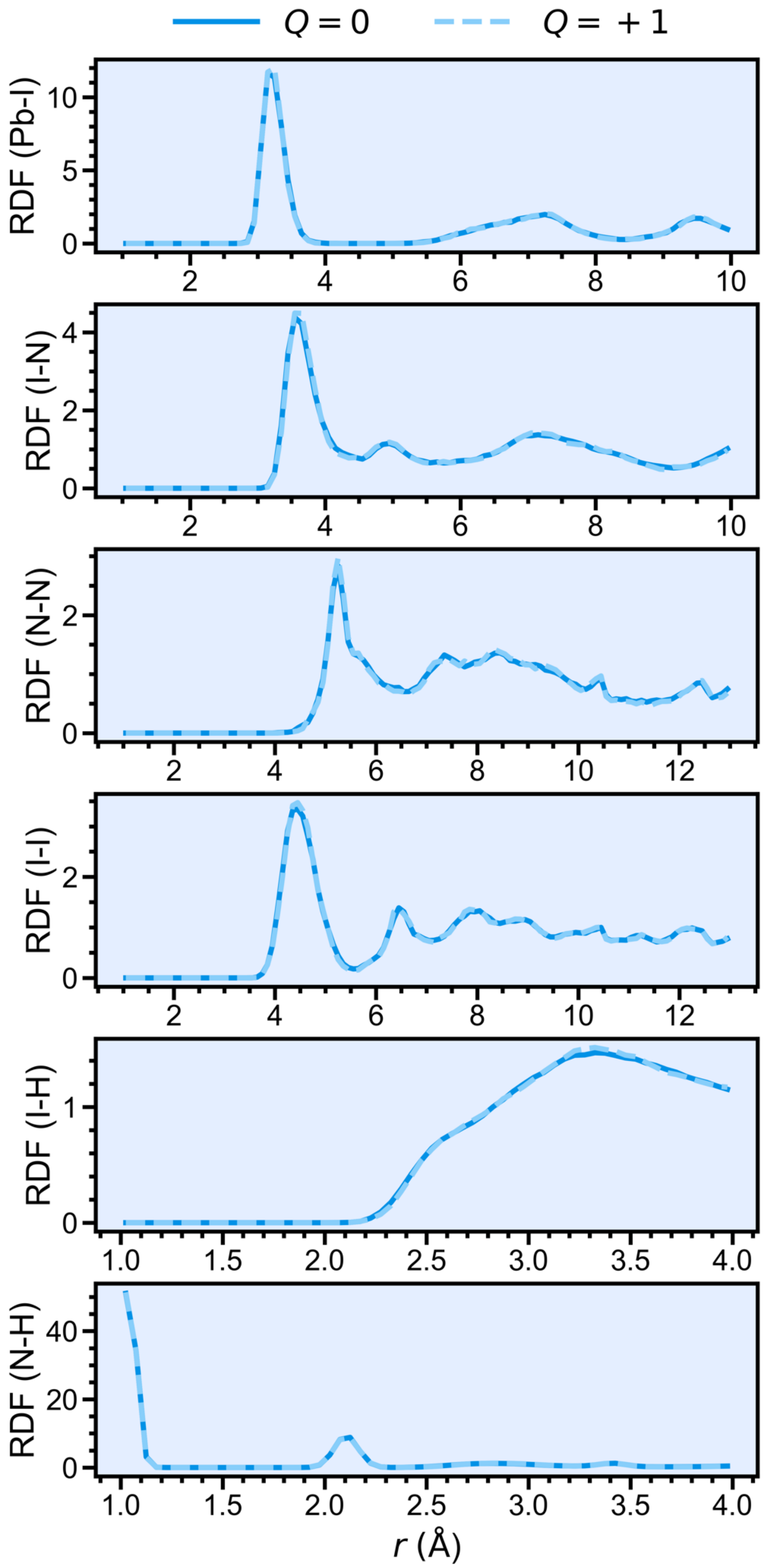

**Figure S23.** Radial distribution functions of different pairs of atomic species for $4AMPPbI_4$. From top to bottom, radial distribution functions for Pb-I, N-I, N-N, I-I, I-H, N-H are shown. Full (dashed) lines show radial distribution functions for a neutral (hole-doped) system.

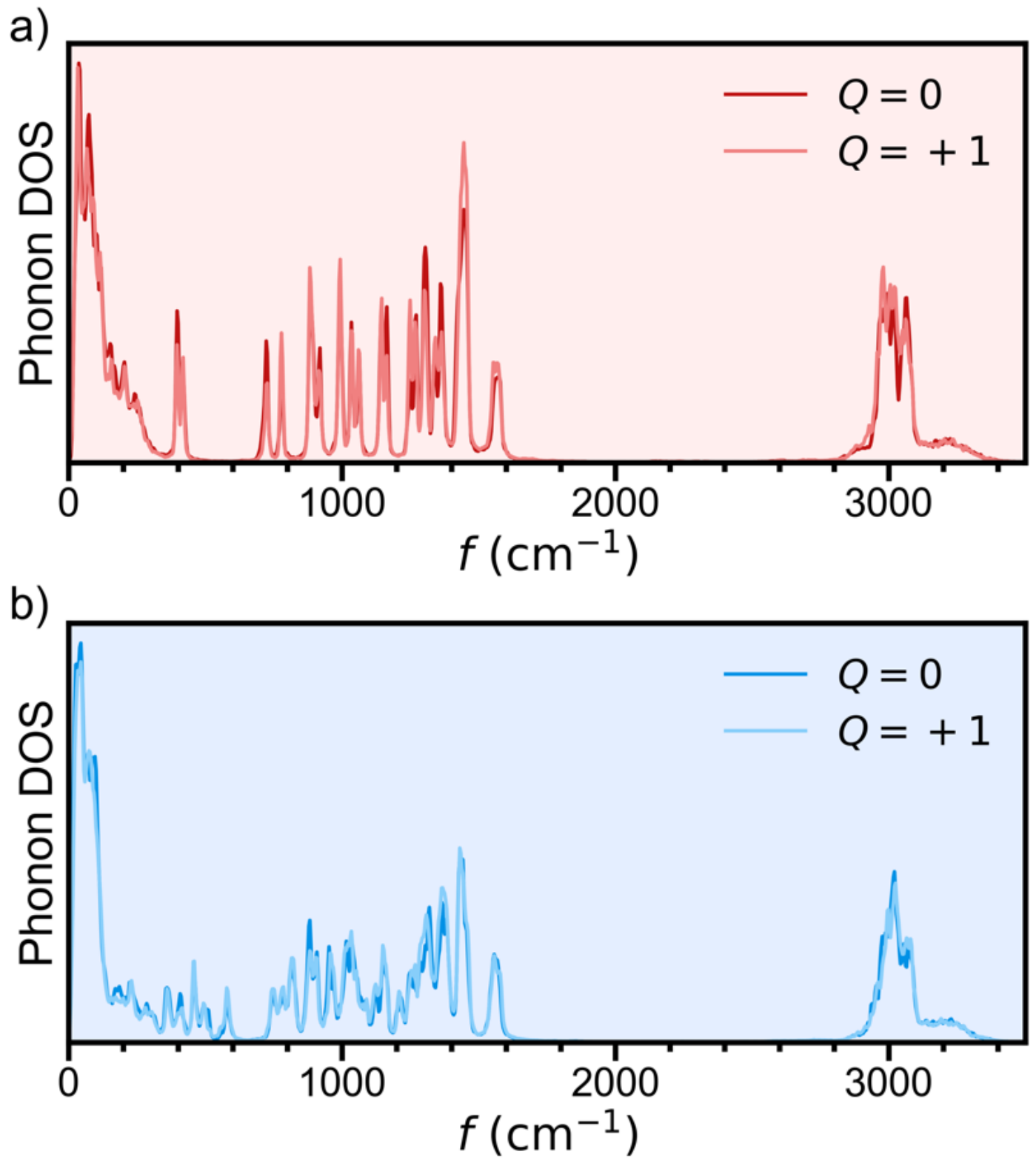

**Figure S24.** Phonon density of states for **a)** $BA_2PbI_4$ and **b)** $4AMPPbI_4$. The phonon densities of states are shown for neutral (Q=0) and hole-doped (Q=+1) systems.

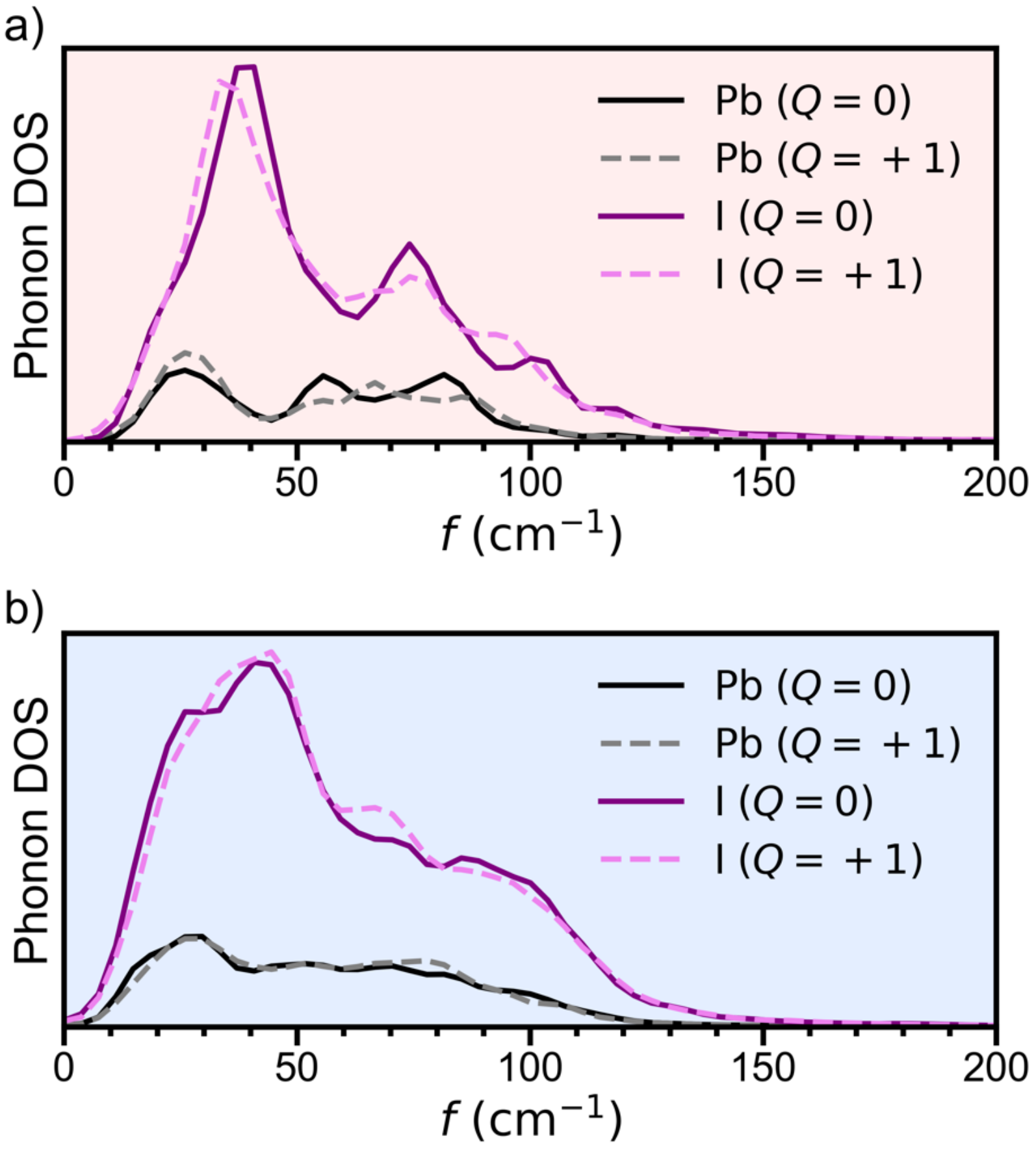

**Figure S25.** Low frequency Pb-I subsystem phonon density of states for **a)** $BA_2PbI_4$ and **b)** $4AMPPbI_4$. Full lines concern neutral systems (Q=0), while dashed lines concern systems doped with 1 hole (Q=+1). At low frequencies, there is a slight softening of Pb and I-related modes in $BA_2PbI_4$ and slight hardening in the case of $4AMPPbI_4$.

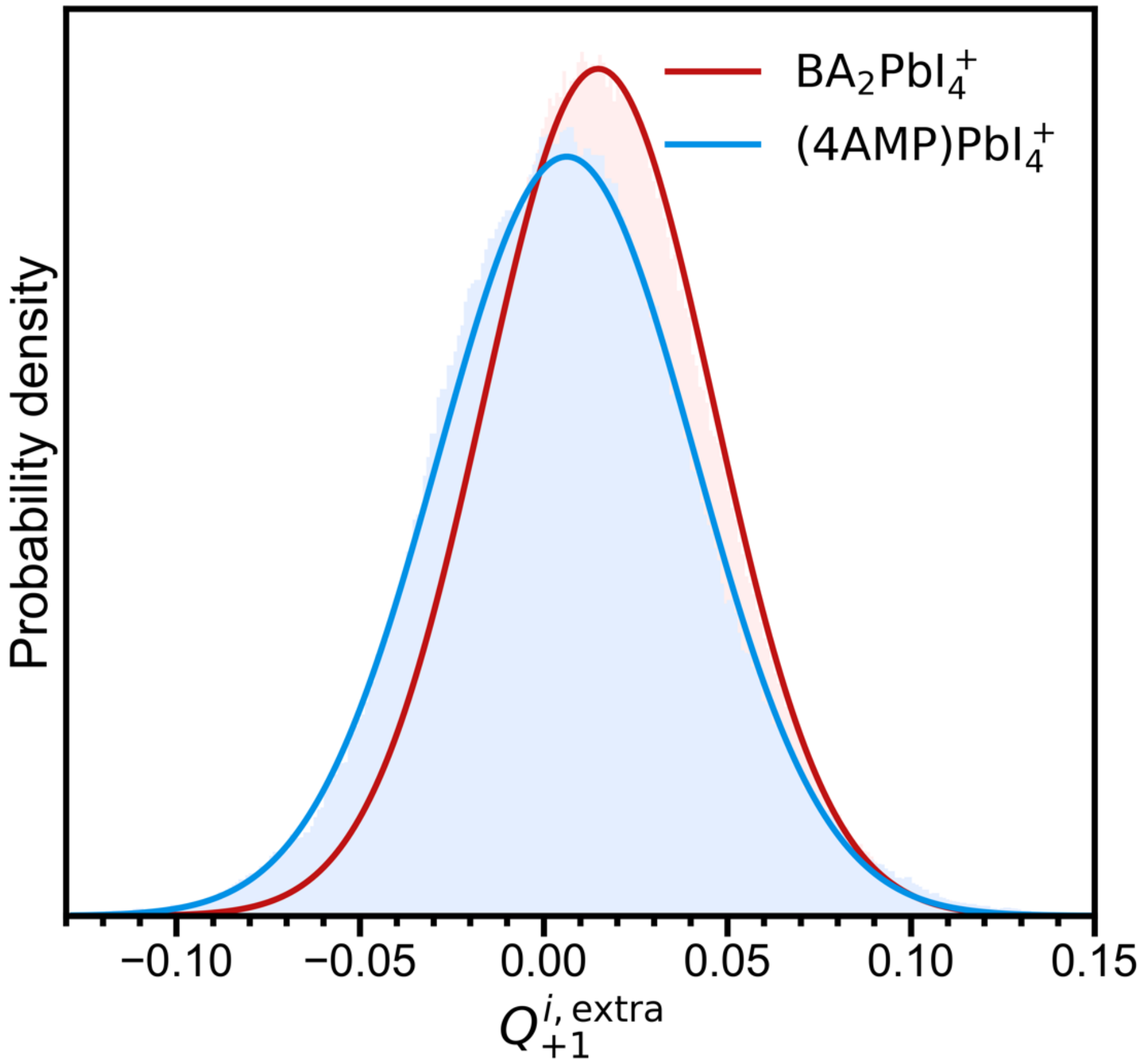

**Figure S26**. Extra Mulliken charge distribution on iodine during MD for systems doped with 1 hole.

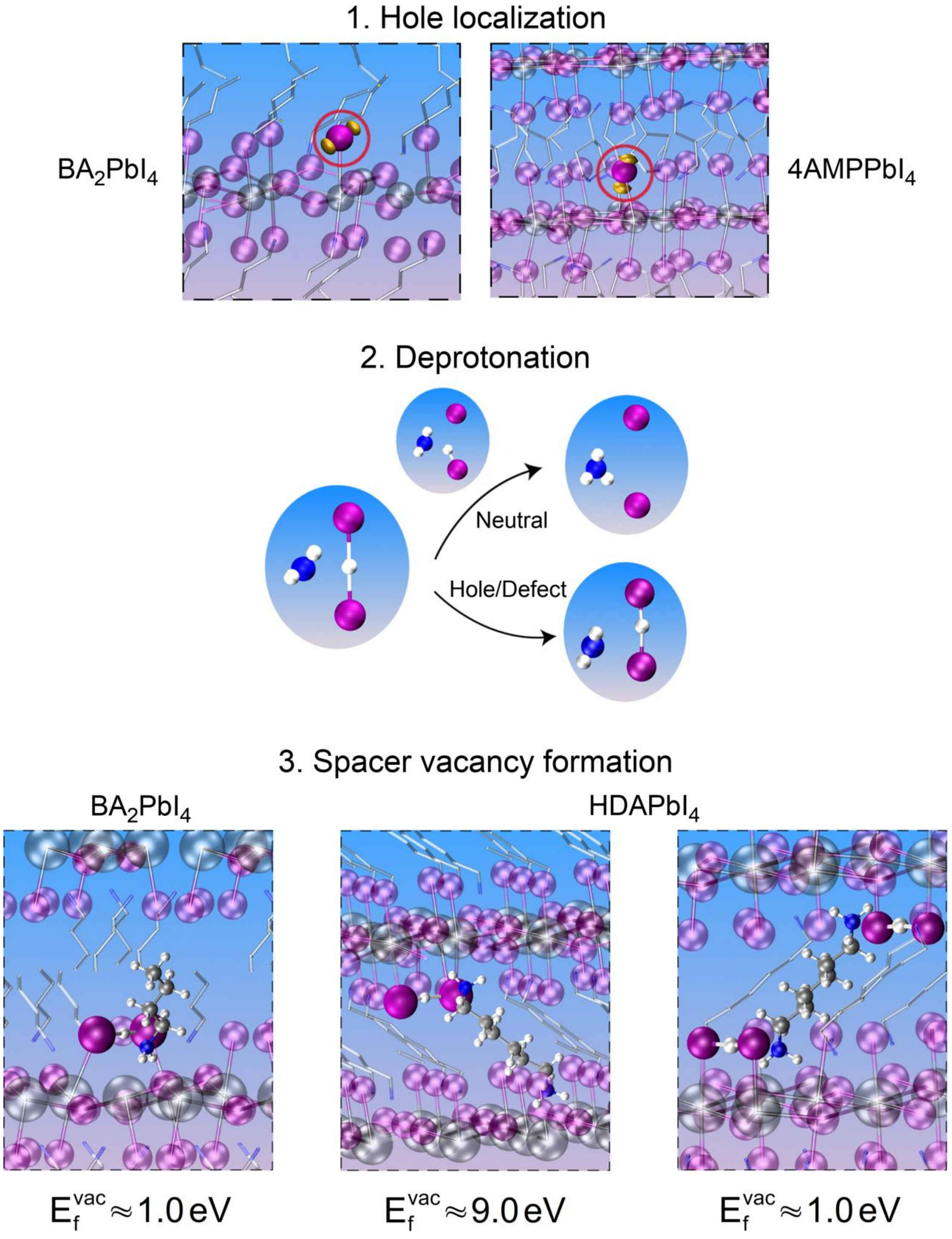

**Figure S27.** Schematic of the proposed degradation pathway. **1.** Hole localization for $BA_2PbI_4$ and $4AMPPbI_4$ observed during MD, shown as a difference between the ground state charge density of the hole doped system and a neutral system with the same geometry, shown at isovalue +0.008. **2.** Creation of a bound deprotonated state in the case of a system doped with a hole or containing a defect. Neutral pristine systems relax through an intermediate state in which H binds to one of the I atoms and is transferred back to the ground (protonated) state, while the deprotonated state remains stable in systems containing a hole or a defect. **3.** Closeups of singly (left and middle panel) or doubly (right panel) deprotonated organic spacers for which the vacancy formation energy is calculated. DJ perovskites must be deprotonated twice to make spacer vacancy formation energetically viable.

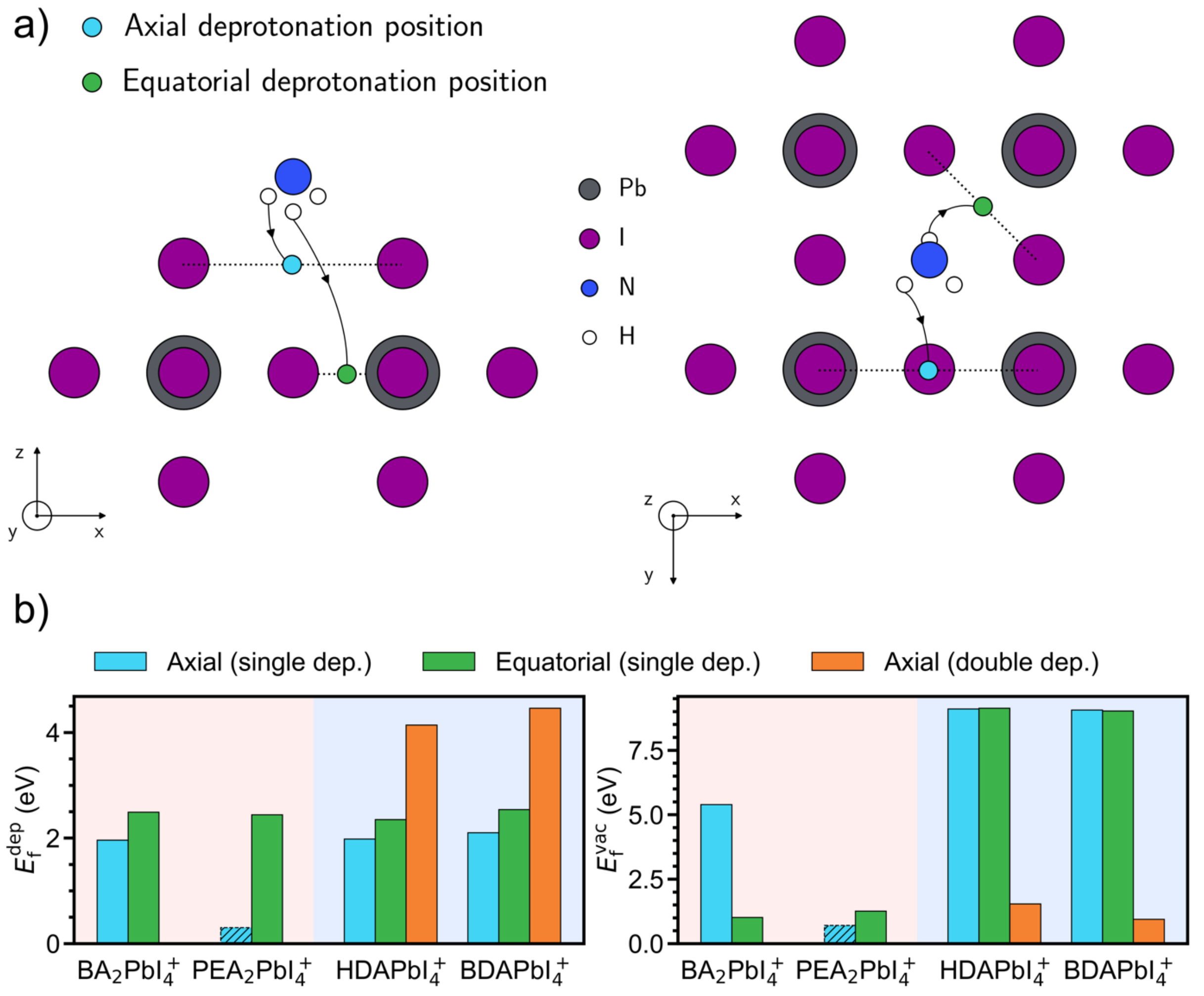

**Figure S28 a)** Schematic diagram of deprotonation positions (left- side view, right-top view). Only one $NH_3$ is shown for clarity. **b)** Minimum energy required to deprotonate an organic spacer ($E_f^{dep}$) and the energy required to create a spacer vacancy ($E_f^{vac}$). The different colors correspond to different positions of the H atom removed from the organic spacer, as shown in a). Only positive systems (with a hole present in the structure) are shown, because the deprotonated state is unstable in neutral systems (the H atom relaxed back to the organic spacer). The crossed-out bar means that the deprotonated state is unstable and the height of the crossed-out bar has no physical meaning (included just for visualisation). We can observe that $E_f^{vac}$ for singly deprotonated DJ perovskite is very large, as expected, since the spacer remains charged and is forming a bond to the perovskite at one end.

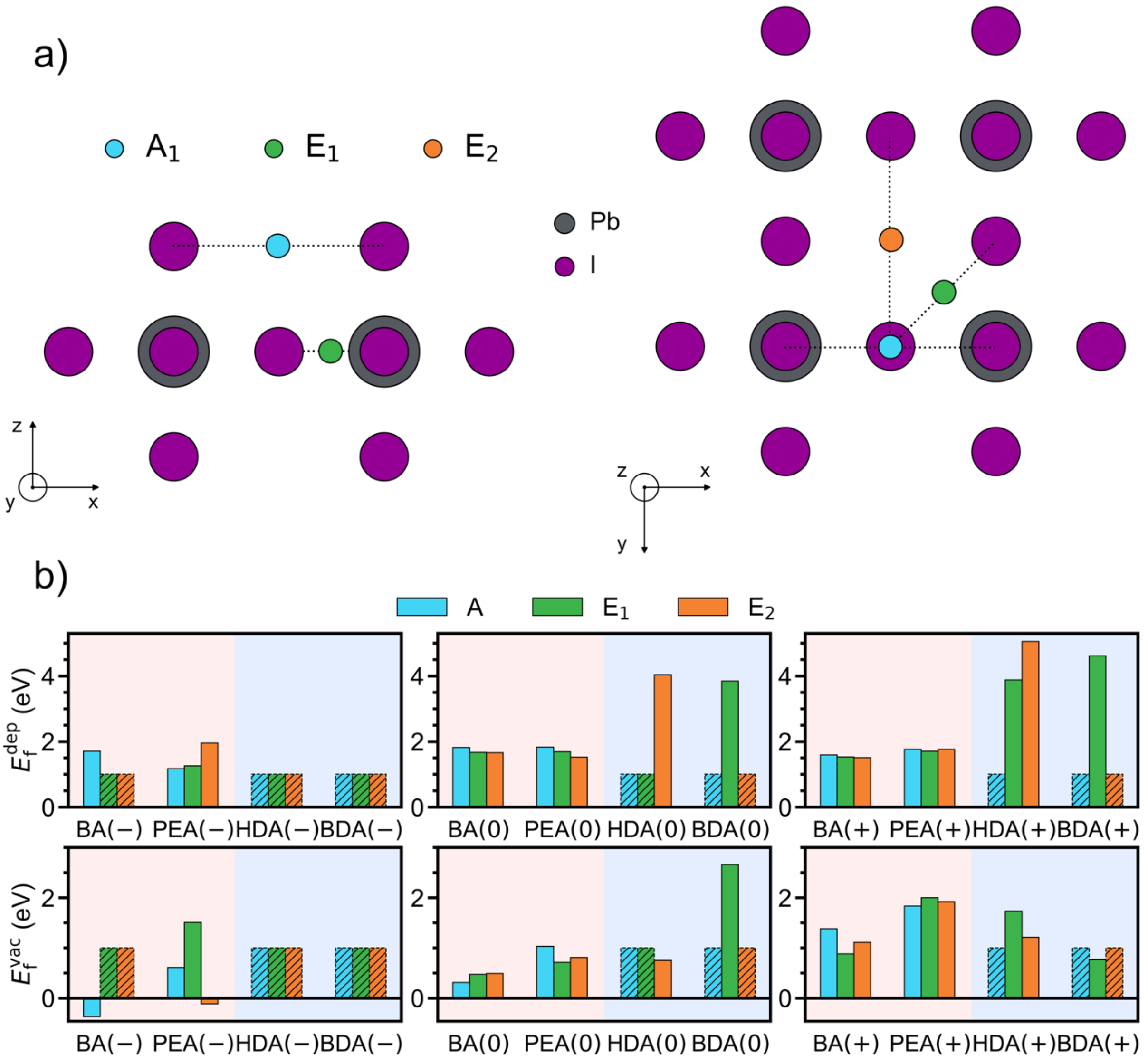

**Figure S29 a)** Schematic diagram of iodine defect positions (left- side view, right-top view). **b)** $E_f^{dep}$ and $E_f^{vac}$ for systems with an added defect iodine atom and different total charges (-1, 0, 1). The different colors correspond to different initial positions of the defect iodine (axial, two different equatorial positions), as shown in a). The crossed-out bars mean that the deprotonated state is unstable (the H atom relaxed back to the organic spacer) and the height of crossed out bar has no physical meaning (included just for visualisation). For visual clarity, the systems are labeled only with the name of their respective organic spacer with the total charge of the system being given in brackets; e.g. BA(+) corresponds to $BA_2PbI_4$ with +1 total charge. For DJ perovskites, only doubly deprotonated cases are shown.

**Table S3**. Summary of the results for deprotonation and spacer vacancy formation energies for all investigated pristine systems. "A" denotes deprotonation to the axial position, "E" denotes deprotonation the equatorial deprotonation position, while "DA" denotes double deprotonation to the axial positions for DJ perovskites (see **SUPPLEMENTARY NOTE 4** and **Figure S28**). The deprotonated state is unstable for all neutral systems, which is denoted using a hyphen (-).

| System | Charge | Deprotonation position | $E_f^{dep}$ (eV) | $E_f^{vac}$ (eV) |
|---|---|---|---|---|
| $BA_2PbI_4$ | 0 | A | - | - |
| | 0 | E | - | - |
| | 1 | A | 1.96 | 5.39 |
| | 1 | E | 2.49 | 1.01 |
| $PEA_2PbI_4$ | 0 | A | - | - |
| | 0 | E | - | - |
| | 1 | A | - | - |
| | 1 | E | 2.44 | 1.26 |
| $HDAPbI_4$ | 0 | A | - | - |
| | 0 | E | - | - |
| | 0 | DA | - | - |
| | 1 | A | 1.98 | 9.10 |
| | 1 | E | 2.35 | 9.13 |
| | 1 | DA | 4.14 | 1.54 |
| $BDAPbI_4$ | 0 | A | - | - |
| | 0 | E | - | - |
| | 0 | DA | - | - |
| | 1 | A | 2.10 | 9.06 |
| | 1 | E | 2.54 | 9.02 |
| | 1 | DA | 4.46 | 0.94 |

**Table S4**. Summary of the results for deprotonation and spacer vacancy formation energies for all investigated systems containing an added interstitial defect I atom. See **Figure S29** for a visualization of the defect positions. The cases in which the deprotonated state was found to be unstable are denoted using a hyphen (-). For DJ perovskites ($HDAPbI_4$ and $BDAPbI_4$), only doubly deprotonated cases are given.

| System | Charge | Defect position | $E_f^{dep}$ (eV) | $E_f^{vac}$ (eV) |
|---|---|---|---|---|
| $BA_2PbI_4$ | -1 | A | 1.71 | -0.37 |
| | -1 | $E_1$ | - | - |
| | -1 | $E_2$ | - | - |
| | 0 | A | 1.82 | 0.31 |
| | 0 | $E_1$ | 1.67 | 0.47 |
| | 0 | $E_2$ | 1.66 | 0.49 |
| | 1 | A | 1.59 | 1.38 |
| | 1 | $E_1$ | 1.53 | 0.88 |
| | 1 | $E_2$ | 1.51 | 1.11 |
| $PEA_2PbI_4$ | -1 | A | 1.17 | 0.61 |
| | -1 | $E_1$ | 1.26 | 1.51 |
| | -1 | $E_2$ | 1.96 | -0.12 |
| | 0 | A | 1.83 | 1.02 |
| | 0 | $E_1$ | 1.69 | 0.71 |
| | 0 | $E_2$ | 1.53 | 0.81 |
| | 1 | A | 1.76 | 1.83 |
| | 1 | $E_1$ | 1.71 | 2.00 |
| | 1 | $E_2$ | 1.76 | 1.92 |
| $HDAPbI_4$ | -1 | A | - | - |
| | -1 | $E_1$ | - | - |
| | -1 | $E_2$ | - | - |
| | 0 | A | - | - |
| | 0 | $E_1$ | - | - |
| | 0 | $E_2$ | 4.04 | 0.75 |
| | 1 | A | - | - |
| | 1 | $E_1$ | 3.88 | 1.73 |
| | 1 | $E_2$ | 5.05 | 1.21 |
| $BDAPbI_4$ | -1 | A | - | - |
| | -1 | $E_1$ | - | - |
| | -1 | $E_2$ | - | - |
| | 0 | A | - | - |
| | 0 | $E_1$ | 3.84 | 2.66 |
| | 0 | $E_2$ | - | - |
| | 1 | A | - | - |
| | 1 | $E_1$ | 4.61 | 0.77 |
| | 1 | $E_2$ | - | - |

## SUPPLEMENTARY NOTE 7 *The effects of organic cations*

In 3D perovskites, it has been recognized that simultaneous halide oxidation (resulting in the formation of $I_2$ and $I_3^-$) and organic cation deprotonation is necessary for the perovskite degradation.[25,40] It was also proposed that methylamine loss is the main driving force for the irreversible device degradation, since it leads to the more pronounced loss of volatile reaction products.[40] In a lateral device geometry, depletion of MA after constant bias was observed near the cathode, confirming that the loss of organic cations is due to electrochemical reduction process.[40,80] In a vertical device geometry, significant redistribution of $MA^+$ and $I^-$ was observed, while for $Pb^{2+}$ only minor changes are found, confirming that main mobile species are halide and organic ions,[40] in agreement with other reports.[81] The hypothesis that the deprotonation of organic cation and generation of organic cation vacancy has detrimental effect on the perovskite stability is also confirmed by the acceleration of bias-induced decomposition at grain boundaries by MAI passivation, and deceleration of decomposition by passivation with KI, which cannot be deprotonated.[82] Furthermore, expulsion of iodide/iodine species in the solution occurs in organic-inorganic perovskites, accompanied by the loss of MA,[29] while no iodine/iodide expulsion occurs for $CsPbI_{1.5}Br_{1.5}$ despite observation of segregation.[27]

The loss of organic cation also affects the halide ion migration. From computational investigations of iodide migration on different surface terminations in $MAPbI_3$, it was found that activation barrier for iodide migration was strongly related to the motion of MA related to formation and disruptions of hydrogen bonds, indicating that iodide migration could be suppressed by stronger hydrogen bonding.[36] Conversely, the absence of organic cation (no hydrogen bonds) is expected to facilitate unimpeded movement of iodide. In agreement with this expectation, *in situ* studies of perovskite degradation under electron beam exposure have demonstrated that the loss of organic cations triggers diffusion of $Pb^{2+}$ and $I^-$, allowing the structure to evolve from corner sharing to edge sharing octahedra, and finally to $PbI_2$.[38] In addition, reduced ion migration (evidenced by lower hysteresis) was found in nonstoichiometric (MA-excess, fewer MA vacancies) $MAPbI_3$ devices, compared to stoichiometric and Pb-excess nonstoichiometric devices.[83] In addition, increased stability of $MAPbBr_3$ compared to $MAPbI_3$ was attributed to the lack of MA migration in the bromide perovskite, since both MA and I migration occurs in $MAPbI_3$, while in $MAPbBr_3$ only slow $Br^-$ migration was observed.[84] The absence of MA migration in $MAPbBr_3$ was attributed to the lattice contraction.[84]

In 2D perovskites, organic cations also play a key role. It has been well accepted that these materials exhibit reduced ion mobility compared to 3D materials, which has been attributed to organic cations acting as barriers to ion hopping in out-of-plane direction,[85,86] and to higher formation energy for vacancies, which inhibits ion migration in in-plane direction as well.[86-88] The reduction of ion migration due to the formation of quasi-2D perovskite[89] or by forming 3D/2D perovskite layers[90] has been demonstrated experimentally. The barrier role of organic cations to halide migration is also confirmed by significantly slower remixing of photosegregated 2D perovskites compared to 3D perovskite.[22,91] For example, in BA-based 2D perovskite demixing occurs within several minutes for BA,[48,92] while full remixing takes ~13 h.[48] For comparison, in 3D perovskites demixing typically occurs in ~20-30 min, while remixing occurs in ~30 min -2 h.[27,93]

Since the bulky spacer cations play an important role in the suppression of ion migration, it would be expected that their deprotonation would result in the acceleration of ion migration. However, investigations of the effect of deprotonation in these materials have been scarce. It was proposed that the deprotonation of spacer cation and subsequent reaction with small organic cation (FA reacting with $PEA^+$ (or $BA^+$) to form $NH_3$ and $PEAMA^+$ ($BAMA^+$) results in the dissolution of the perovskite by $NH_3$.[94] However, the proposed degradation pathway does not consider redox processes resulting from operating under bias and/or accumulation of charge carriers, and thus cannot explain all the observed trends in perovskite degradation. Since different photoinduced halide segregation behavior is observed for different 2D perovskites, which would in turn affect ion migration, understanding the role of bulky organic cations is essential for improving operational stability of devices using 3D/2D and quasi-2D perovskite materials. For example, BA-based perovskites exhibit rapid photoinduced segregation,[48,92] while the suppression of segregation was observed in PEA-based perovskite for illumination times ranging from several minutes up to 3 h.[92,93,95] In contrast, while photoinduced segregation has been reported in a DJ perovskite, timescales have been much longer, 10 h for demixing, 75 h for remixing.[91] This is comparable to segregation timescales (~1 h for demixing and 12 h for remixing) observed for stable perovskite compositions.[30]

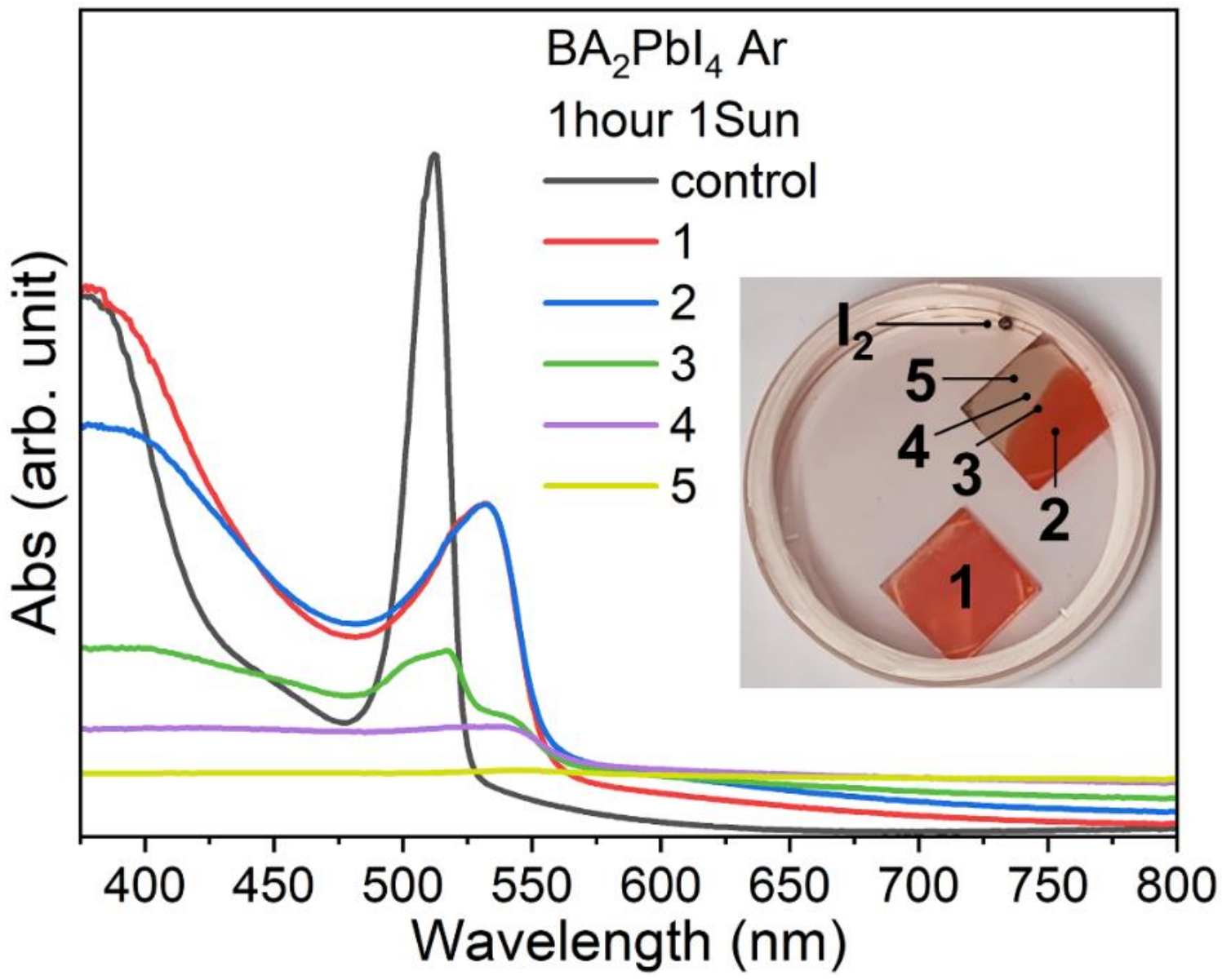

**Figure S30.** Absorption spectra of $BA_2PbI_4$ films exposed to illumination in Petri dish sealed in inert environment in the presence of iodine. More pronounced degradation is observed closer to $I_2$ bead. The control indicates film sealed in inert environment without iodine.

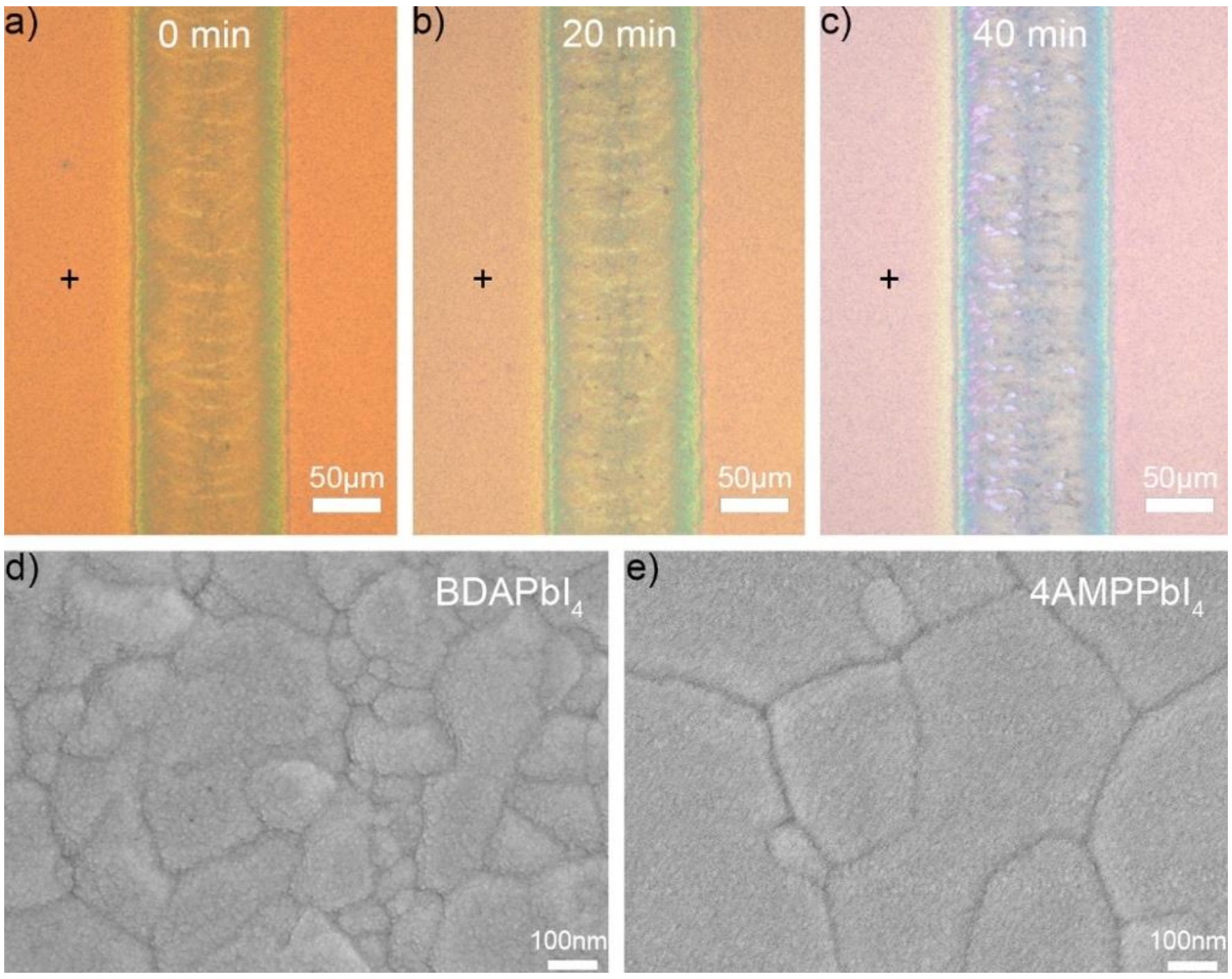

**Figure S31**. Microscope images of encapsulated $BA_2PbI_4$ lateral geometry devices after bias of 10 µA current for **a)** 0 min, **b)** 20 min and **c)** 40 min without illumination. Scale bar is 50 µm. **d), e)** SEM images of BDA and 4AMP for comparison of film quality, reasons for lower stability of BDA. The scale bar is 100 nm.

## SUPPLEMENTARY NOTE 8 *Photo/electrochemical reactions responsible for degradation*

### 3D perovskites

While the bias and illumination degradation trends, as well as degradation trends of different perovskite compositions and different device architectures, are not identical, there are sufficient common features to imply that there is a common degradation mechanism. It has been shown that the performance degradation of perovskite solar cells exhibits similar behavior under open circuit voltage in the dark and under illumination, which was attributed to the presence of electrical charges, in particular holes which contribute to the generation of large number of mobile ions under stress.[50] In general, there is ample evidence for the contribution of electrochemical reactions to the degradation of 3D perovskites (see **SUPPLEMENTARY NOTE 1** – The Critical Role of Holes). These reactions involve oxidation of iodide by the holes and reduction/deprotonation of the organic cation.[25] The trapping of the hole at iodide site, followed by oxidation of iodide and generation of mobile iodide (neutral iodine interstitial and iodide vacancy pair) is the starting point of the process.[21] Several iodine species can coexist in the perovskite,[21,29,32] which can react with both electrons and holes, as well as iodide vacancies and lattice iodide, to form different reaction products,[21,29] such as neutral, positively and negatively charged interstitial iodine, molecular iodine $I_2$, and triodide $I_3^-$.[29] As $I_2$ is not stable in the lattice, it tends to migrate to the surface and then it can be expelled in solution,[29] while triodide $I_3^-$ can readily deprotonate $FA^+$ or $MA^+$, resulting in perovskite decomposition.[34] The catalytic cycle involves formation of $I_3^-$ from $I^-$ and $I_2$ (where $I_2$ can result from various decomposition processes, such as thermal decomposition,[34] bias, or illumination,[21] then $FA^+$ or $MA^+$ form a hydrogen bond to $I_3^-$ resulting in the formation of $HI_3$ (which is in equilibrium with HI and $I_2$) and formamidine (which rapidly decomposes) or methylamine.[34] The loss of various volatile components over time results in the eventual decomposition of the perovskite layer.

### 2D perovskites – RP case

Similar to 3D perovskites, as no significant differences in possible reduction and oxidation processes are anticipated, we propose that electrochemical reactions are responsible for degradation, namely the oxidation of iodide, followed by the deprotonation of organic cation by triiodide $I_3^-$, resulting in the loss of volatile organic reaction products, accelerated migration of iodide species followed by accelerated reactions of triiodide with remaining organic cations, increased loss of volatile reaction products and eventual decomposition of the perovskite layer. As in the case of $MAPbI_3$, each deprotonation event results in the creation of one cation vacancy.

### 2D-perovskites – DJ case

In this case as well, the same types of redox reactions occur as in 3D organic-inorganic perovskites and RP perovskites. However, in this case single deprotonation event does not create cation vacancy due to the hydrogen bonding at the other end. Spacer cation which is missing one proton can get protonated/oxidized again as the reaction is reversible, or it can get deprotonated at the other end which would create a vacancy. It is important to note that the creation of vacancy is therefore a significantly less likely event in DJ perovskites, as it requires two simultaneous deprotonation events on the same cation. Thus, the halide migration remains inhibited due to low number of organic cation vacancies and the perovskite exhibits improved stability.

## SUPPLEMENTARY NOTE 9 *Degradation Products in 3D Perovskites*

The degradation products of 3D perovskites have been studied by different techniques for thermal degradation, as well as degradation under illumination,[96-106] and for both degradation conditions multiple possible pathways have been proposed. Possible reaction pathways under illumination include:[96]

$$CH_3NH_3PbI_3 \overset{h\nu}{\leftrightarrow} CH_3NH_2 + CH_3I + NH_3 + HI + I_2 + PbI_2 + Pb^0 \quad (1)$$

$$CH_3NH_3PbBr_3 \overset{h\nu}{\leftrightarrow} CH_3NH_2 + PbBr_2 + HBr \quad (2)$$

The $MAPbI_3$ reaction consists of the following:

$$CH_3NH_3PbI_3 \overset{h\nu}{\leftrightarrow} CH_3NH_2 + HI + PbI_2 \quad (3)$$

$$CH_3NH_3PbI_3 \overset{h\nu}{\rightarrow} CH_3I + NH_3 + PbI_2 \quad (4)$$

In general, the presence of $CH_3I$ and $NH_3$ as degradation products of MA-containing perovskite, and/or coexistence of decomposition pathways involving methylamine and HI, as well as $CH_3I$ and $NH_3$, have been proposed by multiple studies based on different measurements, including GC-MS (direct detection of gaseous decomposition products, as well as SIMS ($CH_3^+$ and $CH_5N^+$ ions).[42,74,97-99] In the absence of loss of methylamine, Eq.(3) is reversible, while Eq.(4) is irreversible since other reaction products can form from $CH_3I$ and $NH_3$ in addition to MAI.[99] However, even for the MA case the degradation is likely more complex, as a range of decomposition products (methylamine, $CH_3I$, $NH_3$, $I_2$, as well as methane, dimethylamine, trimethylamine, tetramethylhydrazine and tetramethylammonium iodide) was also reported.[106] The formation of complex reaction products was attributed to the fact that $CH_3I$ is a strong alkylating agent that can produce a series of different products in reaction with $NH_3$.[106] As the deprotonation of organic cation and outgassing of $CH_3I$ occurred in the absence of oxygen[106] and the participation of oxygen in the reaction was ruled out based on $^{18}O_2$ isotope experiments,[104] main role of oxygen in the degradation is likely as electron scavenger, and the degradation is then initiated by resulting excess photogenerated holes.

As decomposition products can be complex even for a simple organic cation such as MA due to reactivity of various iodide species,[106] bulky spacer cation degradation is also expected to produce a wide range of possible decomposition products. From the obtained results **(Figure S16),** the limitations of the column used in the detection of HI,[74] and the discussion of possible reactions in 3D perovskites above, we can consider outgassing of amine/$NH_3$ and I-containing products as intrinsic measures of reductive and oxidative decomposition reactions, respectively, as the trends observed are consistent with presence of excess electrons (reductive) or excess holes (oxidative) on different substrates. The experimental results (**Figure 5**, **Figure S16**) are consistent with the observed RP and DJ film degradation patterns, as well as theoretical calculations which predict significant difference in the loss of organic spacer cations between RP and DJ materials and the proposed degradation mechanism. Briefly, the degradation process involves oxidation of iodide (anodic reaction) and reduction/deprotonation of organic cation, with lead reduction having a minor contribution to overall degradation (since lead is far less mobile compared to halide anions and organic cations,[107] and the main decomposition product is $PbI_2$ with only a small fraction of $Pb^0$ detected by XPS.[106] Thus, the main process is electrochemical degradation driven by excess holes generating mobile iodide species which then result in organic cation deprotonation ($MA^+$ can be readily deprotonated by $I_3^-$, and thus iodide chemistry is responsible for the fact that one carrier type, namely holes, has by far more significant contribution to the degradation). The proposed process is also consistent with the fact that excess iodine can not only accelerate the degradation of perovskite under illumination **(Figure S30),** but also cause degradation in the dark.[107]

It should also be noted that while differences in deprotonation and consequently loss of organic spacer explain the observed behavior of RP and DJ perovskites, differences in stability of perovskites within the same class (RP or DJ) require further study to decouple effects of chemical structure and other factors. For example, $BDAPbI_4$ films exhibit smaller grain size compared to $4AMPPbI_4$ (**Figure S31**), and thus would contain more defects due to higher proportion of defective grain boundary regions.

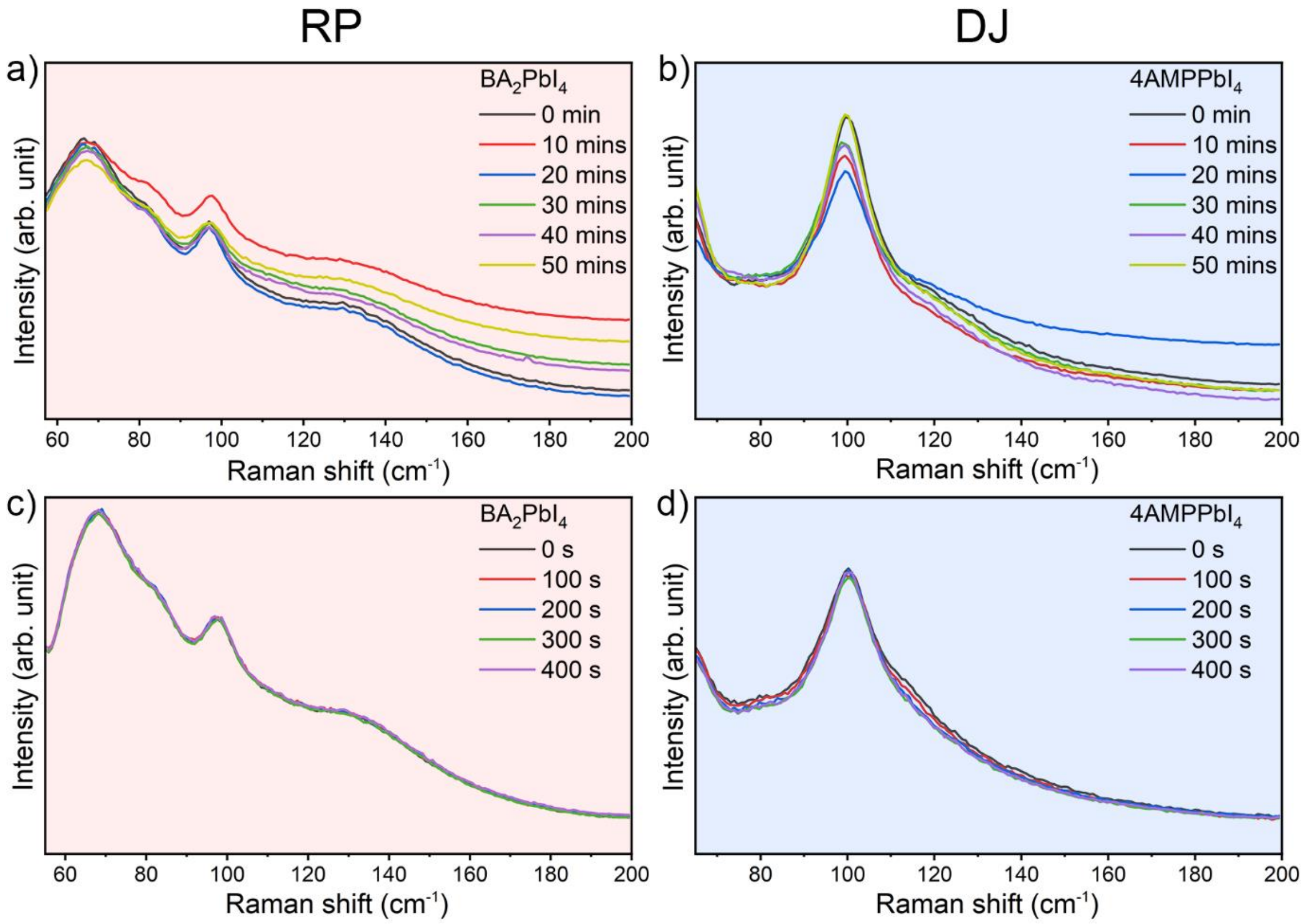

**Figure S32**. Raman spectra of **a), c)** $BA_2PbI_4$ and **b), d)** $4AMPPbI_4$ samples for different solar illumination times and different times of laser illumination for fresh sample, respectively. The low wavenumber region corresponds to peaks due to vibrations in the inorganic part of perovskite lattice and can also contain peaks due to $PbI_2$ and $I_x$,[108-110] and no significant changes are observed in this part of the spectrum, other than a small increase in peak corresponding to $PbI_2$ in $BA_2PbI_4$.

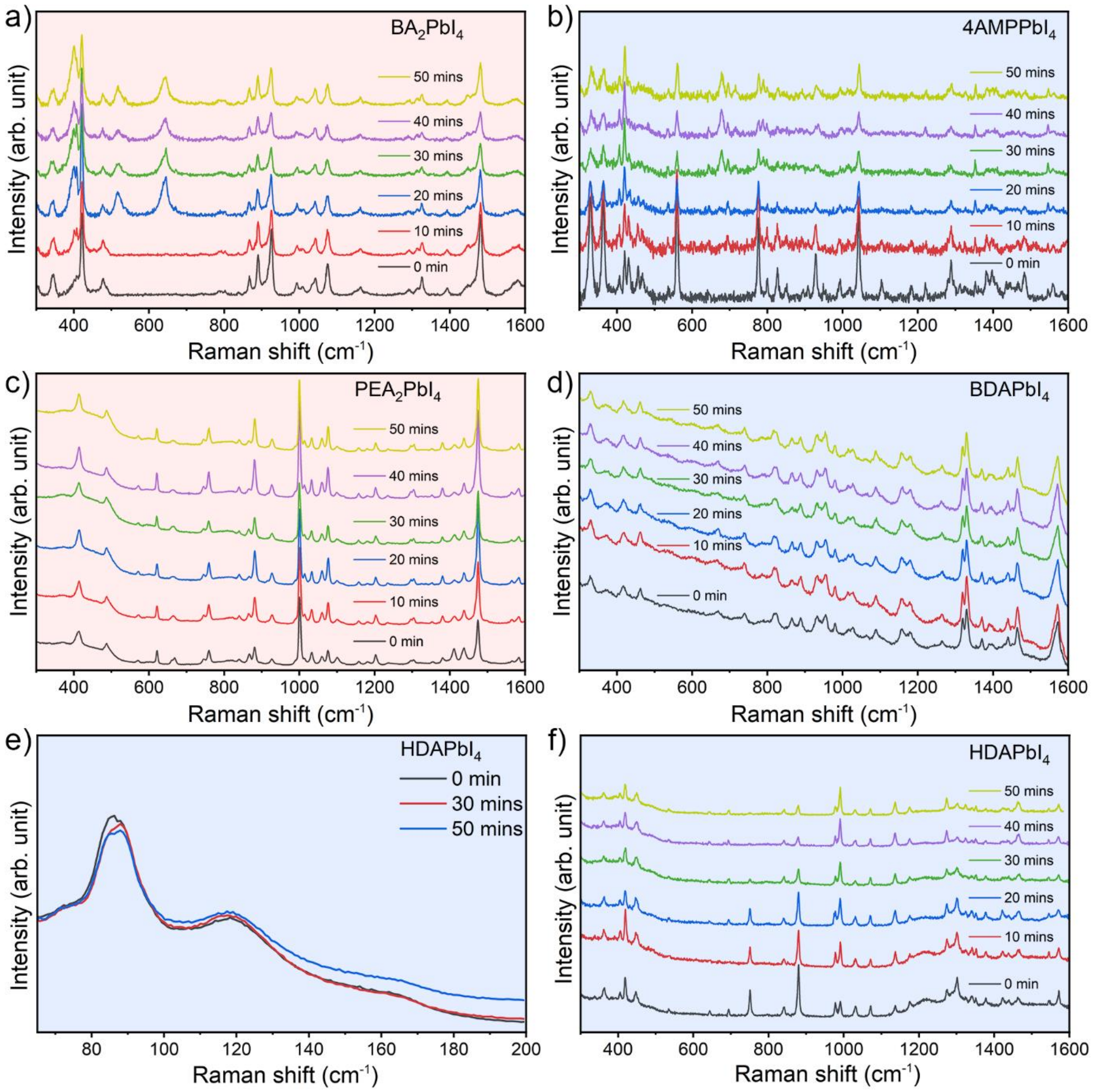

**Figure S33.** Raman spectra of **a)** $BA_2PbI_4$ **b)** $4AMPPbI_4$ **c)** $PEA_2PbI_4$ and **d)** $BDAPbI_4$ for different illumination times in ambient; **e), f)** Raman spectra of $HDAPbI_4$ for different illumination times in ambient at different wavenumber ranges.

For $BA_2PbI_4$, we observe three new peaks after illumination at ~402 $cm^{-1}$, 518 $cm^{-1}$ and 643 $cm^{-1}$. Interestingly, these peaks vanish in the second repeated scan of the same area (**Figure S32**). This phenomenon is reproducible on different samples and different illumination times), suggesting the presence of loosely bound surface species which readily desorbs, with the lower two peaks corresponding to C-C-C and C-C-N deformations in organic cation,[111-113] while the peak at ~643 $cm^{-1}$ could possibly involve $HI_2^-$.[114] The origin of a peak at ~679 $cm^{-1}$ in $4AMPPbI_4$ is less clear, and it could possibly correspond to C-H deformations in 4-monosubstituted pyridine, consistent with likely assignment of several other peaks present in the spectrum[115] or to $HI_2^-$.[114] One possible reason for the appearance of the new peak corresponding to the organic ligand is increased vibrations due to weakened bonding on one side. Since no new peaks in the 600-700 $cm^{-1}$ range are observed for $HDAPbI_4$, but instead we observe changes in ~1100-1300 $cm^{-2}$ range, corresponding to C-C skeletal vibrations,[115] the observed change in $4AMPPbI_4$ corresponds to a change in vibrations of pyridine ring. For $BDAPbI_4$, no significant changes are observed. In contrast, observed changes for $PEA_2PbI_4$ are significant, although smaller compared to BA-based perovskite in agreement with its lower degradation. The changes can be observed for several peaks, one at ~667 $cm^{-1}$ related to ring vibrations and several in the ~1350-1450 $cm^{-1}$ range related to various =C-H and C-H vibrations.[115]

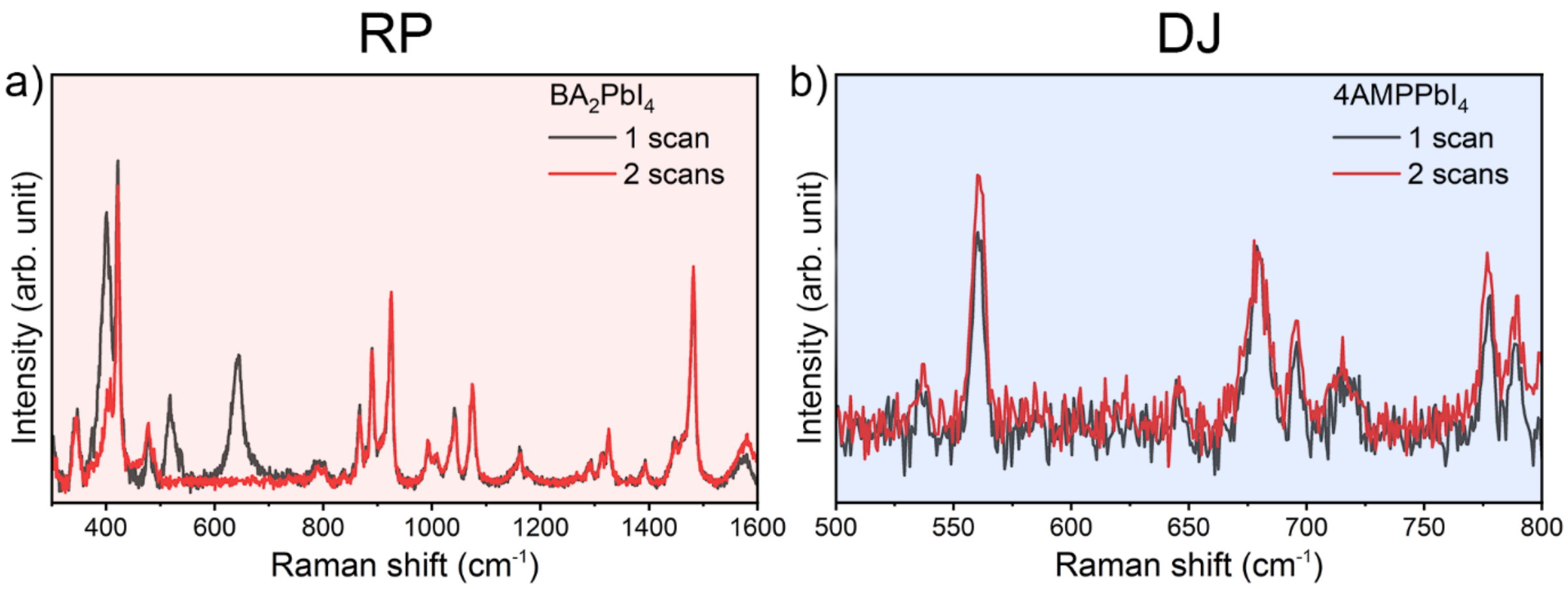

**Figure S34**. Raman spectra (first and second scan) of a) $BA_2PbI_4$ and b) $4AMPPbI_4$ samples after 50 min illumination.

# DEVICE STABILITY

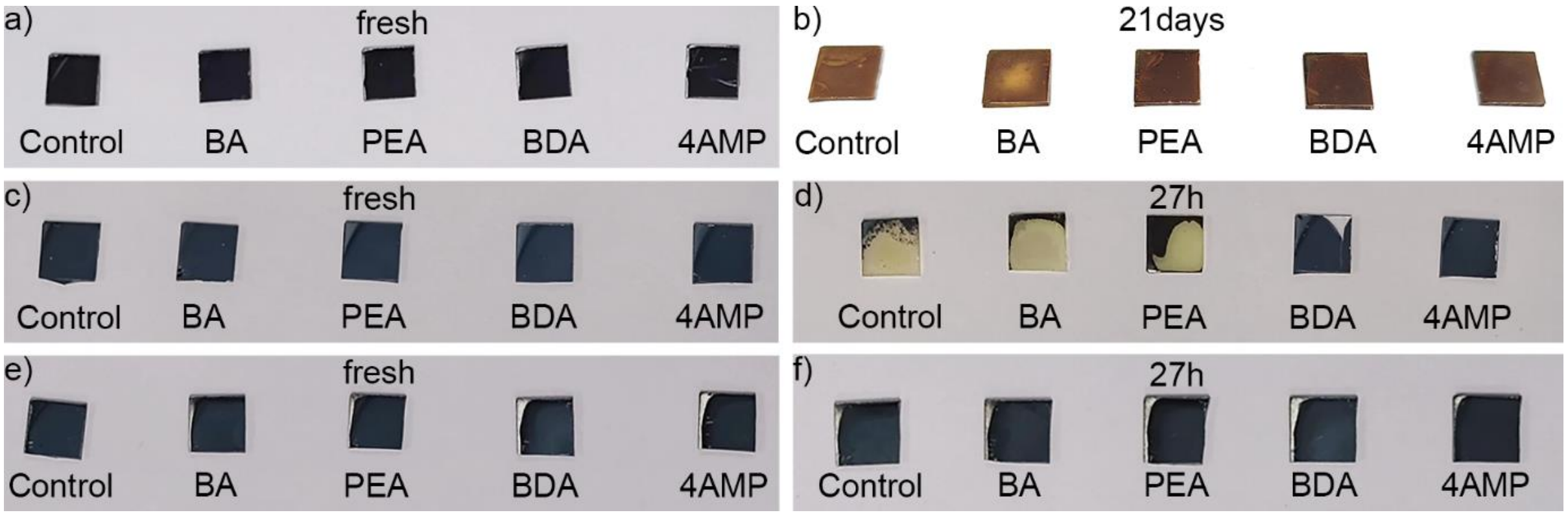

**Figure S35**. Photos of different 3D/2D perovskites after simulated solar illumination (100 mW/cm$^2$) in ambient (RH~60%). **a)** CsFAMA (21 days) **b)** low Br (27 h) **c)** MA-free (27 h). As CsFAMA perovskite serves as a model perovskite for detailed investigations, the film stability test was conducted over longer time period.

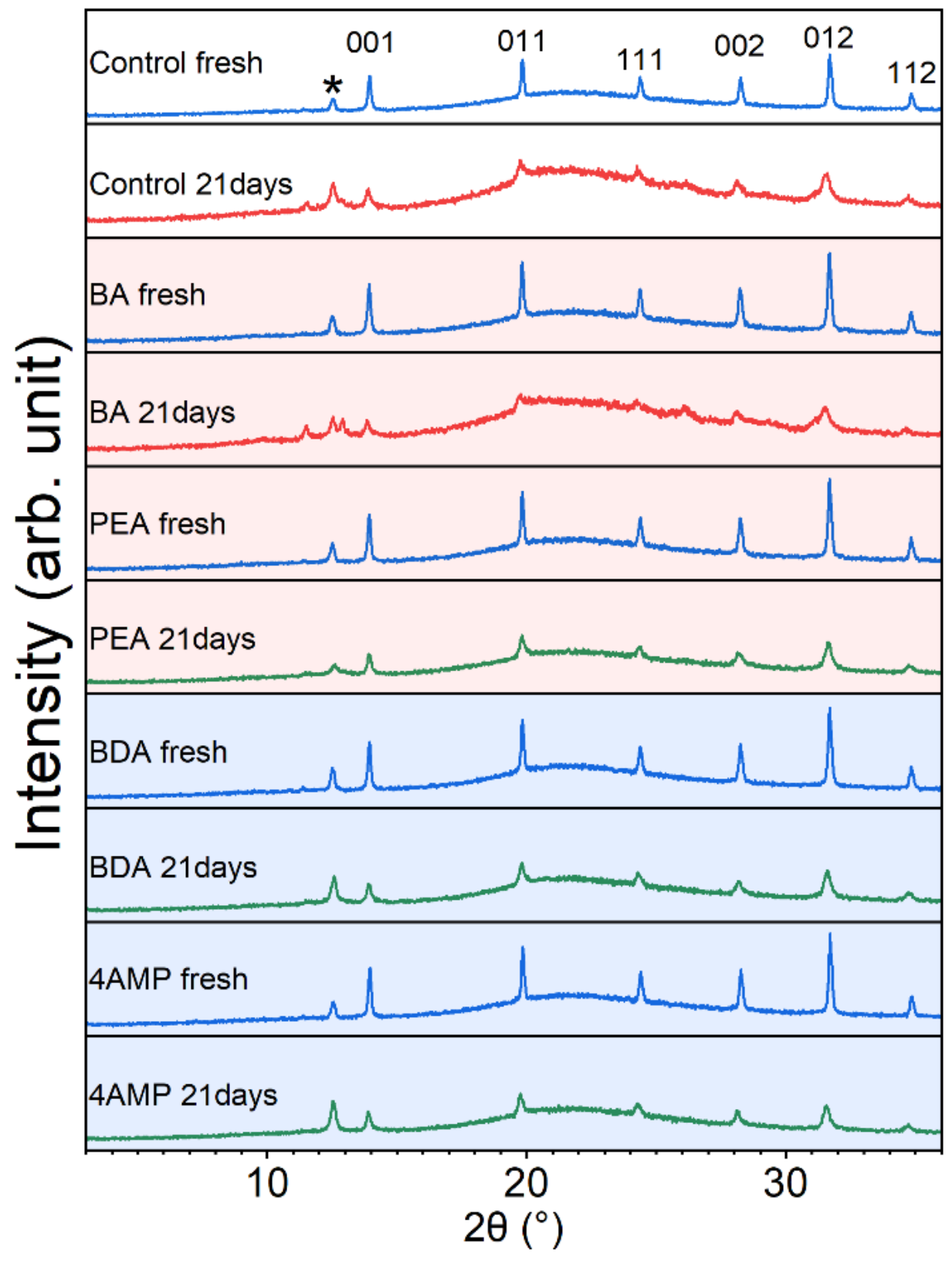

**Figure S36**. XRD patterns of 3D/2D films with CsFAMA perovskite 3D film and different 2D layers before and after 21 days of solar illumination (1 Sun) in ambient (RH ~60%). Asterisk denotes peak corresponding to $PbI_2$.

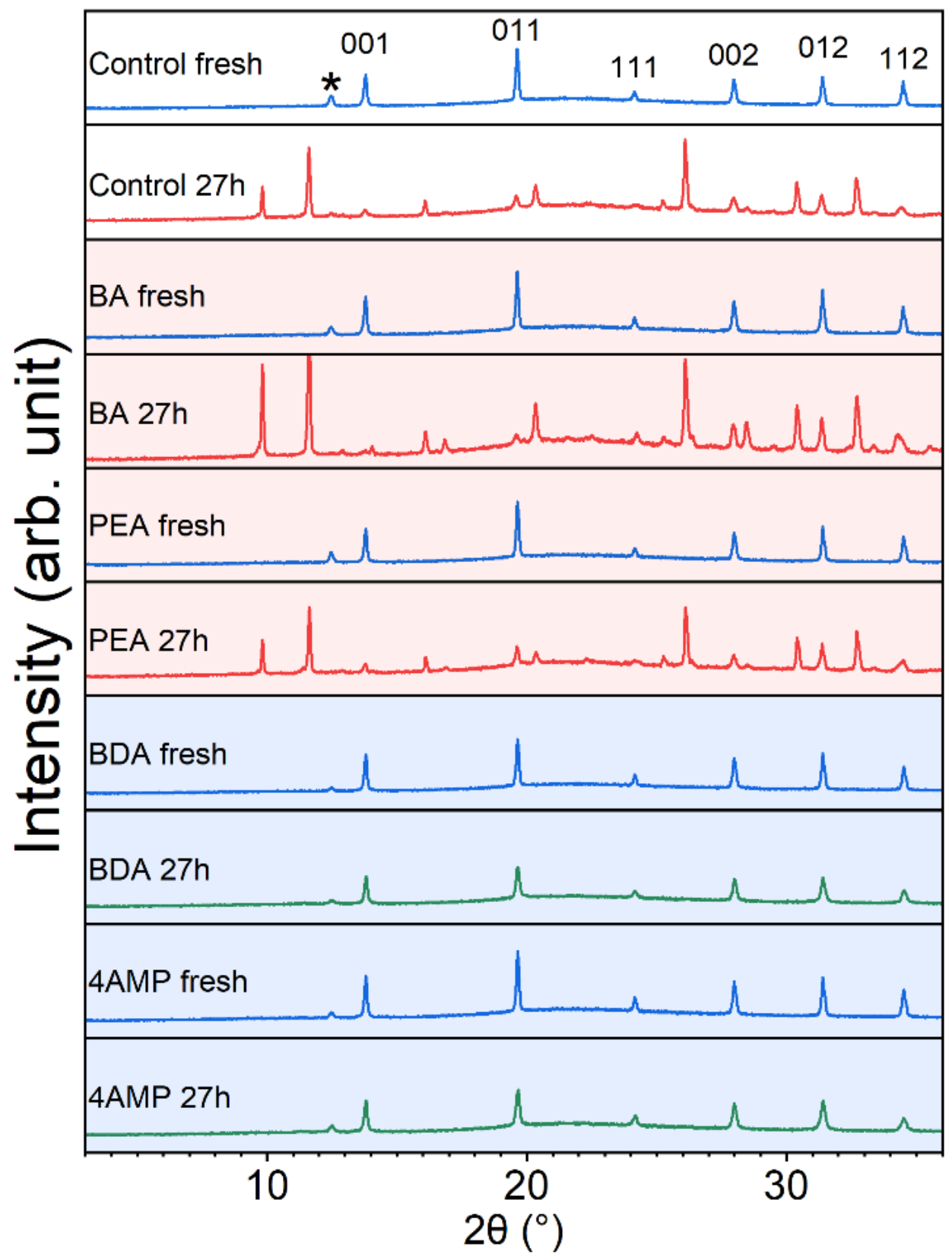

**Figure S37**. XRD patterns of 3D/2D films with low Br perovskite 3D film and different 2D layers before and after 27 h solar illumination (1 Sun) in ambient (RH ~60%). Asterisk denotes peak corresponding to $PbI_2$.

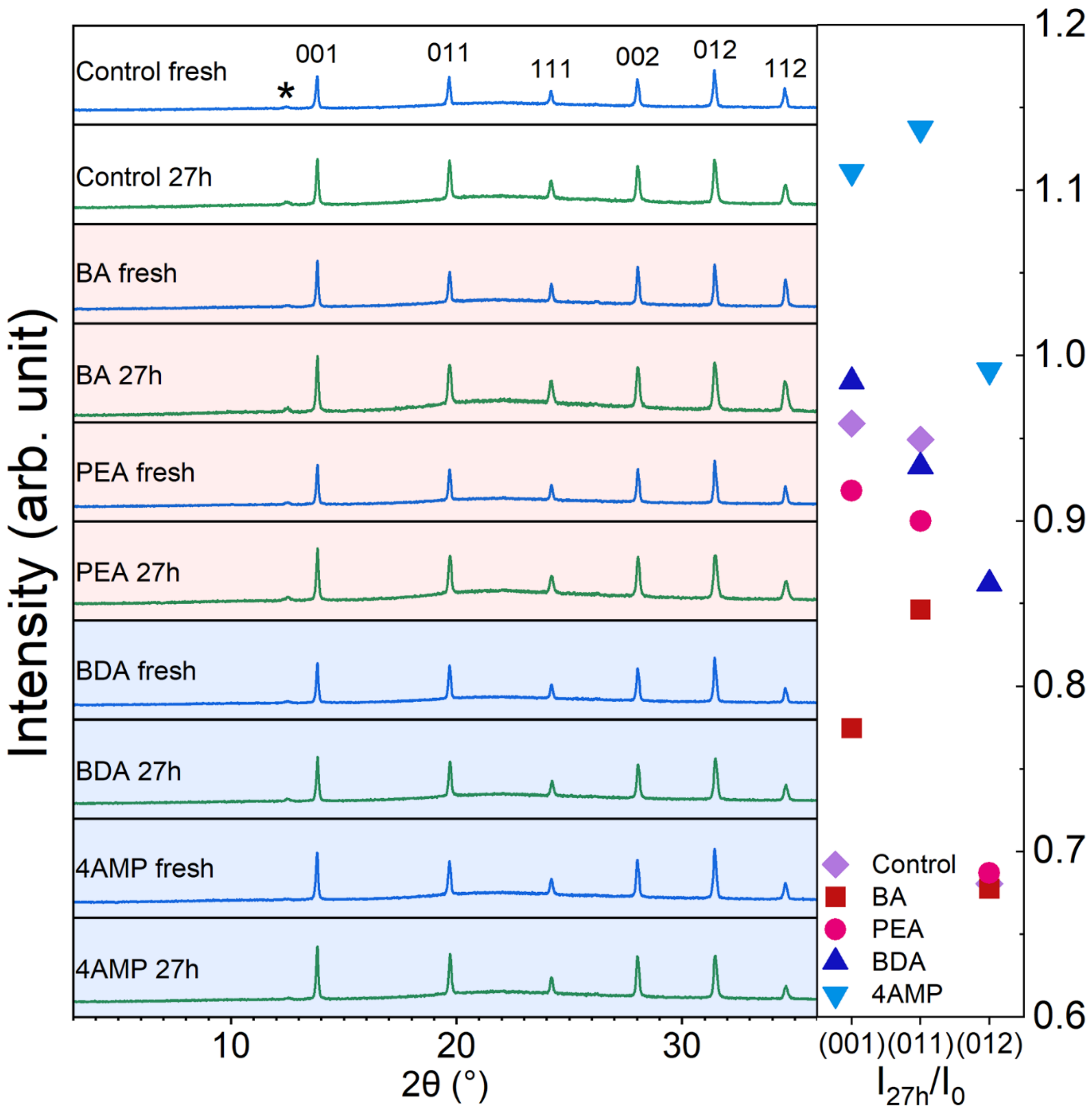

**Figure S38**. XRD patterns of 3D/2D films with MA-free perovskite 3D film and different 2D layers before and after 27 h solar illumination (1 Sun) in ambient (RH ~60%). The panel on the right illustrates ratio of peak intensity before/after illumination for different perovskite peaks. Asterisk denotes peak corresponding to $PbI_2$.

**Table S5**. Solar cell performance parameters of inverted perovskite solar cells. The performance parameters of champion cells are in bold font, and the number of devices for average performance parameters is shown in brackets (15 devices, 3-5 substrates). The device architectures are: ITO/$NiO_x$/ethanolamine/2PACz/$Cs_{0.03}(FA_{0.83}MA_{0.17})_{0.97}Pb(I_{0.83}Br_{0.17})_3$ 3D perovskite/2D perovskite /PCBM/BCP/Ag and ITO/$NiO_x$/Me-4PACz/$Cs_{0.05}(FA_{0.98}MA_{0.02})_{0.95}Pb(I_{0.98}Br_{0.02})_3$ 3D (or $Cs_{0.1}FA_{0.9}PbI_{2.9}Br_{0.1}$ 3D)/2D perovskite/PCBM/BCP/Ag. 3 out of 15 CsFAMA devices were measured under ABET Sun 3000 simulator.

| Perovskite layer/bias | $J_{sc}$ (mA/cm$^2$) | $V_{oc}$ (V) | FF | PCE(%) |
|---|---|---|---|---|
| CsFAMA- BA forward (15 devices) | 22.30±0.42 **23.07** | 1.108±0.007 **1.11** | 77.13±1.06 **78.9** | 19.13±0.52 **20.17** |
| CsFAMA- BA backward (15 devices) | 22.28±0.40 **23.07** | 1.096±0.010 **1.10** | 78.21±0.85 **79.5** | 19.18±0.44 **19.97** |
| CsFAMA- PEA forward (15 devices) | 23.05±0.44 **23.74** | 1.177±0.009 **1.18** | 76.45±1.82 **80.9** | 20.75±0.58 **22.13** |
| CsFAMA- PEA backward (15 devices) | 23.03±0.42 **23.66** | 1.176±0.007 **1.18** | 77.24±2.27 **80.5** | 20.90±0.59 **22.07** |
| CsFAMA- BDA forward (15 devices) | 22.91±0.41 **23.61** | 1.134±0.012 **1.14** | 78.92±1.01 **81.6** | 20.56±0.40 **21.36** |
| CsFAMA- BDA backward (15 devices) | 22.91±0.41 **23.61** | 1.128±0.007 **1.13** | 78.98±0.56 **79.7** | 20.47±0.38 **21.24** |
| CsFAMA- 4AMP forward (15 devices) | 22.50±0.61 **23.71** | 1.163±0.010 **1.20** | 78.34±0.87 **79.4** | 20.50±0.75 **22.17** |
| CsFAMA- 4AMP backward (15 devices) | 22.49±0.61 **23.71** | 1.163±0.005 **1.18** | 79.83±1.47 **81.03** | 20.43±0.73 **22.37** |
| Low Br- BA forward (15 devices) | 23.70±0.74 **24.27** | 1.152±0.006 **1.159** | 82.13±1.68 **84.93** | 22.42±0.53 **22.94** |
| Low Br- BA backward (15 devices) | 24.37±0.52 **25.32** | 1.141±0.010 **1.161** | 78.53±2.02 **81.04** | 21.83±0.78 **22.95** |
| Low Br- PEA forward (15 devices) | 23.94±0.23 **24.41** | 1.144±0.018 **1.166** | 81.79±1.16 **82.78** | 22.38±0.47 **23.28** |
| Low Br- PEA backward (15 devices) | 23.91±0.23 **24.38** | 1.133±0.013 **1.150** | 78.93±1.16 **81.09** | 21.36±0.54 **22.16** |
| Low Br- BDA forward (15 devices) | 24.38±0.16 **24.61** | 1.153±0.009 **1.167** | 82.24±0.76 **83.37** | 23.10±0.20 **23.37** |
| Low Br- BDA backward (15 devices) | 24.38±0.23 **24.78** | 1.136±0.013 **1.157** | 78.70±1.42 **81.34** | 21.79±0.53 **22.63** |
| Low Br- 4AMP forward (15 devices) | 24.00±0.67 **24.74** | 1.147±0.010 **1.163** | 82.49±1.64 **83.72** | 22.69±0.50 **23.83** |
| Low Br- 4AMP backward (15 devices) | 23.97±0.68 **24.74** | 1.133±0.009 **1.156** | 78.56±1.52 **80.30** | 21.34±0.96 **22.57** |
| MA-free - BA forward (15 devices) | 23.87±0.95 **24.84** | 1.061±0.004 **1.065** | 76.33±2.13 **82.41** | 19.32±0.65 **20.37** |
| MA-free - BA backward (15 devices) | 23.77±0.75 **24.60** | 1.064±0.003 **1.067** | 78.20±1.38 **79.95** | 19.77±0.63 **20.76** |
| MA-free - PEA forward (15 devices) | 23.91±0.22 **24.23** | 1.046±0.003 **1.049** | 82.54±0.54 **83.09** | 20.65±0.26 **20.98** |
| MA-free - PEA backward (15 devices) | 23.94±0.21 **24.28** | 1.049±0.003 **1.052** | 82.94±0.47 **83.43** | 20.83±0.22 **21.13** |
| MA-free - BDA forward (15 devices) | 24.33±0.42 **24.86** | 1.044±0.005 **1.053** | 79.76±1.50 **81.51** | 20.26±0.33 **20.80** |
| MA-free - BDA backward (15 devices) | 24.29±0.50 **24.80** | 1.053±0.005 **1.060** | 81.32±1.55 **82.81** | 20.80±0.38 **21.45** |
| MA-free - 4AMP forward (15 devices) | 24.20±0.22 **24.53** | 1.065±0.005 **1.074** | 82.02±0.82 **83.72** | 21.14±0.16 **21.40** |
| MA-free - 4AMP backward (15 devices) | 24.24±0.17 **24.48** | 1.069±0.004 **1.074** | 82.59±0.69 **83.55** | 21.40±0.18 **21.67** |

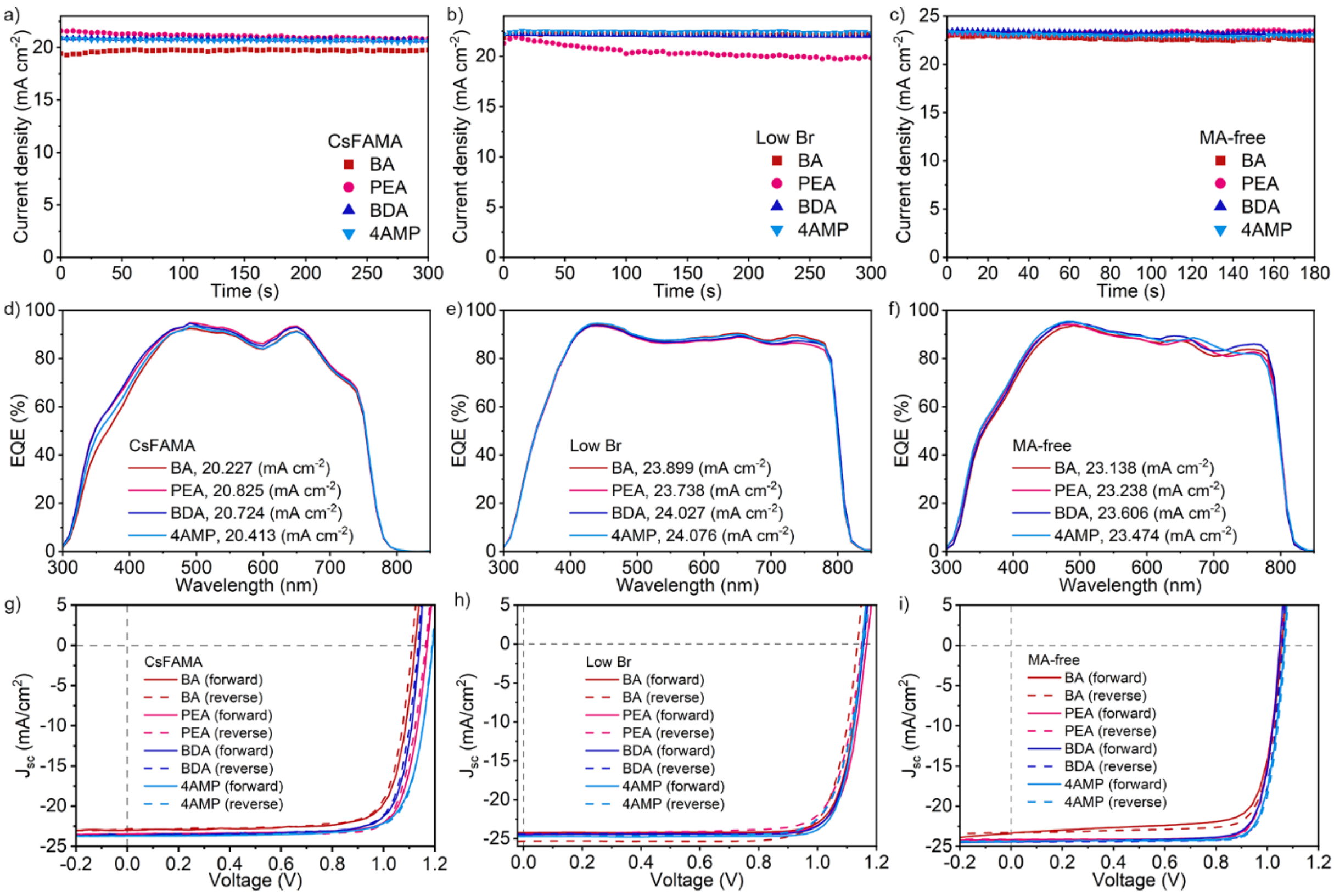

**Figure S39.** Stabilized photocurrent for devices with different spacer cations on **a)** CsFAMA perovskite **b)** low Br perovskite and **c)** MA-free perovskite; EQE of **d)** CsFAMA **e)** low Br and **f)** MA-free perovskite devices; integrated $J_{SC}$ is shown; J-V curves under forward and backward scan corresponding to the best performance devices for **g)** CsFAMA perovskite **h)** low Br perovskite and **i)** MA-free perovskite.

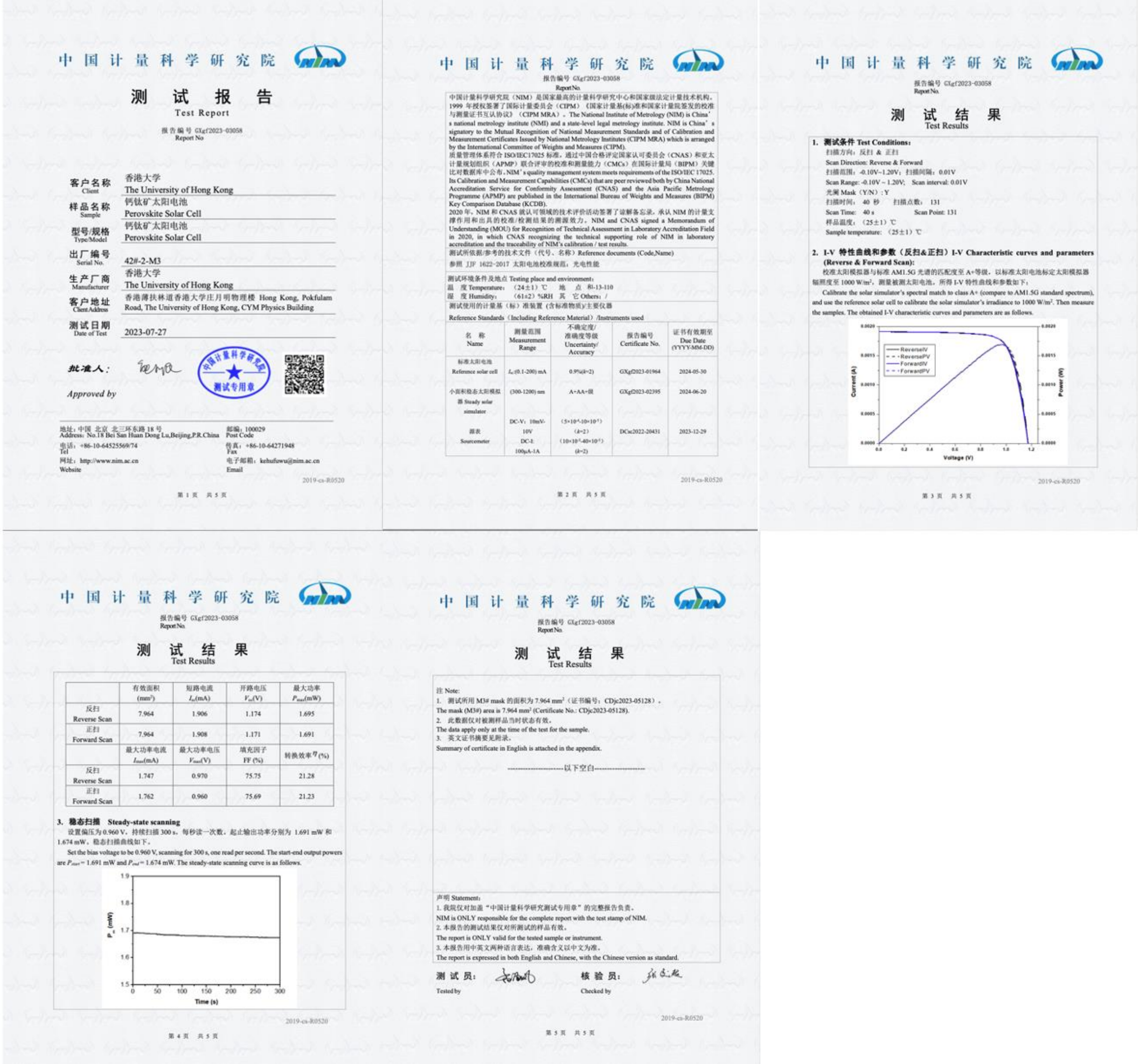

**Figure S40.** Certificate report for CsFAMA devices measured at The National Institute of Metrology of China (NIM). The measurement was performed on encapsulated devices.

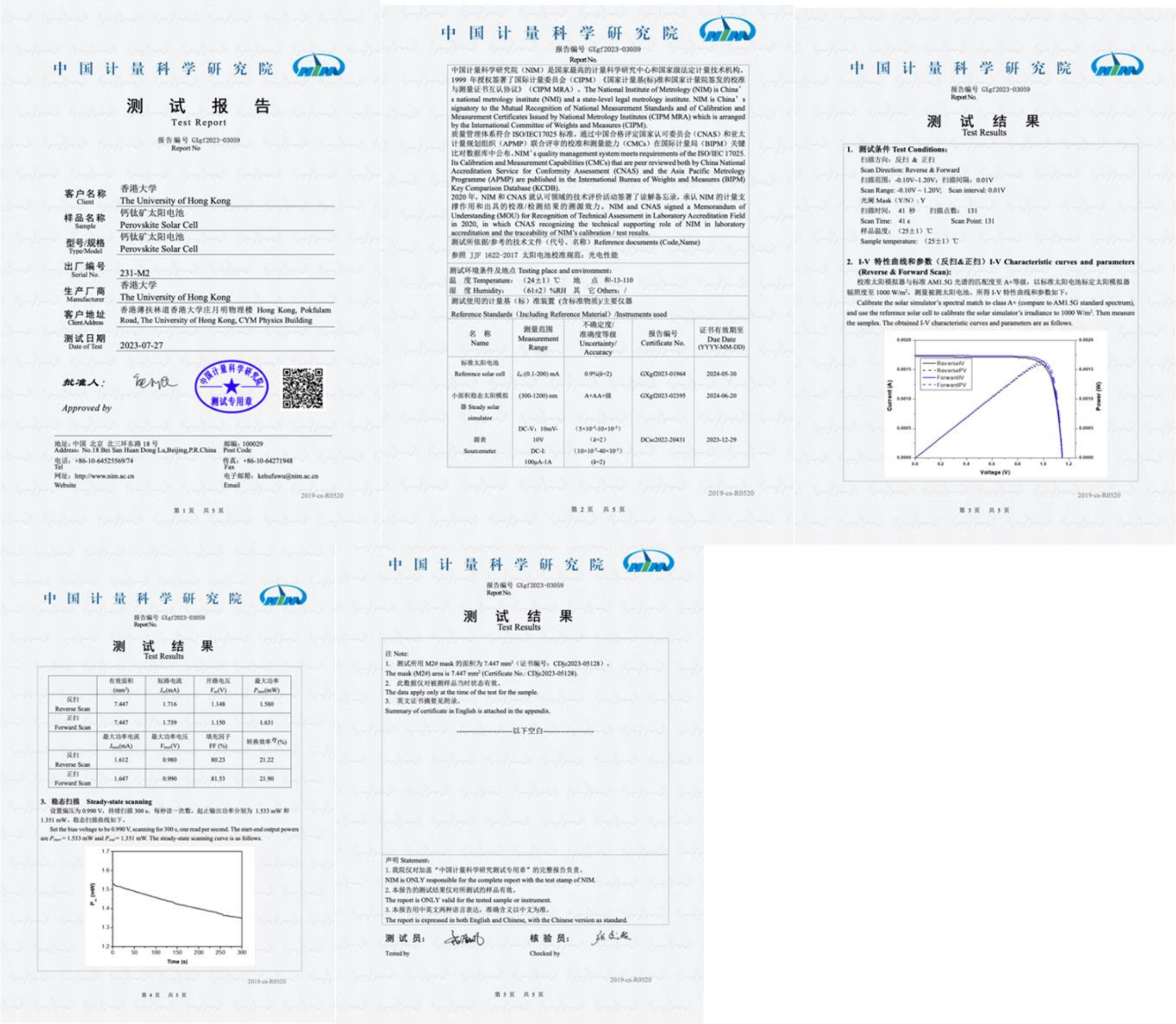

中国计量科学研究院

测 试 报 告

Test Report

报告编号 GXgf2023-03059

Report No.

| | |
|---|---|
| 客户名称 Client | 香港大学 The University of Hong Kong |
| 样品名称 Sample | 钙钛矿太阳电池 Perovskite Solar Cell |
| 型号/规格 Type/Model | 钙钛矿太阳电池 Perovskite Solar Cell |
| 出厂编号 Serial No. | 231-M2 |
| 生产厂商 Manufacturer | 香港大学 The University of Hong Kong |
| 客户地址 Client Address | 香港薄扶林道香港大学庄月明物理楼 Hong Kong, Pokfulam Road, The University of Hong Kong, CYM Physics Building |
| 测试日期 Date of Test | 2023-07-27 |

批准人： Approved by

测试专用章

测 试 结 果

Test Results

| | 有效面积 ($mm^2$) | 短路电流 $I_{sc}$(mA) | 开路电压 $V_{oc}$(V) | 最大功率 $P_{max}$(mW) |
|---|---|---|---|---|
| 反扫 Reverse Scan | 7.447 | 1.716 | 1.148 | 1.580 |
| 正扫 Forward Scan | 7.447 | 1.739 | 1.150 | 1.631 |
| | 最大功率电流 $I_{max}$(mA) | 最大功率电压 $V_{max}$(V) | 填充因子 FF (%) | 转换效率 $\eta$ (%) |
| 反扫 Reverse Scan | 1.612 | 0.980 | 80.23 | 21.22 |
| 正扫 Forward Scan | 1.647 | 0.990 | 81.53 | 21.90 |

3. 稳态扫描 Steady-state scanning

Set the bias voltage to be 0.990 V, scanning for 300 s, one read per second. The start-end output powers are $P_{start}$ = 1.533 mW and $P_{end}$ = 1.351 mW. The steady-state scanning curve is as follows.

**Figure S41.** Certificate report for low Br devices measured at NIM. The measurement was performed in ambient on devices without encapsulation, which is likely the main cause of lower efficiency and decreasing SPO compared to that measured in glove box in Hong Kong.

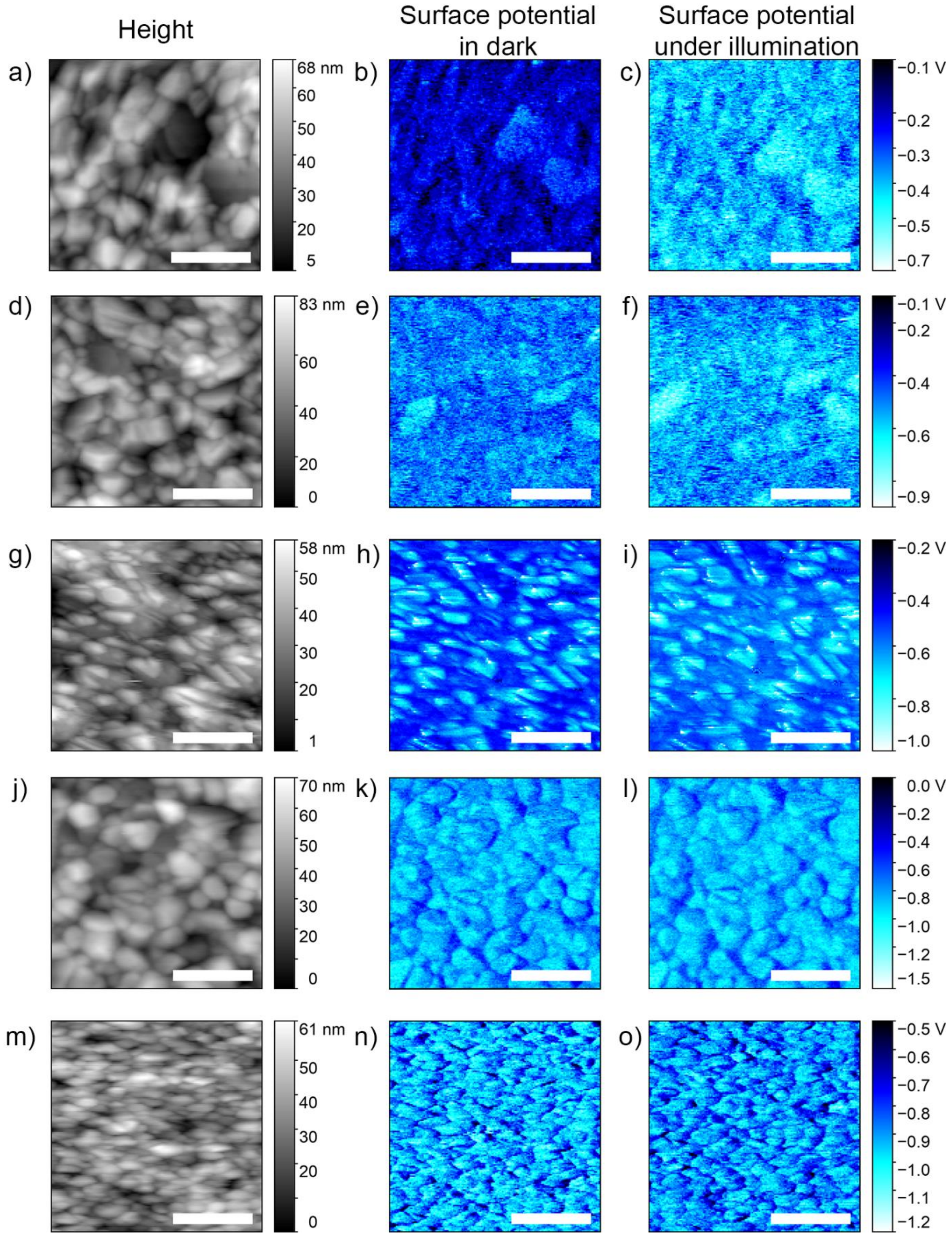

**Figure S42.** Measurement of atomic force microscopy AFM (left) and Kelvin probe force microscopy (KPFM), in dark (middle) and under illumination (right), for the CsFAMA 3D perovskite/SAM/$NiO_x$/ITO sample with different passivating spacers. Control **(a-c),** BA **(d-f),** PEA **(g-i),** BDA **(j-l)** and 4AMP **(m-o)**. All scale bars are 0.5 µm.

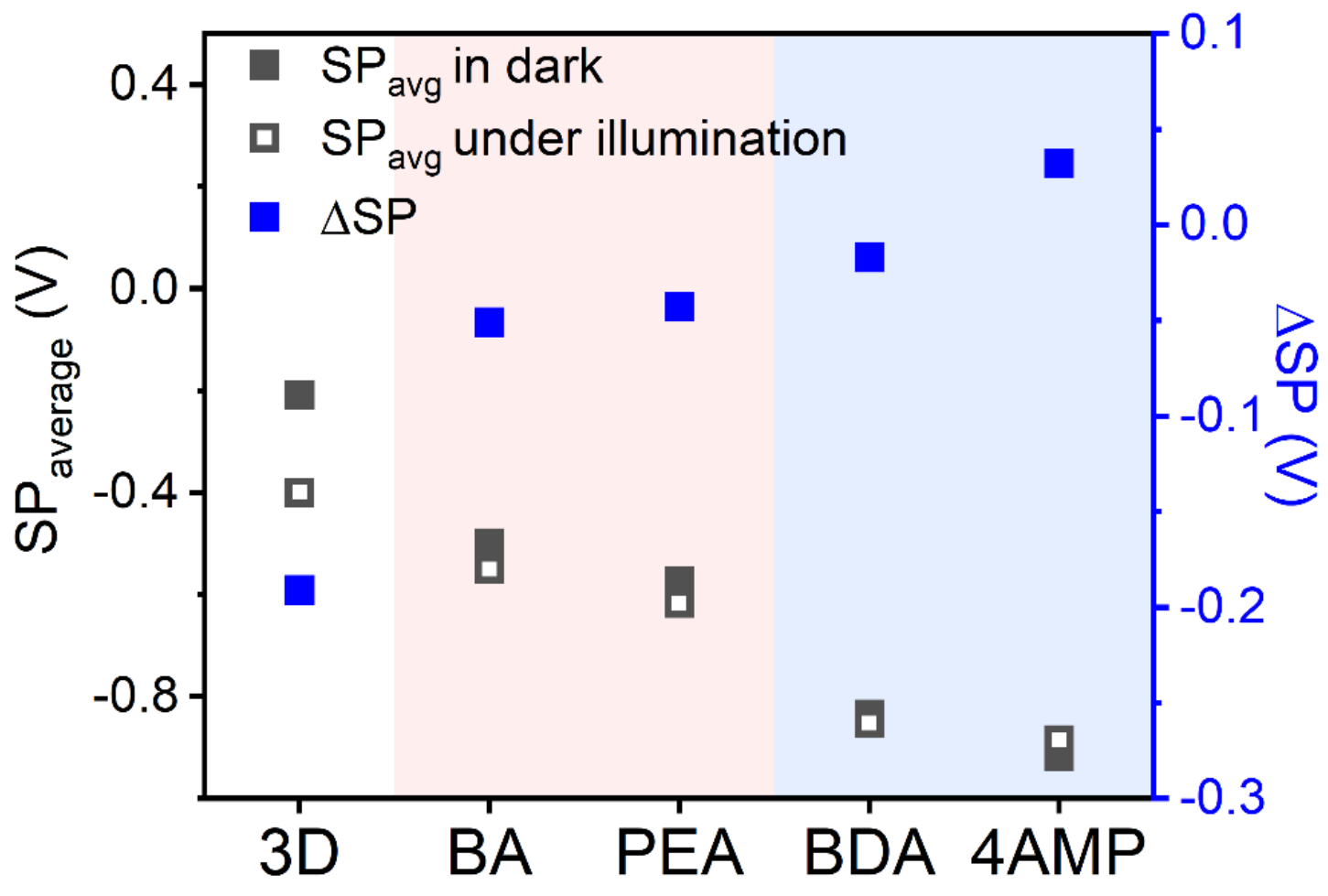

**Figure S43.** Average surface potential ($SP_{average}$, black) and the change in surface potential by illumination (ΔSP, red).

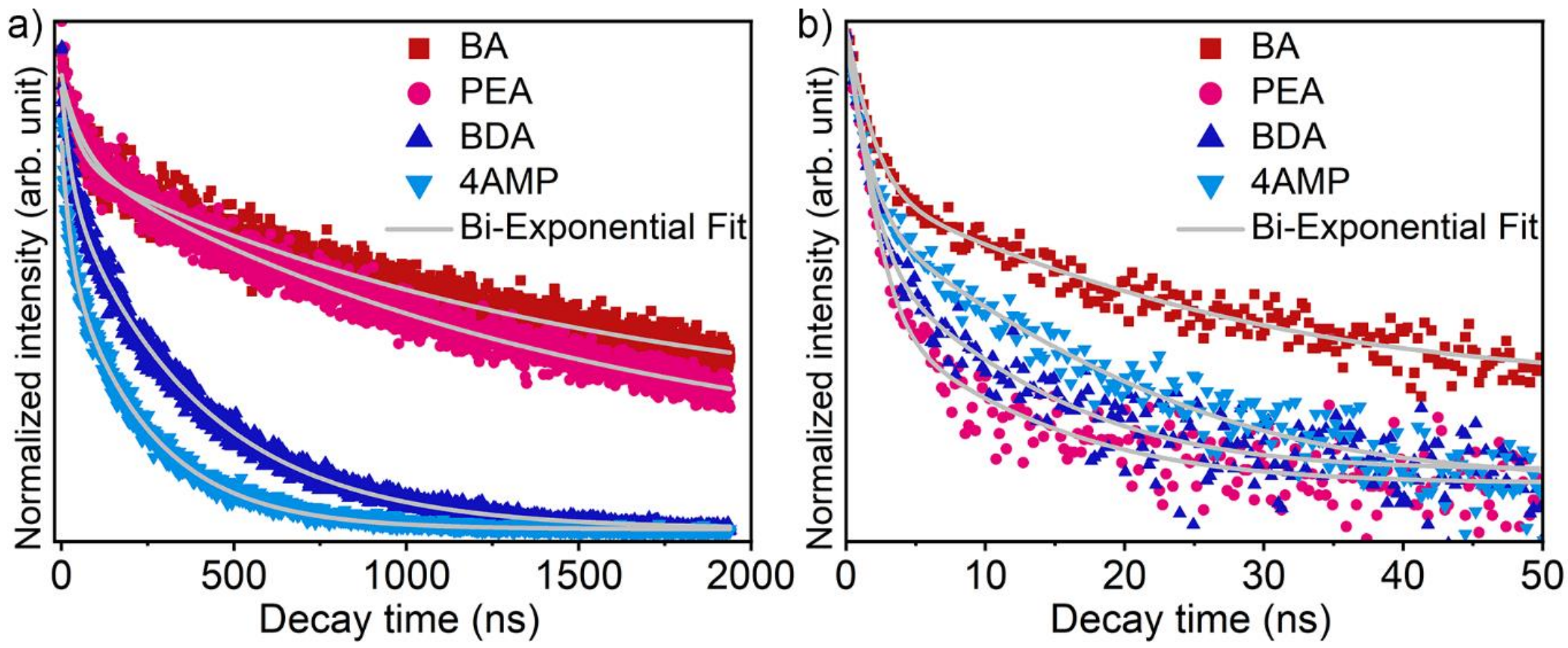

**Figure S44.** TRPL data of **a)** ITO/$NiO_x$/SAM/CsFAMA 3D/2D and **b)** ITO/$NiO_x$/SAM/CsFAMA 3D/2D/PCBM. SAM denotes 2PACz+amine surface modification.

**Table S6.** TRPL data fitting parameters for bi-exponential decay ($A_1 \exp(-t/\tau_1) + A_2 \exp(-t/\tau_2)$) with average decay time calculated as $\tau_{avg} = (A_1 \tau_1 + A_2 \tau_2)/(A_1 + A_2)$. All the samples have structure ITO/$NiO_x$/SAM/CsFAMA 3D/C, where SAM denotes 2PACz+amine surface modification and C denotes layers on top of 3D perovskite.

| Sample | $A_1$ | $\tau_1$ (ns) | $A_2$ | $\tau_2$ (ns) | $\tau_{avg}$ (ns) |
|---|---|---|---|---|---|
| BA | 0.17 | 44 | 0.51 | 1474 | 759 |
| PEA | 0.16 | 65 | 0.61 | 1480 | 773 |
| BDA | 0.19 | 29 | 0.67 | 393 | 422 |
| 4AMP | 0.22 | 26 | 0.54 | 245 | 136 |
| BA/PCBM | 0.74 | 1.33 | 0.23 | 16.1 | 8.7 |
| PEA/PCBM | 0.88 | 1.14 | 0.08 | 8.8 | 4.97 |
| BDA/PCBM | 0.83 | 1.09 | 0.15 | 8.3 | 4.69 |
| 4AMP/PCBM | 0.68 | 1.14 | 0.22 | 11.1 | 6.12 |

## SUPPLEMENTARY NOTE 10 CsFAMA devices with different 2D capping layers

From detailed characterization of CsFAMA-based devices (**Figures S39-41**, **Figure 6** and **Table S5**), we can see that there are differences in charge extraction between different cations, but in all cases photooxidative degradation is expected to be the dominant degradation mechanism as it is entirely unavoidable due to photo/electrochemistry of iodide. BA-based devices exhibit lower efficiency compared to other 2D capping layers, and they also exhibit the worst stability for CsFAMA devices. In terms of efficiency, the performance of PEA-based and 4AMP-based devices is comparable, although we can observe lower $J_{sc}$ values for 4AMP and lower FF values for PEA, indicating differences in charge collection and recombination in these devices. Additional characterization was therefore performed to have better understanding of performance differences. We can observe that passivation of 3D perovskite with 2D perovskite leads to an increase in surface potential (SP) and a decrease in the SP change by illumination (**Figure S42 & S43**). The effects are more pronounced for the 3D perovskite passivated with DJ perovskites than with RP perovskites. From TRPL data (**Figure S44**, **Table S6**), we can observe significantly longer decay times in RP perovskites on HTL, which possibly indicates less efficient hole collection in these materials for energy level alignment in half-devices (HTL/3D/2D). The determined long decay times for 3D/2D perovskites on 2PACz-modified $NiO_x$ are consistent with previous report.[4] However, from observed decay times for samples with PCBM on top of the perovskite, which is the full device structure, there are no clear divisions between DJ and RP materials in terms of charge collection, as the fastest decay times are observed for PEA and 4AMP. In addition, we observe that BA, PEA and BDA exhibit the same sign of SP change, different from 4AMP, and we also observe that devices with 4AMP 2D layers exhibit conventional hysteresis behavior, while devices with BA, PEA and BDA exhibit reverse hysteresis. Thus, there is no clear trend between the type of the 2D perovskite and charge extraction in the devices. The observed trends in decay times are in good agreement with the trends in average $J_{sc}$ (**Table S5**, lower current measured in samples with less efficient charge collection (BA, 4AMP)). For $V_{oc}$ trends, we see more similarity between materials with similar chemical structure (lower values in alkyl chain spacers BA and BDA, vs. high values for PEA and 4AMP). No clear trends can be observed between sample morphology, surface potential, and device performance.

Comparing the stability of different 3D/2D perovskite devices (**Figure 6d-f**), we can observe that overall stability is significantly affected by 3D perovskite used. Differences in stability of 3D perovskites likely occur due to composition differences, namely different proportions of iodide which is readily oxidized, methylammonium which is readily deprotonated, and Cs which cannot be deprotonated. In all cases, devices with DJ capping layers exhibit improved stability compared to those with RP capping layers, as expected since capping layers which are less susceptible to cation vacancy formation present a better barrier to iodide diffusion towards the electrode. The stability of devices with RP and DJ capping layers stored in the dark in inert environment (glovebox) is similar, as shown in **Figure S45**, with ~98% of initial efficiency retained after 1200 h of storage in both cases. This clearly indicates that the stability differences can be attributed to different behavior of RP and DJ capping layers under illumination. Clear deterioration of the electrode attributed to increased ion migration due to increased photooxidative degradation in a device with 2D RP capping layer can be clearly observed from larger increase of I/Ag ratio obtained from XPS measurements of peeled off Ag electrode for fresh and aged devices for the most stable RP (PEA) and DJ (4AMP) capping layers in combination with CsFAMA 3D perovskites (**Table S7**). The obtained results agree with reduced I/Pb ratio in aged devices with 2D RP layer obtained from EDX measurements of perovskite layers after peeling off Ag electrode (**Table S8**). Significant differences in stability of devices with RP and DJ perovskite capping layers under bias are clearly observed in **Figure S46**, consistent with the proposed degradation mechanism. Under forward bias at $J_{sc}$ (**Figure S46a**), which corresponds to significant hole injection into the perovskite, devices with 2RP capping layers rapidly degrade. Similar observation occurs for illumination and reverse bias, which represents a very harsh testing condition (**Figure S46b**).

We can also observe that stability trends between 2 RP and 2DJ perovskites are different for different 3D perovskite compositions, which likely occurs due to differences in excess hole accumulation in different devices, as the energy level alignment across the interfaces is dependent on 3D perovskite used. In CsFAMA-based devices, we also observe a different trend in device stability (BDA > 4AMP), which is different from trend of 2D capping layer stability (4AMP > BDA). This can be explained by differences in charge carrier accumulation, since more efficient charge carrier collection (shorter decay time in TRPL for full device stack, **Table S6**) can be clearly observed for BDA compared to 4AMP. Thus, we can conclude that for optimal device stability it is necessary to choose an intrinsically stable capping layer, and to optimize device architecture to minimize charge accumulation in the devices.

To further verify the observed stability trends, multiple devices were tested (MPPT, in dry air, under 100 $mW/cm^2$ simulated solar illumination) for the most stable 3D perovskite (MA-free perovskite) with BA and with BDA capping layers. Obtained results, shown in **Figure S47a**, are in good agreement with the observed trends in **Figure 6f**. Furthermore, another clear difference can be observed between devices with RP and DJ capping layers. When illuminated under OC condition in dry air (which lead to faster degradation compared to MPPT condition), the electrode damage after prolonged testing in the case of RP capping layer compared to DJ capping layer can also be clearly observed, as shown in **Figure S47b**. This is consistent with the proposed degradation mechanism. We expect that the DJ perovskite will retain the reduced iodide diffusion properties due to fewer spacer cation vacancies, resulting in reduced iodide transport to the electrode and consequently reduced electrode degradation and less transport of silver ions to the counter electrode.

**Table S7.** XPS of Ag electrode before and after MPPT test in dry air for two device architectures considered.

| Device | Element | Atomic (%) | I/Ag (%) |
|---|---|---|---|
| CsFAMA-4AMP fresh | I<br>Ag | 1.03<br>98.97 | 1.04 |
| CsFAMA-4AMP aged | I<br>Ag | 2.88<br>97.12 | 2.97 |
| CsFAMA-BDA fresh | I<br>Ag | 0.7<br>99.3 | 0.70 |
| CsFAMA-BDA aged | I<br>Ag | 3.4<br>96.6 | 3.52 |
| CsFAMA-PEA fresh | I<br>Ag | 0.51<br>99.49 | 0.51 |
| CsFAMA-PEA aged | I<br>Ag | 8.06<br>91.94 | 8.77 |
| CsFAMA-BA fresh | I<br>Ag | 1.3<br>98.7 | 1.32 |
| CsFAMA-BA aged | I<br>Ag | 9.8<br>90.2 | 10.86 |

**Table S8.** Pb and I content in perovskite layer for the two device architectures considered before and after MPPT test in dry air determined by EDX.

| Device | Element | Atomic% | I/Pb |
|---|---|---|---|
| CsFAMA-4AMP fresh | Pb<br>I | 3.42<br>13.69 | 4.0 |
| CsFAMA-4AMP aged | Pb<br>I | 0.66<br>2.31 | 3.5 |
| CsFAMA-BDA fresh | Pb<br>I | 2.82<br>11.51 | 4.1 |
| CsFAMA-BDA aged | Pb<br>I | 2.39<br>8.56 | 3.6 |
| CsFAMA-PEA fresh | Pb<br>I | 2.92<br>11.40 | 3.9 |
| CsFAMA-PEA aged | Pb<br>I | 2.52<br>7.30 | 2.9 |
| CsFAMA-BA fresh | Pb<br>I | 2.12<br>7.48 | 3.5 |
| CsFAMA-BA aged | Pb<br>I | 3.38<br>8.47 | 2.5 |

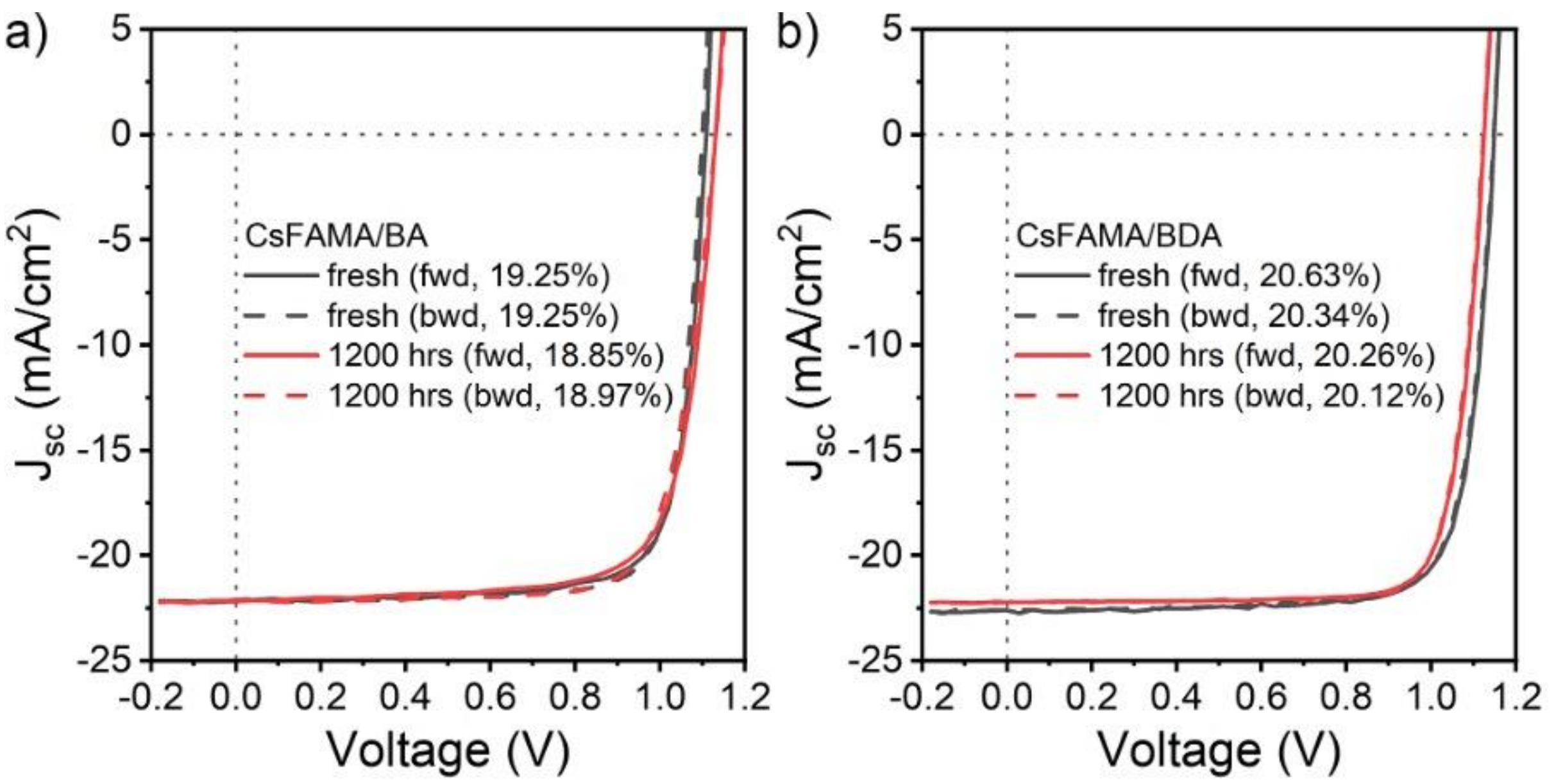

**Figure S45.** J-V curves for CsFAMA 3D perovskite fresh devices and devices stored in glovebox for 1200 h for **a)** BA and **b)** BDA 2D capping layers.

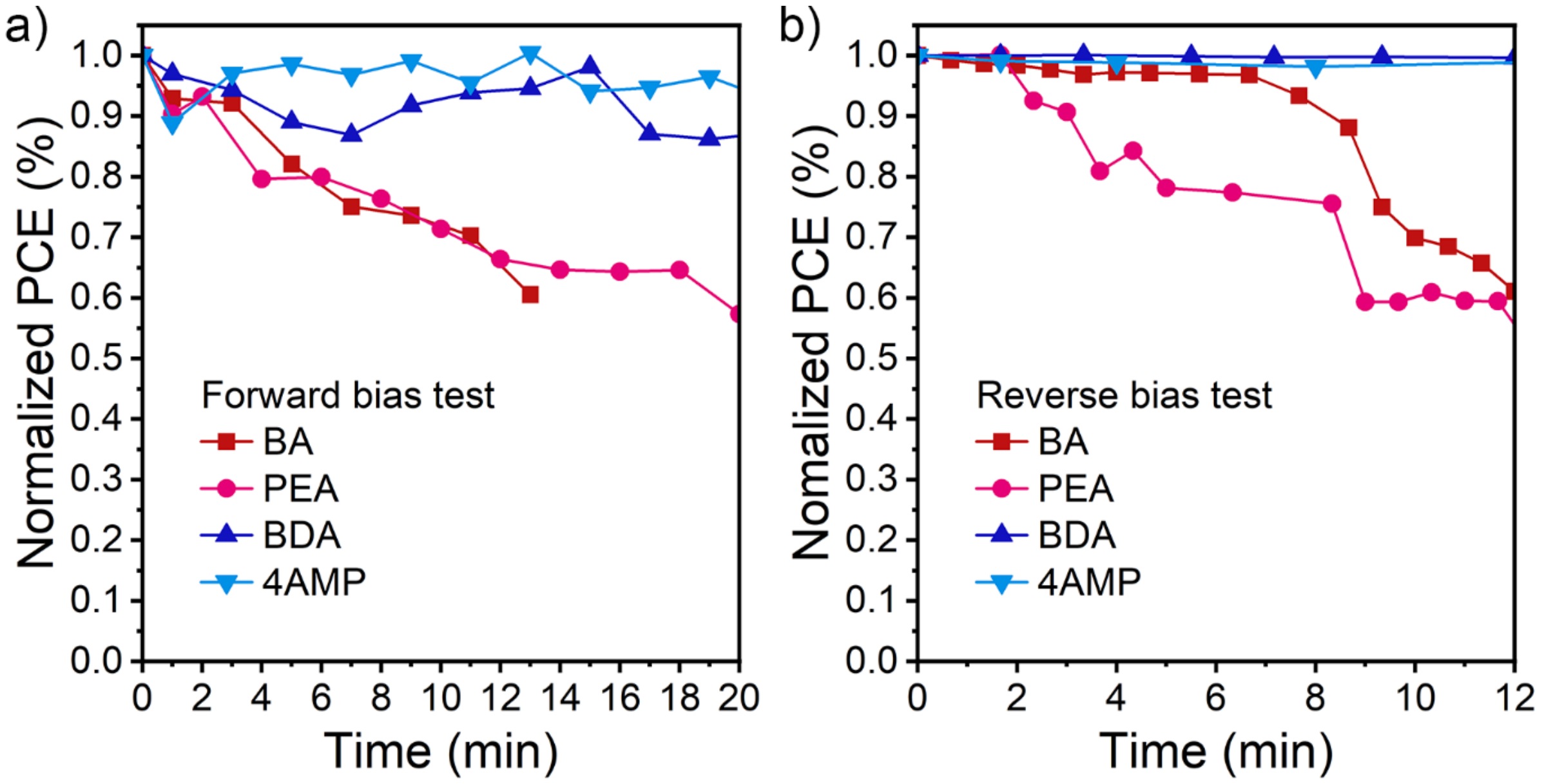

**Figure S46.** Stability of triple cation CsFAMA devices with different 2D capping layer: **a)** under forward bias at $J_{sc}$ without illumination (edge encapsulation used to allow outgassing,[38] devices encapsulated with epoxy and cover glass); **b)** Stability of triple cation CsFAMA devices with different 2D capping layer under reverse bias at –0.5 V under simulated solar illumination (100 mW/cm$^2$).

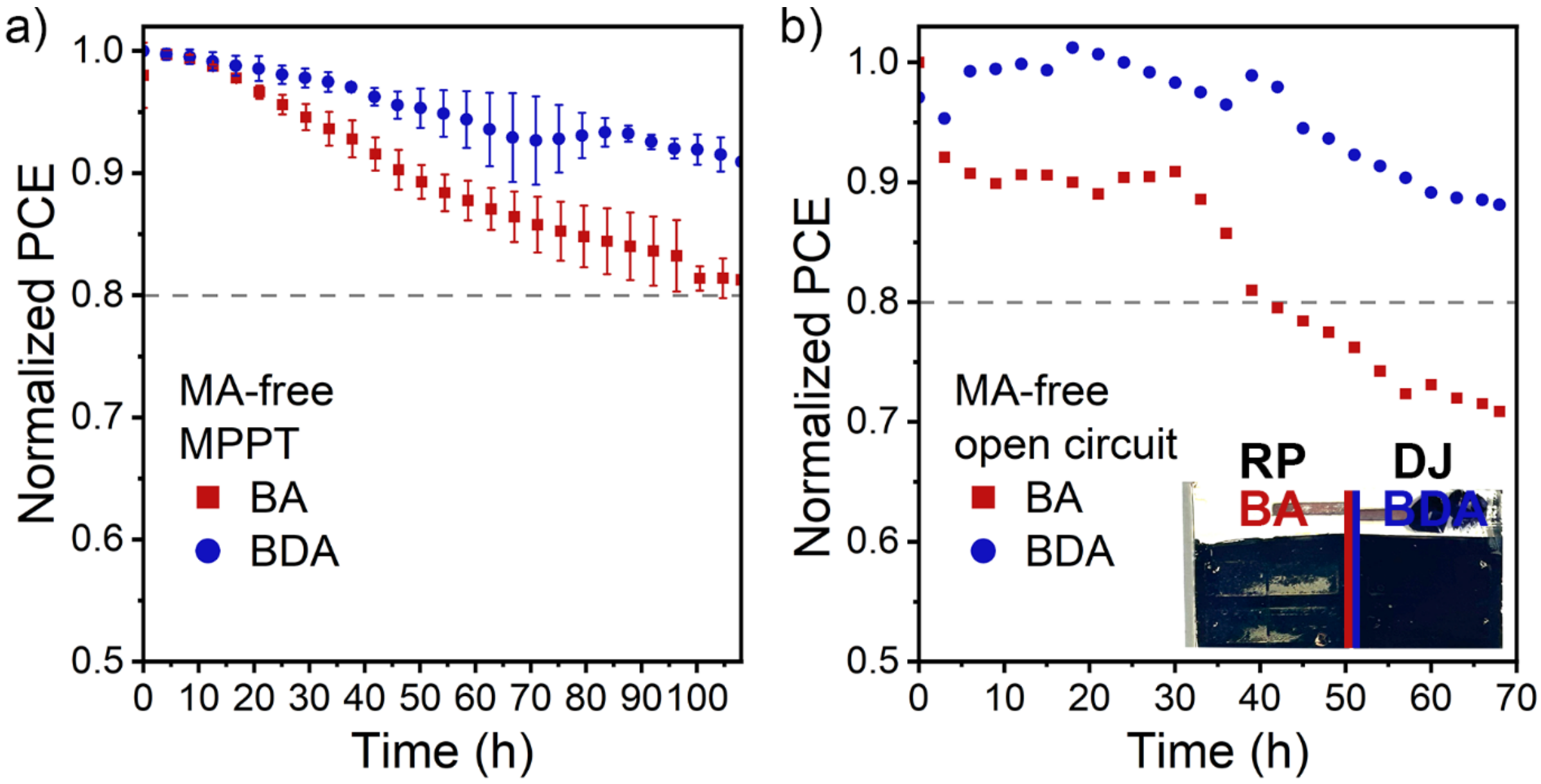

**Figure S47. a)** Normalized PCE as a function of time of simulated solar illumination (100 mW/cm$^2$) for MPPT test of MA-free devices with different 2D capping layers. Average values and error bars (standard deviation) are estimated from 3 devices. **b)** Normalized PCE as a function of time of simulated solar illumination (100 mW/cm$^2$) for OC test of MA-free devices with different 2D capping layers. Photos of the electrodes after 1 week of illumination under OC condition are shown in the inset. Clear diffusion of metal can be observed for the devices with BA 2D layer. Observed faster degradation under OC compared to MPPT condition, as well as migration of the metal to the opposite electrode,[61] is consistent with the expectations for degradation caused by electrochemical redox reactions (**Supplementary Note 2**).

## SUPPLEMENTARY NOTE 11 – *Data supporting proposed mechanisms*

### Direct comparisons of DJ and RP

1. Significant difference in absorption spectra, FTIR spectra (organic cation bands) and XRD patterns after illumination (**Figure 1, Figure 4, Figures S1-S10).** In total, 6 DJ 2D perovskites degrade significantly slower than 6 2D RP perovskites, for both iodides and bromides.
2. No significant expulsion of iodide into solution under illumination (**Figure 2b**) and bias (**Figure 2d**) from DJ perovskites, significant expulsion from RP perovskites.
3. No significant degradation under bias in lateral device geometry for DJ perovskites, significant degradation near positive electrode for RP perovskites (**Figure 5a-d**)
4. Significantly lower (BDA) or negligible (4AMP) outgassing from DJ perovskites under illumination; significant outgassing of corresponding amine and $CH_3I$ from RP perovskites (**Figure 5e, f**).
5. Significant improvements in stability of 3D/2D perovskite solar cells with 2D perovskite layer under illumination or bias (**Figure 6d-f, Figure S46&S47**).
6. Significantly smaller changes in Pb:I ratio after bias and lack of visible changes in the electrode for DJ compared to RP perovskites (**Table S2, Figure S17**).
7. Differences in reduction potentials from C-V scans (**Figure S18**) and small to negligible degradation of DJ perovskites compared to completely degraded RP perovskites after measurement (**Figure S19**).
8. Differences in deprotonation/spacer cation vacancy formation energies (significantly higher for DJ, **Figures S27-29, Tables S3&S4**).
9. Differences in *in situ* Raman spectra of DJ and RP perovskites under illumination (**Figure S32-S34**)

### Additional evidence for the role of holes in degradation of RP perovskites

1. Differences in degradation rates when deposited on HTL and on ETL layers (**Figure 2c, Figures S14-16**)
2. Degradation near positive electrode in lateral bias devices (**Figure 5 a, b**).
3. Iodide expulsion under both bias (**Figure 2d**) and illumination (**Figure 2a**)
4. Film degradation after C-V scan (**Figure S19**).
5. Excess iodine contributing to degradation (**Figure S30**)